\documentclass[12pt]{article}
\pdfoutput=1
\usepackage{amsfonts,amssymb,amsmath,mathtools,graphics,slashed}
\usepackage{hyperref,hep,subfig}
\usepackage{rotating}
\makeatletter

\newcommand{\Rmnum}[1]{\expandafter\@slowromancap\romannumeral #1@}
\makeatother

\newcommand{\nn}{\notag \\}

\newcommand{\Mdef}{X}

\begin{document}

\makeatletter
\renewcommand{\theequation}{\thesection.\arabic{equation}}
\@addtoreset{equation}{section}
\makeatother

\baselineskip 18pt

\begin{titlepage}

\vfill

\begin{flushright}
Imperial/TP/2014/JG/04\\
\end{flushright}

\vfill

\begin{center}
   \baselineskip=16pt
   {\Large\bf The thermoelectric properties of \\inhomogeneous holographic lattices}
  \vskip 1.5cm
  \vskip 1.5cm
      Aristomenis Donos$^1$ and Jerome P. Gauntlett$^2$\\
   \vskip .6cm
    \vskip .6cm
      \begin{small}
      \textit{$^1$DAMTP, 
       University of Cambridge\\ Cambridge, CB3 0WA, U.K.}
        \end{small}\\
      \vskip .6cm
      \begin{small}
      \textit{$^2$Blackett Laboratory, 
        Imperial College\\ London, SW7 2AZ, U.K.}
        \end{small}\\*[.6cm]

\end{center}

\vfill

\begin{center}
\textbf{Abstract}
\end{center}

\begin{quote} 
We consider inhomogeneous, periodic, holographic lattices of $D=4$ Einstein-Maxwell theory.
We show that the DC thermoelectric conductivity matrix can be expressed analytically in
terms of the horizon data of the corresponding black hole solution. We numerically construct such black hole solutions for
lattices consisting of one, two and ten wave-numbers. 
We numerically determine the AC electric conductivity which reveals Drude physics as well as resonances associated with sound modes. No evidence for an intermediate frequency scaling regime is found.
All of the monochromatic lattice black holes that
we have constructed exhibit scaling behaviour 
at low temperatures 
which is consistent with the appearance of
$AdS_2\times\mathbb{R}^2$ in the far IR at $T=0$.
\end{quote}

\vfill

\end{titlepage}
\setcounter{equation}{0}

%%%%%%%%%%%%%%%%%%%%%%%%%%%%%%%%%%%%%%%%%%%%%%%%%%%%%%%%%%%%%%%%%%%%%%%
%\tableofcontents
%%%%%%%%%%%%%%%%%%%%%%%%%%%%%

\section{Introduction}

In considering possible applications of holography to real world systems, the electrical conductivity is an interesting observable to focus on.
Indeed many exotic materials, which are known to be strongly coupled, exhibit striking and poorly understood phenomena. For example,
the strange metallic phase, arising in the cuprates and heavy fermion systems, has a DC resistivity which scales linearly
in temperature, in contrast to ordinary Fermi liquids were it scales quadratically. 

For systems at finite charge density, the
electric and heat currents mix and so one should
consider the thermoelectric conductivity matrix.
The electrically charged AdS-RN black holes of Einstein-Maxwell theory provide a natural starting point to
investigate thermoelectric conductivities using holography.
These black holes describe CFTs at finite charge density with unbroken translation invariance. 
However, the latter implies that momentum is conserved and hence the application of an external electric
field gives rise to infinite thermoelectric DC conductivities.
More precisely, the real part of the AC conductivity has a delta function for all temperatures \cite{Hartnoll:2007ih,Hartnoll:2009sz,Herzog:2009xv}. 
In order to alleviate this feature one needs a mechanism to dissipate momentum. This can be achieved by considering ``holographic lattice" black holes
where one 
explicitly breaks the translation invariance using UV deformations\cite{Horowitz:2012ky,Horowitz:2012gs,Horowitz:2013jaa,Donos:2012js,Ling:2013nxa,Chesler:2013qla,Donos:2013eha,Andrade:2013gsa,Donos:2014uba,Balasubramanian:2013yqa}.

Recently it has been shown how to obtain the thermoelectric DC conductivity in terms of black hole horizon data for 
a general class of homogeneous Q-lattices \cite{Donos:2014uba,Donos:2014cya}. Recall that Q-lattices exploit a global symmetry in the bulk
space-time in order to break translation invariance while maintaining a homogeneous metric \cite{Donos:2013eha}. 
At a technical level this is significant because the holographic black holes can be constructed by solving ODEs rather than PDEs, and this simplification
helped in obtaining the results in \cite{Donos:2014uba,Donos:2014cya}.
The basic strategy of \cite{Donos:2014uba,Donos:2014cya} is to consider linearised perturbations about the black holes with 
sources for the electric and heat currents that are linear in time. By then manipulating the equations of motion to obtain expressions for the electric and heat currents in terms of horizon data, and also demanding regularity of the perturbation
at the black hole horizon, leads to the final result.

In the first part of this paper we show that the techniques\footnote{For other work on the electric DC conductivity for 
various holographic black holes and using different approaches, see \cite{Iqbal:2008by,Davison:2013jba,Blake:2013bqa,Blake:2013owa,Davison:2013txa}\cite{Andrade:2013gsa}\cite{Gouteraux:2014hca,Mefford:2014gia,Taylor:2014tka}. The methods of 
\cite{Donos:2014uba,Donos:2014cya} were recently used to obtain the electric DC conductivity in the presence of a magnetic field \cite{Blake:2014yla}
and for a class of helical lattices \cite{Donos:2014oha}. They were also used
to obtain the thermoelectric DC conductivity in the context of massive gravity \cite{Amoretti:2014mma}.} of \cite{Donos:2014uba,Donos:2014cya} can also be applied in the context of inhomogeneous
holographic latices. More specifically we obtain an analytic result for the thermoelectric DC conductivity for holographic lattices associated with an arbitrary periodic chemical potential depending on one of the spatial coordinates, in the context of $D=4$ Einstein-Maxwell theory. Our final results, which are summarised in section \eqref{sumsec}, are remarkably similar
to those obtained in \cite{Donos:2014uba,Donos:2014cya}. In particular, the electric DC conductivity is naturally written as a sum of two terms, one
of which is precisely the electric conductivity with vanishing heat current and hence can be thought of, loosely, as being associated with 
the evolution of charged particle-hole pairs (possibly pair produced). 
The other term can be thought of as arising
from momentum dissipation processes. 
We also find a result for the ``figure of merit" $ZT$, which provides a measure of the efficiency
of a thermoelectric engine (e.g. see \cite{MRS:7962323}), and show that it can become arbitrarily large at low temperatures (the maximum known value for real materials is less than three.)

We also find in the high temperature limit that the electrical DC conductivity saturates to a constant value, set by the details of the UV deformation, generically with\footnote{Note, by contrast, that this is not the same as the $\omega\to\infty$ limit of the optical conductivity, $\sigma(\omega)$, which approaches unity.}
$\sigma>1$. This saturation of the DC conductivity is reminiscent of the Mott-Ioffe-Regel bound \cite{Gunnarsson:2003zz,takmir} of real metals, but here it
is arising in a strongly coupled setting. Note, by contrast, that the $T\to\infty$ limit of
the electrical conductivity for the Q-lattices diverges, except in the special case that the UV deformation is a marginal
operator as in \cite{Andrade:2013gsa}, for example. 

In the second part of this paper we construct fully back reacted black hole solutions of $D=4$ Einstein-Maxwell theory corresponding to various holographic lattices by numerically solving PDEs. 
We will consider monochromatic lattices with a single wave-number $k$ as well as dichromatic lattices
with wave-numbers $k$ and $2k$ with the same phase. We also consider
an example of a ``dirty lattice" built from many wave-numbers (ten) and random phases.
In the monochromatic case, such black holes 
were first constructed in \cite{Horowitz:2012gs}, building on the pioneering work \cite{Horowitz:2012ky} and while we recover many of the results of that paper, we also find some important differences.
We will calculate the optical conductivity and observe the appearance of Drude-type peaks that are broadly
similar to what was seen in \cite{Horowitz:2012gs}.
We use our AC results to obtain the limiting DC conductivity and we find excellent agreement (better than $10^{-4}\%$)
when we compare with the results using our new analytic formula. 
This provides an excellent test of our numerics and the fit to Drude physics.

A striking claim of \cite{Horowitz:2012ky,Horowitz:2012gs} was the existence of an
intermediate frequency scaling regime for the optical conductivity for various holographic lattices, including the lattices we will construct here.
More precisely, for the monochromatic case, the optical conductivity was reported to have the form 
$|\sigma(\omega)|\sim B\omega^{-2/3}+C$, where $B,C$ are
frequency independent constants within the range $2<\omega \tau<8$, where $\tau$ is the characteristic time scale obtained from the Drude peak.
Since similar behaviour is seen for the high $T_c$ cuprate superconductors, albeit with $C=0$ and a frequency independent phase
(e.g. \cite{2003Natur.425..271M,2006AnPhy.321.1716V}), it is important to further investigate this issue. The experimental data is plotted on a log-log diagram
and, similar looking plots were presented in \cite{Horowitz:2012ky,Horowitz:2012gs}, based on their results for the AC conductivity. While we find some discrepancy with the AC conductivity plots in \cite{Horowitz:2012gs} a more important
point is that if such an intermediate power-law is present it should be manifest using more refined measures. In \cite{Donos:2013eha}, it was suggested that a sharp diagnostic for such intermediate scaling is to plot the quantity
$1+(\omega/\mu)|\sigma|''/|\sigma|'$ and look for a range of $\omega/\mu$ in which
this quantity is constant. Doing this we will find no evidence for such an intermediate scaling regime for the black holes that we construct here.
Indeed, the intermediate behaviour for the optical conductivity is broadly similar to what was seen for 
the homogeneous Q-lattices constructed in \cite{Donos:2013eha}\footnote{An intermediate scaling was also not seen for a different class of lattices in the recent constructions of \cite{Taylor:2014tka}.}.

At very low temperatures, all of the black holes associated with monochromatic lattices that we have constructed appear to approach $AdS_2\times\mathbb{R}^2$ in the far IR.
More precisely, as $T\to 0$ the DC conductivity of the black holes exhibit a scaling behaviour consistent with the
$T=0$ black holes being domain walls interpolating between an irrelevant deformation of
$AdS_2\times\mathbb{R}^2$ in the far IR and $AdS_4$ in the UV, as first envisaged by \cite{Hartnoll:2012rj}.
We find no evidence for the new ``floppy" ground states that were discussed in \cite{Hartnoll:2014gaa}; it is logically possible that they
appear at even lower temperatures than what we have considered, but the robustness of the scaling behaviour makes this
seem unlikely to us. It is an open possibility whether stronger lattice deformations than we have constructed and/or different types of lattice deformations
 will lead to 
a transition to new IR behaviour, as in the metal-insulator transitions of \cite{Donos:2012js,Donos:2013eha,Donos:2014uba} or the metal-metal transitions of \cite{Donos:2014uba}.

The plan of the rest of the paper is as follows. In section 2 we describe the inhomogeneous lattice black holes 
of Einstein-Maxwell theory that we will be considering. 
The derivation of the thermoelectric DC conductivity is presented in section 3. For readers who are just interested in the final analytic results, 
we point them to sections \ref{sumsec} and the subsequent discussion in section \ref{highlow}.
In section 4, 
following the approach of \cite{Headrick:2009pv},
we describe the numerical methodology that we employ to solve the PDEs which leads to the holographic lattice black holes. We also explain how we obtain the AC conductivity. The main results of our numerical
constructions are presented in section 4.3. 
We show that the electrical conductivity satisfies a standard type of sum rule, following \cite{Gulotta:2010cu}, and 
a second sum rule which is associated with the electromagnetic duality of the $D=4$ Einstein-Maxwell theory
\cite{WitczakKrempa:2012gn}.
We briefly conclude in section 5. 
We have three appendices. In appendix \ref{stressheat} we discuss the derivation of the 
stress tensor and heat current, while appendix \ref{convgtest} describes some aspects of the implementation of our numerics
as well as some of the convergence checks that we used. In appendix \ref{floppy} we make some additional comments 
on the relation of our work to that of \cite{Hartnoll:2014gaa}.

\section{Inhomogeneous lattices}
We will focus on Einstein-Maxwell theory in four bulk dimensions, which is a minimal and rather universal setting to study holographic lattices.
In particular, it can be obtained as a consistent Kaluza-Klein truncation associated with an arbitrary $AdS_4\times M$
solution of string/M-theory, where $M$ is a compact manifold with an isometry. 
An interesting class of examples is provided by the infinite class of $AdS_4\times SE_7$ solutions, 
where $SE_7$ is a seven-dimensional Sasaki-Einstein space, dual
to CFTs with $N=2$ supersymmetry in $d=3$ space-time dimensions \cite{Gauntlett:2007ma}.

The action is given by
\begin{align}\label{eq:bulk_action}
S=\int d^4 x \sqrt{-g}\,\left(R+6-\frac{1}{4}\,F^{2} \right)\,,
\end{align}
with $F=dA$ being the field strength of the gauge field $A$ and $F^2=F_{\mu\nu}F^{\mu\nu}$. 
The equations of motion can be written in the form
\begin{align}\label{eq:eom}
 E_{\mu\nu}\equiv R_{\mu\nu}+{3}g_{\mu\nu}-\frac{1}{2}\left(
F_{\mu\rho}F_{\nu}{}^{\rho}- \frac{1}{4}g_{\mu\nu}\,F^2\right)&=0\,,\nn
 \nabla_{\mu}F^{\mu\nu}&=0\,.
\end{align}
Note that we have chosen the cosmological constant
so that a unit radius $AdS_4$ solves the equations of motion.
We have also set $16\pi G=1$ in order not to clutter up various equations.

The electrically charged AdS Reissner-Nordstr\"om (AdS-RN) black brane solution solves the equations of motion and 
is the bulk dual of a CFT held at temperature $T$ and deformed by a constant chemical potential $\mu$. 
Recall that at $T=0$ the solution interpolates between $AdS_4$ in the UV and the electrically charged $AdS_2\times\mathbb{R}^2$ solution in the IR.
The AdS-RN black hole preserves translation invariance and hence there is no mechanism for momentum to dissipate upon adding an external electric field. 
This gives rise to infinite DC conductivity, or more precisely a delta function in the optical conductivity at zero frequency.
This feature can be eliminated
by studying more general black holes in which the
chemical potential has a periodic dependence on one of the spatial dimensions, $x$, 
with period $L$.
%:  $\mu\left(x\right)=\mu\left(x+L\right)$. 
We can write 
\begin{align}\label{genmu}
\mu\left(x\right)=\mu_{0}+\bar\mu\left(x\right)\,,
\end{align}
with $\mu_0$ a constant, and $\bar \mu\left(x\right)=\bar \mu\left(x+L\right)$ is a periodic function which averages 
to zero over a period. Note that when $\mu_{0}\ne 0$, a simple scaling argument reveals
that true UV parameters are $T/\mu_{0}$ combined with the function $\bar\mu\left(x/\mu_{0} \right)/\mu_{0}$ with period $L\,\mu_{0}$. We also note that in the figures that appear later in
the paper we have dropped the subscript from $\mu_0$ for clarity.

Some special examples of these holographic lattice black holes have been studied previously, for the special case of monochromatic sources.
Specifically, black holes associated with deformations of the form $\mu=\mu_0+V \cos(k x)$ were constructed
for $\mu_{0}\ne 0$ in \cite{Horowitz:2012gs} (and will be reconstructed here in section \ref{sec:background}) 
and for $\mu_{0}= 0$ in \cite{Chesler:2013qla}. In the $T=0$ limit the black holes with $\mu_{0}\ne 0$ in \cite{Horowitz:2012gs} approach
$AdS_2\times\mathbb{R}^2$ in the far IR, perturbed by irrelevant deformations, and we will find the same feature here (in contrast to the more recent
claims of \cite{Hartnoll:2014gaa}),
while those with $\mu_{0}= 0$ that were constructed\footnote{Note that we have also constructed some black holes with $\mu_0=0$ numerically, as well as
calculated the optical conductivity, and our results are in agreement with \cite{Chesler:2013qla}.} in \cite{Chesler:2013qla} approach $AdS_4$.

Our new analytic results for the DC thermoelectric conductivity, described in the next subsections, will be valid for an arbitrary periodic
chemical potential deformation
of the form \eqref{genmu}. An ansatz that is general enough to cover the relevant black hole solutions of interest is given by
\begin{align}\label{eq:DC_ansatz}
ds^{2}&=-U\,H_{tt}\,dt^{2}+\frac{H_{rr}}{U}\,dr^{2}+\Sigma\,\left[e^{B} dx^{2}+e^{-B}\,dy^{2} \right]\,, \nn
A&=a_{t}\,dt\,,
\end{align}
where $U=U(r)$, while $H_{tt}, H_{rr},\Sigma,e^{B}$ and $a_t$ are all functions of both $r$ and $x$.

The boundary conditions at the asymptotic $AdS_4$ boundary, which we take to be located at $r\to \infty$, are given by
$U,\Sigma\to r^2$, $H_{tt}, H_{rr},e^{B} \to 1$ and $a_t\to \mu\left(x\right)$ as in \eqref{genmu}. The black hole 
horizon is taken to be located at $r=0$ and regularity of the solution implies that we can expand the functions in powers of $r$ as
\begin{align}\label{nhexpbh}
U\left(r\right)&=4\pi\,T\,r+U^{(2)}\left(x\right)r^2+\dots\,,\nn%+\cdots\ ,\nn
a_{t}(r,x)&=r\left(a^{(0)}_{t}\left(x\right)+a^{(1)}_{t}\left(x\right)r+\dots\right)\,,\nn
H_{tt}(r,x)&=H^{(0)}_{tt}\left(x\right)+H_{tt}^{(1)}\left(x\right)r+\dots\,,\nn
H_{rr}(r,x)&=H^{(0)}_{tt}\left(x\right)+H_{rr}^{(1)}\left(x\right)r+\dots\,,\nn
\Sigma(r,x)&=\Sigma^{(0)}\left(x\right)+\Sigma^{(1)}\left(x\right)r+\dots\,,\nn
B(r,x)&=B^{(0)}\left(x\right)+B^{(1)}\left(x\right)r+\dots\,,
%&a_{t}=a^{(0)}_{t}\left(x\right)\,r+a^{(1)}_{t}\left(x\right)\,r^{2}+\cdots
\end{align}
where $a^{(0)}_{t},H^{(0)}_{tt},\Sigma^{(0)}$ and $B^{(0)}$ are all periodic functions of $x$, as are the higher order terms in the expansion in $r$.
Indeed, regularity of the solutions as $r\to 0$ is easily seen by replacing the $t$ coordinate with the in-going 
Eddington-Finklestein coordinate $v$ defined
by
\begin{align}\label{iefv}
v=t+(4\pi T)^{-1}\ln r+\mathcal{O}(r)\,.
\end{align}

The current density $J^a\equiv \{J^t,J^x,J^y\}$ in the dual field theory takes the form
\begin{align}\label{defcurrent}
J^a=\sqrt{-g} F^{ar}\,,
\end{align}
where the right hand-side is evaluated at the boundary $r\to\infty$. With this definition $J^a$ has a finite limit as $r\to\infty$ (see the discussion
in appendix \ref{stressheat}).
The total constant charge, $q$, of the background black holes is given by $q\equiv \int J^t$, where
we have introduced the notation 
\begin{align}\label{intdef}
\int \qquad \leftrightarrow\qquad L^{-1}\,\int_{0}^{L}dx\,,
\end{align} with $L$ the period of $x$. We can obtain an expression for $q$ in terms of horizon data by using the gauge-equations of motion. Indeed
the only non-zero component of the gauge-field equation of motion is the $t$ component which we can write as
$\sqrt{-g}\nabla_\mu F^{\mu t}=\partial_r(\sqrt{-g} F^{r t})+\partial_x(\sqrt{-g} F^{x t})=0$.
Since $\sqrt{-g} F^{x t}$ depends on $\partial_x a_t$, after integrating over a
period of $x$ the second-term vanishes and we deduce that
\begin{align}
q&=\int \frac{\Sigma\partial_r a_t}{(H_{rr}H_{tt})^{1/2}}\,,\nn
&=\int\frac{\Sigma^{(0)}a^{(0)}_t}{H_{tt}^{(0)}}\,,
\end{align}
where the second line follows by evaluating the constant at the horizon.

\section{The thermoelectric DC conductivity}
\subsection{Calculating $\sigma$ and $\bar\alpha$}
In this subsection we calculate the DC conductivities associated with switching on 
a constant electric field on the boundary theory in the $x$ direction, the
direction in which the background lattice breaks translational invariance. Recall, by definition,
that the linear response is given by
\begin{align}
J=\sigma E,\qquad Q =\bar\alpha T E\,,
\end{align}
where $J\equiv J^x$ is the electric current and $Q\equiv T^{tx}-\mu J$ is the heat current, both in the $x$ direction,
and $\sigma$, $\bar\alpha$ are the electric and thermoelectric DC conductivities. We will show how $\sigma$, $\bar\alpha$
can be expressed in terms of horizon data of the unperturbed black hole.

We first introduce gauge field perturbations of the form 
\begin{align}\label{eq:gauge_pansatz}
\delta A=\delta a_\mu(r,x) dx^\mu-E\,t\,dx\,,%=\delta a_{t}\,dt+\delta a_{r}\,dr+\delta a_{x}\,dx-E\,t\,dx
\end{align}
where $E$ is the constant magnitude of the linearised electric field in the $x$ direction
and $\delta a_\mu$, whose non-vanishing components lie in the set $\left\{\delta a_{t},\delta a_{r}, \delta a_{x}\right\}$,
are functions of $r,x$ and are periodic in $x$. 
This is
supplemented with metric perturbations $\delta g_{\mu\nu}$, with non-vanishing components lying in the set
$\left\{\delta g_{tt}, \delta g_{tr},\delta g_{rr},\delta g_{rx}, \delta g_{xx}, \delta g_{tx}, \delta g_{yy}\right\}$, which are again functions of $r,x$ and again are
periodic in $x$. 
It will be convenient to not fully fix our gauge and coordinate dependence apart from requiring that some components
fade sufficiently fast close to the $AdS_{4}$ boundary.

The next step is to use the equation of motion for the gauge-field to show that $J$ is 
constant and moreover to obtain an expression in terms of horizon data.
Specifically, the $r$ and the $x$ components of the gauge field equation of motion imply that $\partial_x(\sqrt{-g}F^{xr})=0$ and $\partial_r(\sqrt{-g}F^{rx})=0$,
respectively,
and hence $ J=\sqrt{-g}F^{xr}$ is a constant. Thus, we can write
\begin{align}\label{jex}
 J=\frac{e^{-B}}{\sqrt{H_{rr}H_{tt}}}\,\left[  \partial_{x}a_{t}\delta g_{tr}- \partial_{r}a_{t}\delta g_{tx}+ H_{tt} U \left( \partial_{x}\delta a_{r}-\partial_{r}\delta a_{x}\right)\right]\,,
\end{align}
where the right hand side can be evaluated at any value of $r$, including at the black hole horizon.

The next key step is to obtain a similar result for the heat current in the $x$ direction, $Q$, induced by $E$. Following
\cite{Donos:2014cya} we first observe that 
if $k$ is any Killing vector satisfying $L_k F=0$, then we can define a two-form $G$ by
 \begin{align}
 G^{\mu\nu}=\nabla^\mu k^\nu+
\frac{1}{2} k^{[\mu}F^{\nu]\sigma}A_\sigma+\frac{1}{4}({\psi}-2\theta)F^{\mu\nu}\,,
 \end{align}
 where $\psi$ and $\theta$ are defined by $L_k A=d\psi$ and $i_k F=d\theta$. The two-form $G$ has the important 
 property that 
\begin{align}\label{divG}
 \nabla_\mu G^{\mu\nu}={3} k^\nu\,,
 \end{align}
when the equations of motion \eqref{eq:eom} are satisfied (see appendix B of \cite{Donos:2014cya}).
Focussing on the Killing vector $k=\partial_t$, if we consider the $r$ and $x$ components of \eqref{divG}
we deduce that $\partial_x(\sqrt{-g}G^{xr})=\partial_r(\sqrt{-g}G^{rx})=0$
and hence that $\sqrt{-g}G^{rx}$ is a constant. Choosing $\theta=-EX-a_t-\delta a_t$ and $\psi=-Ex$, we conclude that at linearised order we can write 
\begin{align}\label{pexp}
Q&\equiv 2\sqrt{-g}G^{rx}\,,\nn&
=2\sqrt{-g}\nabla^{r} k^x+a_t\sqrt{-g}F^{rx}\,,\nn
&=\frac{e^{-B}U^2  H_{tt}^{3/2}}{\sqrt{H_{rr}}}\,\Bigg[ 
\partial_r\left(\frac{\delta g_{tx}}{U  H_{tt}}\right)  
-\partial_x\left(\frac{\delta g_{tr}}{U  H_{tt}}\right) 
%UH_{tt}(\partial_x\delta g_{tr}-\partial_{r}\delta g_{tx})+\partial_r(UH_{tt})\delta g_{tx}
%-\partial_x(U  H_{tt})\delta g_{tr}\Bigg]
\Bigg]- a_t J\,,
%&=\frac{e^{-B}}{2\sqrt{H_{rr}H_{tt}}}\,\Bigg[ 
%UH_{tt}(\partial_x\delta g_{tr}-\partial_{r}\delta g_{tx})+\partial_r(UH_{tt}-(a_t)^2)\delta g_{tx}\nn
%&-\partial_x(U  H_{tt}-(a_t)^2)\delta g_{tr}+2 UH_{tt}a_t(\partial_x\delta a_{r}-\partial_{r}\delta a_{x})
%\Bigg]
\end{align}
and we can evaluate the right hand side at any value of $r$.
In particular, if we evaluate at the boundary $r\to\infty$
we find, as we explain in appendix \ref{stressheat},
\begin{align}
Q=(T^{tx}-\mu J)\,.
\end{align}

To proceed we now need to ensure that the perturbation is regular at the horizon, after switching
to the Eddington-Finklestein coordinate $v$ given in \eqref{iefv}.
Near $r=0$ we demand that we can expand
\begin{align}\label{eq:nh_exp}
\delta g_{tt}&=U\left(r \right)\,\left(\delta g^{(0)}_{tt}\left(x\right)+{\cal O}(r) \right),\quad
\delta g_{rr}=\frac{1}{U}\,\left( \delta g_{rr}^{(0)}\left(x\right)+{\cal O}(r)\right),\nn
\delta g_{tr}&=\delta g_{tr}^{(0)}\left(x\right)+{\cal O}(r),
\quad
\delta g_{xx}=\delta g_{xx}^{(0)}\left(x\right)+{\cal O}(r),\quad \delta g_{yy}=\delta g_{yy}^{(0)}\left(x\right)+{\cal O}(r)\,,\nn
\delta g_{tx}&=e^{B^{(0)}}\,\left(\delta g_{tx}^{(0)}\left(x\right)+{\cal O}(r)\right),\quad
\delta g_{rx}=\frac{e^{B^{(0)}\left(x\right)}}{U}\,\left( \delta g_{rx}^{(0)}\left(x\right)+{\cal O}(r) \right)\,,\nn
\delta a_{t}&=\delta a_{t}^{(0)}\left(x\right)+{\cal O}(r),\quad
\delta a_{r}=\frac{1}{U}\,\left(\delta a_{r}^{(0)}\left(x\right)+{\cal O}(r)\right)\,,\nn
\delta a_{x}&=\ln{U}\, \delta a_{x}^{(0)}\left(x\right)+{\cal O}(r)\,,
\end{align}
with the following constraints on the leading functions of $x$: 
\begin{align}\label{eq:nh_constr}
&\delta g_{tt}^{(0)}+\delta g_{rr}^{(0)}-2\,\delta g_{rt}^{(0)}=0,\qquad \delta g^{(0)}_{rx}=\delta g^{(0)}_{tx}\,,\nn
&\delta a_{r}^{(0)}=\delta a_{t}^{(0)}, \qquad\delta a_{x}^{(0)}=- \frac{E}{4\pi\,T}\,.
\end{align}
Observe, in particular, that the expression for $\delta a_{x}^{(0)}$ involving the UV data $E$ arises as a direct
consequence of the way in which we switched on the background electric field in \eqref{eq:gauge_pansatz}. 

Expanding out the right hand side of \eqref{jex} at the black hole horizon we find
that at leading order in $r$ we must have
\begin{align}\label{jexph}
%&\partial_{x}\delta a^{(0)}_{t}=-E-e^{B^{(0)}}\,J+\frac{e^{B^{(0)}}\,a^{(0)}_{t}}{2H^{(0)}_{tt}}\delta g^{(0)}_{tx}\\
&J=e^{-B^{(0)}}\left(E+\partial_{x}\delta a^{(0)}_{t}\right)-\frac{\,a^{(0)}_{t}}{H^{(0)}_{tt}}\delta g^{(0)}_{tx}\,,
\end{align}
where the right hand side must be a constant.
We can also evaluate the right hand side of the expression for $Q$ in \eqref{pexp} at the horizon. At leading order in $r$ we deduce that
\begin{align}
&Q=-4\pi T \delta g_{tx}^{(0)}\,,
\end{align}
and hence we obtain the important condition
\begin{align}\label{constcons}
\delta g_{tx}^{(0)}=constant\,.
\end{align}
By expanding to next order in $r$, at fixed temperature, we obtain another constraint on the horizon boundary data: 
\begin{align}\label{eq:nh_derconstraints1}
%&\partial_{x}\delta g_{tx}^{(0)}=0\nn
&\partial_{x}\,\left(4 \pi T\,\frac{\delta g_{tr}^{(0)}}{H_{tt}^{(0)}} \right)+\frac{a_{t}^{(0)}}{H_{tt}^{(0)}}\,\left(E+ \partial_{x}\delta a^{(0)}_{t}\right)
\nn
&+\frac{\delta g_{tx}^{(0)}   }{(H_{tt}^{(0)})^2}\left( 
(a_t^{(0)})^2+2\pi T\left(H_{rr}^{(1)}+2H_{tt}^{(0)}B^{(1)}-3H_{tt}^{(1)}\right)-2H_{tt}^{(0)}U^{(2)}
\right) =0\,.
\end{align}
Remarkably, using the background equations of motion we can rewrite this in the following useful form
\begin{align}\label{eq:nh_derconstraints}
%&\partial_{x}\delta g_{tx}^{(0)}=0\nn
&\partial_{x}\,\left(4 \pi T\,\frac{\delta g_{tr}^{(0)}}{H_{tt}^{(0)}} 
-\frac{1}{\Sigma^{(0)}} 
\partial_x\left[  B^{(0)}-\ln(   H_{tt}^{(0)}  \Sigma^{(0)})         \right]     \delta g_{tx}^{(0)}
\right)\nn
&\qquad\qquad +\frac{a_{t}^{(0)}}{H_{tt}^{(0)}}\,\left(E+ \partial_{x}\delta a^{(0)}_{t}\right)
+\frac{1}{\Sigma^{(0)}}\,\left[\partial_{x}\ln{\frac{e^{B^{(0)}}}{\Sigma^{(0)}}} \right]^{2}\delta g_{tx}^{(0)}\,=0\,.
\end{align}

The constraints \eqref{eq:nh_constr}, \eqref{jexph}, \eqref{constcons} and \eqref{eq:nh_derconstraints} %and \eqref{eq:nh_derconstraints}
are sufficient to get a consistent set of ODEs for the expansion parameters in the falloff \eqref{eq:nh_exp}.
In particular, by expanding the right hand side of \eqref{jex} and \eqref{pexp} in higher powers of $r$ at the black hole horizon 
do not lead to additional constraints.

We have now assembled the ingredients to obtain the DC conductivities $\sigma$ and $\bar\alpha$. We 
multiply equation \eqref{jexph} by $e^{B^{(0)}}$ and then integrate over a period of $x$ to obtain an equation involving $E$, $J$ and $\delta g_{tx}^{(0)}$. 
Equation \eqref{jexph} can also be used in \eqref{eq:nh_derconstraints} which we then integrate to give a relation between $J$ and $\delta g_{tx}^{(0)}$. 
We can solve for $J$ in terms of $E$, and hence obtain an expression for $\sigma=J/E$.
We find
\begin{align}\label{eq:DC_analytic}
\sigma
&=\frac{\int \left(e^{B^{(0)}}\left(\frac{a_{t}^{(0)}}{H_{tt}^{(0)}}\right)^{2}+ \frac{1}{\Sigma^{(0)}} \left[\partial_{x}\ln{\frac{e^{B^{(0)}}}{\Sigma^{(0)}}} \right]^{2}\right)}{\int e^{B^{(0)}} \int \left(e^{B^{(0)}}\left(\frac{a_{t}^{(0)}}{H_{tt}^{(0)}}\right)^{2}+\frac{1}{\Sigma^{(0)}} \left[\partial_{x}\ln{\frac{e^{B^{(0)}}}{\Sigma^{(0)}}} \right]^{2}\right) - \left(\int e^{B^{(0)}}\frac{a_{t}^{(0)}}{H_{tt}^{(0)}}\right)^{2}}\,,
\end{align}
where we remind the reader that the notation $\int$ means $L^{-1}\,\int_{0}^{L}dx$, with $L$ the period of $x$. 
As advertised this formula for $\sigma$ only depends on the near horizon data
of the unperturbed black hole. The Schwarz inequality implies that
\begin{align}\label{schineq}
\left( \int e^{B^{(0)}}\frac{a_{t}^{(0)}}{H_{tt}^{(0)}}\right)^{2}\leq  \int e^{B^{(0)}}\,\int e^{B^{(0)}}\,\left(\frac{a_{t}^{(0)}}{H_{tt}^{(0)}}\right)^{2}\,,
\end{align}
and hence we deduce that $\sigma>0$.

We also find an expression relating $Q=-4\pi T \delta g_{tx}^{(0)}$ and $E$ and we deduce that $\bar\alpha\equiv \frac{Q}{ T E}$ can be written
as
\begin{align}\label{albar}
\bar\alpha=
\frac{4\pi\int e^{B^{(0)}}\frac{a_{t}^{(0)}}{H_{tt}^{(0)}}}{\int e^{B^{(0)}} \int \left(e^{B^{(0)}}\left(\frac{a_{t}^{(0)}}{H_{tt}^{(0)}}\right)^{2}+\frac{1}{\Sigma^{(0)}} \left[\partial_{x}\ln{\frac{e^{B^{(0)}}}{\Sigma^{(0)}}} \right]^{2}\right) - \left(\int e^{B^{(0)}}\frac{a_{t}^{(0)}}{H_{tt}^{(0)}}\right)^{2}}\,.
\end{align}

\subsection{Calculating $\bar\kappa$ and $\alpha$}

In this section we will introduce a source for the heat current. Following \cite{Donos:2014cya} 
we consider the following time dependent perturbation around the 
background \eqref{eq:DC_ansatz}:
\begin{align}\label{eq:thermal_pert_ansatz}
\delta ds^{2}&=\delta g_{\mu\nu}\,dx^{\mu}\,dx^{\nu}-2t(UH_{tt}\zeta)\,dt\,dx\notag\,,\\
\delta A&=\delta a_{\mu}\,dx^{\mu}+t(a_{t}\,\zeta)dx\,.
\end{align}
The non-zero static perturbations are in the sets
$\left\{\delta g_{tt}, \delta g_{tr},\delta g_{rr},\delta g_{rx}, \delta g_{xx}, \delta g_{tx}, \delta g_{yy}\right\}$ and $\left\{\delta a_{t},\delta a_{r}, \delta a_{x}\right\}$ and they depend on $r$ and periodically on $x$. It is important to emphasise that the terms that are linear in $t$, parametrised by $\zeta$,
have been chosen so that all time dependence drops out after we substitute into the equations of motion. As discussed in \cite{Donos:2014cya} 
they provide a source for the heat current.

The near horizon expansion for the perturbation is very similar to \eqref{eq:nh_exp}
\begin{align}\label{eq:nh_exp_2}
\delta g_{tt}&=U\left(r \right)\,\left(\delta g^{(0)}_{tt}\left(x\right)+{\cal O}(r) \right),\quad
\delta g_{rr}=\frac{1}{U}\,\left( \delta g_{rr}^{(0)}\left(x\right)+{\cal O}(r)\right),\nn
\delta g_{tr}&=\delta g_{tr}^{(0)}\left(x\right)+{\cal O}(r),
\quad
\delta g_{xx}=\delta g_{xx}^{(0)}\left(x\right)+{\cal O}(r),\quad \delta g_{yy}=\delta g_{yy}^{(0)}\left(x\right)+{\cal O}(r)\,,\nn
\delta g_{tx}&=e^{B^{(0)}}\,\left(\delta g_{tx}^{(0)}\left(x\right)+\delta  g_{tx}^{(l)}\left(x\right)\,U\,\ln U+{\cal O}(r)\right),\quad
\delta g_{rx}=\frac{e^{B^{(0)}\left(x\right)}}{U}\,\left( \delta g_{rx}^{(0)}\left(x\right)+{\cal O}(r) \right)\,,\nn
\delta a_{t}&=\delta a_{t}^{(0)}\left(x\right)+{\cal O}(r),\quad
\delta a_{r}=\frac{1}{U}\,\left(\delta a_{r}^{(0)}\left(x\right)+{\cal O}(r)\right)\,,\nn
\delta a_{x}&= \delta a_{x}^{(0)}\left(x\right)+{\cal O}(r)\,,
\end{align}
where once again regular in-falling boundary conditions require 
\begin{align}\label{eq:nh_constr_2}
&\delta g_{tt}^{(0)}+\delta g_{rr}^{(0)}-2\,\delta g_{rt}^{(0)}=0,\qquad \delta g^{(0)}_{rx}=\delta g^{(0)}_{tx}, \qquad \delta a_{r}^{(0)}=\delta a_{t}^{(0)}\,.
\end{align}
The extra logarithmic term appearing in the expansion of $\delta g_{tx}$ in \eqref{eq:nh_exp_2}, when compared to \eqref{eq:nh_exp}, is fixed by expanding the equations of motion of the fluctuations near the horizon at $r=0$. More specifically we find
\begin{align}
\delta g_{tx}^{(l)}=-\frac{e^{-B^{(0)}}}{4\pi\,T}\,H_{tt}^{(0)}\,\zeta\,.
\end{align}
This is precisely of the form needed to make the perturbation regular after combining with the time dependent term in \eqref{eq:thermal_pert_ansatz}.

Once again we find that $J$, given by \eqref{jex}, is a constant.
Furthermore, expanding the equations of motion close to the horizon we find once more that
\begin{align}
\delta g_{tx}^{(0)}=constant\,,
\end{align}
\begin{align}
&J=e^{-B^{(0)}}\partial_{x}\delta a^{(0)}_{t}-\frac{\,a^{(0)}_{t}}{H^{(0)}_{tt}}\delta g^{(0)}_{tx}\,,
\end{align}
and
\begin{align}
&\partial_{x}\,\left(4 \pi T\,\frac{\delta g_{tr}^{(0)}}{H_{tt}^{(0)}} 
-\frac{1}{\Sigma^{(0)}} 
\partial_x\left[  B^{(0)}-\ln(   H_{tt}^{(0)}  \Sigma^{(0)})         \right]     \delta g_{tx}^{(0)}
\right)\nn
&\qquad\qquad +\frac{a_{t}^{(0)}}{H_{tt}^{(0)}}\,\partial_{x}\delta a^{(0)}_{t}
+\frac{1}{\Sigma^{(0)}}\,\left[\partial_{x}\ln{\frac{e^{B^{(0)}}}{\Sigma^{(0)}}} \right]^{2}\delta g_{tx}^{(0)}+4\pi T\,\zeta=0\,.
\end{align}
The expression for $Q$ given in \eqref{pexp} is again a constant and expanding near the horizon
we have, as before,
\begin{align}
&Q=-4\pi T \delta g_{tx}^{(0)}\,.
\end{align}

As in \cite{Donos:2014cya} we find that the heat current has a time-independent piece, given by $Q$, and a time-dependent piece:
\begin{align}
T^{tx}-\mu J^{x}=Q-\zeta t T^{xx}\,,
\end{align}
as we discuss in appendix \ref{stressheat}.
As explained in appendix C of \cite{Donos:2014cya}, the time-dependent piece is associated with a static susceptibility for the $Q$$Q$ correlator, given by $T^{xx}$ of the background\footnote{The absence of analogous time-dependent pieces in $J$ in this sub-section and in both $Q$ and $J$ in
the last sub-section, imply that the static susceptibilities for the $J$$Q$ correlator and the $J$$J$ correlator vanish.}. On the other hand
the time independent piece is associated with the DC conductivity.

Using almost identical manipulations of the previous section we 
deduce the DC conductivities:
\begin{align}
\alpha\equiv\frac{J}{T\zeta}=\frac{4\pi\int e^{B^{(0)}}\frac{a_{t}^{(0)}}{H_{tt}^{(0)}}}{\int e^{B^{(0)}} \int \left(e^{B^{(0)}}\left(\frac{a_{t}^{(0)}}{H_{tt}^{(0)}}\right)^{2}+\frac{1}{\Sigma^{(0)}} \left[\partial_{x}\ln{\frac{e^{B^{(0)}}}{\Sigma^{(0)}}} \right]^{2}\right) - \left(\int e^{B^{(0)}}\frac{a_{t}^{(0)}}{H_{tt}^{(0)}}\right)^{2}}\,,
\end{align}
and
\begin{align}
\bar\kappa\equiv\,\frac{Q}{T\zeta}=\frac{(4\pi)^2 T \int e^{B^{(0)}} }{\int e^{B^{(0)}} \int \left(e^{B^{(0)}}\left(\frac{a_{t}^{(0)}}{H_{tt}^{(0)}}\right)^{2}+\frac{1}{\Sigma^{(0)}} \left[\partial_{x}\ln{\frac{e^{B^{(0)}}}{\Sigma^{(0)}}} \right]^{2}\right) - \left(\int e^{B^{(0)}}\frac{a_{t}^{(0)}}{H_{tt}^{(0)}}\right)^{2}}\,.
\end{align}
Comparing with \eqref{albar}, it is satisfying that we have $\alpha=\bar\alpha$. Indeed this 
is expected since the lattice deformation does not break time-reversal 
invariance.

\subsection{Summary of DC conductivity}\label{sumsec}
We now summarise the results of the previous two subsections. To do so it is helpful to
define the following quantity, constructed from the horizon data of the background black holes
given in \eqref{nhexpbh}:
\begin{align}\label{defM}
\Mdef=\int e^{B^{(0)}} \int \left(e^{B^{(0)}}\left(\frac{a_{t}^{(0)}}{H_{tt}^{(0)}}\right)^{2}+\frac{1}{\Sigma^{(0)}} \left[\partial_{x}\ln{\frac{e^{B^{(0)}}}{\Sigma^{(0)}}} \right]^{2}\right) - \left(\int e^{B^{(0)}}\frac{a_{t}^{(0)}}{H_{tt}^{(0)}}\right)^{2}\,,
\end{align}
where $\int$ is defined in \eqref{intdef}.
Using the Schwarz inequality \eqref{schineq} we have $\Mdef\ge0$.
The DC thermoelectric conductivities can then be written in the form:
\begin{align}\label{finform}
\sigma&=\frac{1}{\int e^{B^{(0)}}}+\frac{\left(\int e^{B^{(0)}}\frac{a_{t}^{(0)}}{H_{tt}^{(0)}}\right)^{2}}{\Mdef\int e^{B^{(0)}}}\,,
\nn
\bar\alpha=\alpha&=4\pi\frac{\int e^{B^{(0)}}\frac{a_{t}^{(0)}}{H_{tt}^{(0)}}}{\Mdef}\,,
\nn
\bar\kappa&=\frac{(4\pi)^2 T\int e^{B^{(0)}}}{\Mdef}\,.
\end{align}
We have shown that these results are valid for all black hole solutions within the ansatz \eqref{eq:DC_ansatz}, with near horizon behaviour given by \eqref{nhexpbh} and approaching $AdS_4$ in the UV with chemical potential $\mu(x)$. In fact we can show that they are also valid for more general black hole solutions provided that they have the same near horizon and asymptotic limits. For example, we have explicitly carried out the derivation for the black holes that we construct numerically in section 4 which have $g_{rx_1}\ne 0$.

We first observe that $\sigma$ and $\bar\kappa$ are both positive, as expected. We next note the similarity
of these expressions with those obtained for the homogeneous lattices of \cite{Donos:2014cya}. 
In particular, the electric conductivity appears as the sum of two positive terms. The first term
has a precise interpretation as the conductivity with zero heat current. Recalling the definition
\begin{align}
\sigma_{Q=0}\equiv\left(\frac{J}{E}\right)_{Q=0}=\sigma-\frac{\alpha^2 T}{\bar\kappa} \,,
\end{align}
which is guaranteed to be positive because it is proportional to the determinant of the positive definite thermo-electric matrix, we find
\begin{align}
\sigma_{Q=0}=\frac{1}{\int e^{B^{(0)}}}\,.
\end{align}
Thus, very roughly, we can interpret the first term in $\sigma$ as being associated with the evolution of 
charged particle-hole pairs. The second term in $\sigma$ is then $\alpha^2 T/\kappa$ which is obviously positive.
For the special case of the neutral AdS Schwarzschild black hole the second term in $\sigma$ vanishes
and the first term gives unity, and so we recover the result of \cite{Iqbal:2008by}. 
For the AdS Schwarzschild black hole we will also have $\alpha=0$, but
a divergent $\bar\kappa$ or, more precisely, a delta function in the AC thermal conductivity, since there is no momentum dissipation.

Observe that in general we have
\begin{align}\label{kapalrat}
\frac{\bar \kappa}{\alpha}=\frac{4\pi T\int e^{B^{(0)}}}{\int e^{B^{(0)}}\frac{a_{t}^{(0)}}{H_{tt}^{(0)}}}\,,
\end{align}
which is similar to an expression for the homogeneous lattices in \cite{Donos:2014cya}, but unlike \cite{Donos:2014cya} the right-hand side is not simply given by $Ts/q$. 

We next introduce $\kappa$, the thermal conductivity at zero electric current. We find
\begin{align}
\kappa\equiv \bar\kappa-\frac{\alpha^2 T}{\sigma}&=\frac{(4\pi)^2T\int e^{B^{(0)}}}{\Mdef+\left(\int e^{B^{(0)}} \frac{a_{t}^{(0)}}{H_{tt}^{(0)}}\right)^{2}}\,.
\end{align}
%with $\kappa=T\bar L/\int e^{B^{(0)}}$.
We also obtain the following expressions for the Lorenz factors:
\begin{align}
\bar L&\equiv \frac{\bar\kappa}{\sigma T}=\frac{(4\pi)^2\left(\int e^{B^{(0)}}\right)^2}{\Mdef+\left(\int e^{B^{(0)}} \frac{a_{t}^{(0)}}{H_{tt}^{(0)}}\right)^{2}}\,,\nn
L&\equiv \frac{\kappa}{\sigma T}=\frac{(4\pi)^2\left(\int e^{B^{(0)}}\right)^2\Mdef}{\left[\Mdef+\left(\int e^{B^{(0)}} \frac{a_{t}^{(0)}}{H_{tt}^{(0)}}\right)^{2}\right]^2}\,.
\end{align}
Generically $L$, $\bar L$ are neither equal nor constant and the Wiedemann-Franz law is violated.

Finally we recall the definition\footnote{To avoid confusion, in our notation the Seeback coefficient, $S$, is given by $\alpha/\sigma$.} of the dimensionless ``figure of merit", $ZT$, 
\begin{align}\label{fofm}
ZT\equiv \frac{\alpha^2 T}{\kappa\sigma}\equiv\frac{\alpha^2 T}{\bar \kappa \sigma_{Q=0}}\,.
\end{align}
The figure of merit provides a measure of the efficiency of thermoelectric engines. There is no upper bound on $ZT$
and when $ZT$ approaches infinity the efficiency approaches the Carnot limit.
The maximum value of $ZT$ for any known material is less than three. For our holographic lattice
we find that
\begin{align}
ZT=\frac{\left(\int e^{B^{(0)}}\frac{a_{t}^{(0)}}{H_{tt}^{(0)}}\right)^2}{X}\,.
\end{align}
We will see in the next subsection that holographic lattices can have arbitrarily high figures of merit at low temperatures.

\subsection{High and low temperature behaviour}\label{highlow}
It is interesting to examine the high temperature behaviour of the DC conductivity.
More precisely we are interested in the limit $T$ much greater than $\mu_0$ and $1/L$ where $\mu_0$ is the
constant term in the modulated chemical potential \eqref{genmu} and 
$L$ is the period of the lattice. In this limit, the black hole background is approximated by
the AdS-Schwarzschild black hole metric:
\begin{align}
ds^2=-Udt^2+U^{-1}dr^2+r^2 (dx^2+dy^2)
\end{align}
with $U=r^2-r_+^3/r^2$ and $ r_+=4\pi T/3$. Note that here the horizon is located 
at $r=r_+$ (and not at $r=0$ as above).
Furthermore, in the high temperature limit the leading term in the solution for the
gauge-field equations of motion is given by
$a_t=(1-\frac{r_+}{r})\mu(x)$ for arbitrary periodic $\mu(x)$.
Using \eqref{finform} we obtain, as $T\to\infty$,
\begin{align}\label{hightsc}
\sigma&=1+\frac{(\int\mu)^2}{\int\mu^2-(\int\mu)^2}\,,\qquad
\alpha=\frac{(4\pi)^2T}{3}\frac{\int\mu }{\int\mu^2-(\int\mu)^2}\,,\nn
\bar\kappa&=\frac{(4\pi)^4T^3}{9}\frac{1}{\int\mu^2-(\int\mu)^2}\,,\qquad
\kappa=\frac{(4\pi)^4T^3}{9}\frac{1}{\int\mu^2}\,.
\end{align}
It is interesting that as $T\to\infty$ we have $\sigma$ approaching a constant value, with $\sigma\ge 1$ (recall that for the optical conductivity 
we have $\lim_{\omega\to\infty}\sigma(\omega)=1$). 
This is reminiscent of the Mott-Ioffe-Regel bound \cite{Gunnarsson:2003zz,takmir} of metals (see figure \ref{fig:Snew}), though here, of course,
there are no quasi-particles.

We can also consider the low-temperature behaviour of the DC conductivity. This will obviously depend on the precise nature of the
zero temperature ground states. As we will discuss in the next section, all of the black holes that we have constructed 
which are associated with monochromatic lattices with $\mu=\mu_0+V \cos(k x)$ and $\mu_{0}\ne 0$,
seem to approach
$AdS_2\times\mathbb{R}^2$ in the far IR at $T=0$, perturbed by an irrelevant deformation.
For these black holes we can obtain the low-temperature behaviour as follows.
We have $U\to 6r^2$,
$H_{tt},H_{rr}\to 1$, $a^{(0)}_t\to 2\sqrt{3}$ and $e^{B_0}, \Sigma$ approach constants that depend
on the UV lattice data.
We immediately conclude from \eqref{defM} that $\Mdef\to 0$ and hence the
second term in the DC electric conductivity
in \eqref{finform} dominates the first. More precisely, 
using the analysis of \cite{Hartnoll:2012rj,Edalati:2010pn}
as $T\to 0$ we find that the 
DC conductivity scales as 
\begin{align}\label{dcscal}
\sigma \sim T^{2-2\Delta(\bar k)},\quad \alpha \sim T^{2-2\Delta(\bar k)},\quad \bar\kappa\sim T^{3-2\Delta(\bar k)}\,,
\end{align}
where
\begin{align}\label{defDel}
\Delta(\bar k)=\frac{1}{2}+\frac{1}{2}\left[5+4\bar k^2-4\sqrt{1+2\bar k^2}\right]^{1/2}\,,
\end{align}
and $\bar k$ is the renormalised wave-number,
$\bar k\equiv k/(6\Sigma^{(0)}e^{B^{(0)}})^{1/2}$, which depends on the UV wave-number $k$ and the zero temperature domain wall solution. It is worth emphasising that for the $T=0$ AdS-RN black brane we have 
$\Sigma^{(0)}e^{B^{(0)}}=\mu^2/12$. Therefore we can define a length renormalisation factor, $\bar\lambda$, via
\begin{align}\label{renscale}
\bar k=\frac{k\sqrt{2} }{\mu \bar\lambda}
\end{align}
with, in general, $\bar\lambda\ne 1$ for the lattice black holes. As $T\to 0$ we note that
the DC conductivity $\sigma$ diverges and this is associated with the Drude peak in the optical conductivity
turning into a delta function at $T=0$.
The scaling of $\sigma$, omitting the issue of $\bar\lambda$, 
was pointed out in \cite{Hartnoll:2012rj}, who obtained
it in the limit of small lattice strengths by using the memory matrix formalism
and also by taking a limit of the optical conductivity obtained from a matching argument. The
issue of length renormalisation was discussed in \cite{Donos:2012js,Donos:2013eha}.

It is also interesting that we have the scaling $\kappa\sim T$. In particular, while $\kappa$ is going to zero, $\bar\kappa$ diverges if $\bar k^2>3/4+\sqrt3/2$, goes to a constant
if $\bar k^2=3/4+\sqrt3/2$ and vanishes if $0\le \bar k^2<3/4+\sqrt3/2$. 
We also note that as $T\to 0$ we have $\bar L \to \frac{4\pi^2}{3}=\frac{s^2}{q^2}$ and
$\bar\kappa/\alpha\to Ts/q$, in this limit, independent of the lattice deformation. 
Finally, we note that the figure of merit is diverging as $T\to 0$ with $ZT\sim T^{2-2\Delta(\bar k)}$.

Although we will not be discussing them further in this paper, we pause to comment upon the DC results for the lattices 
$\mu=\mu_0+V \cos(k x)$ with $\mu_{0}= 0$ that were studied in \cite{Chesler:2013qla}. In the $T=0$ limit these black holes approach $AdS_4$ in
the far IR.  It is straightforward to see that in the black hole solutions the $x$ dependence of the gauge field can be expanded in terms of Fourier modes that are odd multiples of $k$ whereas for the metric functions they will be even multiples. Hence, we can deduce that for these
black holes we have, for all temperatures,
\begin{align}\label{finformcs}
\sigma=\frac{1}{\int e^{B^{(0)}}}\,,
\qquad
\bar\alpha=\alpha=0\,,
\qquad
\bar\kappa=\frac{(4\pi)^2 T}{\int \Big(e^{B^{(0)}}\left(\frac{a_{t}^{(0)}}{H_{tt}^{(0)}}\right)^{2}+\frac{1}{\Sigma^{(0)}} \left[\partial_{x}\ln{\frac{e^{B^{(0)}}}{\Sigma^{(0)}}} \right]^{2}\Big)}\,.
\end{align}
%Clearly these black holes saturate the bound \eqref{anotherbd}.

\section{Numerical construction of inhomogeneous lattices and the AC conductivity}\label{sec:numerics}
In this section we will numerically construct black holes corresponding to inhomogeneous lattices. We can then immediately obtain the
DC conductivity using the results of the last section. We will also numerically determine the
AC conductivity.
We will see the appearance of a coherent Drude-type peak in the AC conductivity, as in \cite{Horowitz:2012gs}, but we will
not see any evidence for an intermediate scaling regime that was reported in \cite{Horowitz:2012gs}.
On the other hand we observe see an interesting resonance phenomenon, also seen in \cite{Horowitz:2012gs}, 
which we associate with sound modes. We also carry out a detailed check of a sum-rule satisfied by the optical conductivity
and also a sum-rule associated with electromagnetic duality \cite{WitczakKrempa:2012gn}.
Finally, we show that the low temperature black holes exhibit scaling behaviour consistent
with them approaching $AdS_2\times\mathbb{R}^2$ in the IR, in contrast to \cite{Hartnoll:2014gaa}.

In section 4.1 and 4.2 we will describe the numerical construction and in section 4.3 we will present
the main results.

\subsection{The backgrounds}\label{sec:background}

To construct the black hole geometries that we are interested in, we will make the following
 ansatz\footnote{In fact, instead of using the $Q$ variables in our numerics, we have found it slightly more accurate to use $F$ variables defined through $Q_{ii}=1+z\, F_{ii}$ and $Q_{zx}=z\,F_{zx}$.}:
\begin{align}\label{eq:ansatz}
ds^{2}&=\frac{1}{z^{2}}\left[-f\,Q_{tt}\,dt^{2}+\frac{Q_{zz}}{f}\,dz^{2}+Q_{xx}\,\left(dx+Q_{zx}\,dz \right)^{2}+Q_{yy}\,\,dy^{2}\right]\,,  \nn
A&=\left(1-z\right)\,a_{t}\,dt \,,
\end{align}
where
\begin{align}
f&=\left(1-z\right)\,\left(1+z+z^{2}-\frac{\mu^{3}z^{3}}{4} \right)\,,
\end{align}
and $\mathcal{F}=\left\{Q_{tt}, Q_{rr},Q_{xx},Q_{zx},Q_{yy},a_{t} \right\}$ are all functions of the radial coordinate $z$ and $x$. 
In this section, the $AdS_4$ boundary will be located at $z=0$ and the black hole horizon at $z=1$. Notice that the function $f$, with
$Q_{tt}= Q_{rr}=Q_{xx}=Q_{yy}=1$, $Q_{zx}=0$ and $a_{t}=\mu$
gives the standard electrically charged AdS-RN black hole. We also notice that we have not fixed the diffeomorphism invariance
in the $\left( z, x\right)$ coordinates, for reasons we now explain. 
 
Substituting the ansatz \eqref{eq:ansatz} into the equations of motion \eqref{eq:eom} one finds a consistent set of PDEs for the functions $\mathcal{F}$ but, due to diffeomorphism invariance of Einstein's equations, the boundary value problem is underdetermined \cite{Headrick:2009pv}. Similar problems can 
arise due to the gauge invariance of the Maxwell field, but the specific 
electric ansatz \eqref{eq:ansatz} leads to just a second-order equation for the function $a_{t}$ without any constraints on it.

In order to deal with the diffeomorphism invariance of Einstein's equations we will follow the approach of Headrick, Kitchen and Wiseman \cite{Headrick:2009pv} (see also \cite{Adam:2011dn,Wiseman:2011by}).
The key step is to modify Einstein's equation from $E_{\mu\nu}=0$ in \eqref{eq:eom} to 
$E_{\mu\nu}=\nabla_{\left( \mu\right.}\,\xi_{\left. \nu\right)}$ where the ``DeTurck term"
on the right-hand side
is defined by the vector 
$\xi^{\mu}=g^{\nu\lambda}\,\left( \Gamma_{\nu\lambda}^{\mu}\left(g\right)-\bar{\Gamma}_{\nu\lambda}^{\mu}\left(\bar{g}\right)\right)$ 
and $\bar{g}$ is a fixed reference metric. The addition of this term 
transforms Einstein's equations into an elliptic set of equations for the metric functions, for arbitrary reference metric $\bar{g}$. Indeed it is the first order term 
$g^{\nu\lambda}\Gamma_{\nu\lambda}^{\mu}\left(g\right)$ in $\xi^{\mu}$ that modifies the character of Einstein's equations.

The role of the second term in $\xi$, involving the reference metric, is as follows. We want solutions of $E_{\mu\nu}=\nabla_{\left( \mu\right.}\,\xi_{\left. \nu\right)}$ to be solutions satisfying $E_{\mu\nu}=\xi^\mu=0$, a point which we will return to below. That this might be possible relies on interpreting
$\xi^{\mu}=0$ as a gauge-fixing condition, and this is where the reference metric is important.
For the special case of the ansatz \eqref{eq:ansatz} we still have diffeomorphisms in $z$ and $x$ leaving us with two gauge conditions to be imposed. 
Assuming that the reference metric lies within the ansatz \eqref{eq:ansatz},
it is easy to check that only non-trivial components of the vector $\xi^{\mu}$ are the $z$ and the $x$ components. 
Thus, at the level of counting constraints, the condition $\xi^\mu=0$ matches the number of gauge conditions left to be imposed in order to obtain a 
gauge-fixed black hole solution.

For the case of Einstein's equations with a negative cosmological constant, it has been shown that there are no solutions of $E_{\mu\nu}=\nabla_{\left( \mu\right.}\,\xi_{\left. \nu\right)}$ with a non-trivial $\xi$, provided that $\xi=0$ on the boundary of the given problem \cite{Figueras:2011va}. 
A similar general statement for Einstein-Maxwell theory with a negative cosmological constant is still lacking. Our approach here, therefore, 
will be to check that in the continuum limit our solutions are converging towards $\xi^{\mu}=0$, or equivalently, since $\xi$ is a space-like vector, $\xi^{2}=0$. 
In fact, we have been able to achieve a resolution of at least
$\xi^{2}<10^{-19}$ for all of the background geometries that we consider in this paper. 
We further discuss the implementation of our numerics and convergence tests in appendix \ref{convgtest}. 
Our results, and also those in \cite{Horowitz:2012gs}, 
constitute some evidence that the theorem of \cite{Figueras:2011va} can be extended to the case of Einstein-Maxwell theory with a cosmological constant.

It is clear from the above discussion that the choice of reference metric is important since it is ultimately part of the gauge fixing procedure. 
The holographic lattice black holes that we are interested in can be viewed as deformations of the AdS-Reissner-Nordstr\"om black hole mentioned above. Guided by this, we will take the reference metric $\bar{g}$ to simply be that of the AdS-RN black hole.

In order for the two dimensional elliptic problem at hand to have a unique solution, we need to impose appropriate boundary conditions. We will choose the coordinate $x$ to be periodic and we are therefore left with the boundary conditions that need to be imposed on the black hole horizon and on
the $AdS_{4}$ boundary, both of which are singular points of the PDEs.

On the horizon at $z=1$, we will impose that the functions $\mathcal{F}$ are analytic, with an expansion of the form
\begin{align}
\mathcal{F}\,\left(r,x\right)=\mathcal{F}\left(1,x\right)-\partial_{z}\mathcal{F}\left(1,x\right)\,\left(1-z\right)+\mathcal{O}\left(\left(1-z\right)^{2}\right)\,.
\end{align}
After substituting into the equations we obtain a total of six sets of constraints on the values of
$\mathcal{F}\left(1,x\right)$, and the normal derivatives, $\partial_{z}\mathcal{F}\left(1,x\right)$. The simplest amongst these is that
surface gravity should be constant, which simply reads
\begin{align}
Q_{tt}\left(1,x\right)=Q_{zz}\left(1,x\right)\,.
\end{align}
It is precisely these six constraints that we will be imposing as boundary conditions at the $z=1$ surface.

We now turn to the $AdS_{4}$ boundary at $z=0$. Demanding that the only deformations of the CFT are temperature and the inhomogeneous chemical potential $\mu\left(x\right)$ we are led to the asymptotic expansion
\begin{align}\label{eq:background_as_exp}
a_{t}\left(z,x\right)&=\mu\left(x\right)+q_{t}\left(x\right)\,z+\mathcal{O}\left(z^{2}\right)\,,\nn
Q_{tt}\left(z,x\right)&=1+q_{tt}\left(x\right)\,z^{3}+\frac{1}{4}\,\left( -\mu^{2}+\left( q_{t}\left(x\right)-\mu\left(x\right)\right)^{2}\right)\,z^{4}+g_{1}\left(x\right)\,z^{\left(3+\sqrt{33} \right)/2}+\mathcal{O}\left(z^{5}\,\ln z\right))\,,\nn
Q_{zz}\left(z,x\right)&= 1+\frac{1}{4}\,\left( \mu^{2}-\left(q_{t}\left(x\right)-\mu\left(x\right) \right)^{2}+\mu^{\prime}\left(x\right){}^{2}\right)\,z^{4}+g_{2}\left(x\right)\,z^{\left(3+\sqrt{33} \right)/2}+\mathcal{O}\left(z^{5}\,\ln z\right))\,,\nn
Q_{xx}\left(z,x\right)&= 1+q_{xx}\left(x\right)\,z^{3}+g_{1}\left(x\right)\,z^{\left(3+\sqrt{33} \right)/2}+\mathcal{O}\left(z^{5}\,\ln z\right))\,,\nn
Q_{zx}\left(z,x\right)&= q_{zx}\left(x\right)\,z^{4}+\frac{1}{5}\,\left[\left(q_{t}\left(x\right)-\mu\left(x\right)\right)\,\mu^{\prime}\left(x\right)-2\,q_{xx}^{\prime}\left(x\right) \right]\,z^{4}\,\ln z+\mathcal{O}\left(z^{5}\,\ln z\right))\,,\nn
%&\qquad\qquad+2\,\frac{g^{\prime}_{2}\left(x\right)-g_{1}^{\prime}\left(x\right)}{3+\sqrt{33}}\,z^{\left(3+\sqrt{33} \right)/2+1}
Q_{yy}\left(z,x\right)&= 1+q_{yy}\left(x\right)\,z^{3}-\frac{1}{4}\mu\left(x\right)^{\prime\,2}\,z^{4}+g_{1}\left(x\right)\,z^{\left(3+\sqrt{33} \right)/2}+\mathcal{O}\left(z^{5}\,\ln z\right))\,,
\end{align}
with
\begin{align}\label{condofqs}
q_{tt}\left(x\right)+q_{xx}\left(x\right)+q_{yy}\left(x\right)=0\,.
\end{align}
The functions $\left\{q_{t},\,q_{tt},\,q_{xx},q_{zx},\, g_{1},\,g_{2} \right\}$ are arbitrary functions which will be fixed by solving the PDEs with a regular horizon at $z=1$.

It is worth highlighting the terms parametrised by $g_1$ and $g_2$ that arise from solving
the modified Einstein equations. The condition $\xi^\mu=0$ implies 
$g_2= -\frac{1}{2} \,( 3+\sqrt{33}) g_1$.
If one considers $g_i$ as parametrising a linearised
perturbation about the $\mu(x)$ deformed $AdS_4$ space, one can see that these conditions
imply that the $g_i$ can be absorbed into a redefinition of the $z$ coordinate via $z(1-g_1 z^\Delta/2)=\bar z$, and hence are
pure gauge. 
It is also worth mentioning here that the appearance of the non-analytic terms, which are appearing at order higher than $z^4$,
will affect the convergence rates of the numerical scheme, locally in $z$, as we discuss further in appendix B.
 
It is clear from the asymptotic expansion \eqref{eq:background_as_exp} that a suitable set of boundary conditions on the $AdS_{4}$ boundary are
\begin{align}\label{eq:background_as_bc}
&Q_{tt}\left(0,x\right)=Q_{zz}\left(0,x\right)=Q_{{x}{x}}\left(0,x\right)=Q_{{y}{y}}\left(0,x\right)=1\,,\nn
&Q_{zx}\left(0,x\right)=0,\quad a_{t}\left(0,x\right)=\mu\left(x\right)\,.
\end{align}
We will be choosing 
\begin{align}
\mu\left(x\right)=\mu_0+\bar\mu\left(x\right) 
\end{align}
with $\bar \mu\left(x\right)$ averaging to zero over a period in $x$.

Observe that $\partial_{y}$ is a Killing vector for our geometry \eqref{eq:ansatz} which also preserves the gauge-field.
Since it has no fixed points in the bulk, following the general arguments of \cite{Donos:2013cka}, we can conclude that
our solutions should satisfy the Smarr-type relation:
\begin{align}\label{smarrform}
\int\,\left[T^{tt}\left(x\right)+T^{yy}\left(x\right)-\mu\left(x\right) J^{t}\left(x\right) \right]=T\,S\,,
\end{align}
where the charge density, $J^{t}$, and the entropy, $S$, are given by\footnote{The origin of the
shift by $\mu(x)$ in $J^t$ is the factor of $(1-z)$ in \eqref{eq:ansatz}.}
\begin{align}
J^{t}\left(x\right)=-q_{t}\left(x\right)+\mu\left(x\right),\quad S=4\pi\,\int Q^{1/2}_{xx}\left(1,x\right)\,Q^{1/2}_{yy}\left(1,x\right)\,,
\end{align}
and the components of the stress tensor (following from a similar analysis to appendix \ref{stressheat})
are given by
 \begin{align}\label{teecomp}
T^{tt}\left(x\right)&=2+\frac{\mu^{2}}{2}-3\,q_{tt}\left(x\right),\nn 
T^{xx}\left(x\right)&=1+\frac{\mu^{2}}{4}+3\,q_{xx}\left(x\right),\nn
 T^{yy}\left(x\right)&=1+\frac{\mu^{2}}{4}+3\,q_{yy}\left(x\right)\,.
\end{align}
The Smarr-relation \eqref{smarrform} provides a check for the numerical error of our solutions. 
Observe, from \eqref{condofqs}, that the stress tensor is traceless.
Also, on-shell, with vanishing deTurck vector, we obtain the Ward identity
$\nabla_\mu T^{\mu\nu}+J^\mu F_{\mu}{}^\nu= 0$,
which we have also verified in our numerical solutions (at the order of $10^{-3}\%$ error).

In order to numerically integrate the system of PDEs in the bulk, subject to the boundary conditions we have just described, we discretise the problem in the $z$ and $x$ directions. This leads to a non-linear algebraic system of equations which we solve using Newton's method.

Since the $x$ direction is periodic and we expect all of the functions to be smooth away from the two boundaries, we find it appropriate to use spectral methods for that direction. More specifically we will use a Fourier decomposition in order to approximate the partial derivatives along the $x$ direction and an equi-spaced grid is appropriate. We will denote the number of grid points in the $x$ direction by $N_{x}$.
For the monochromatic and dichromatic lattices, described
at the beginning of section 4.3, we have taken $N_x=45$ and $N_x=90$, respectively. 
For the dirty lattices, described in section \ref{sec:dirty}, for which the 
memory requirement of our numerical computation is significantly higher, we take $N_x=150$.

A little more care is required for the discretisation of the radial direction $z$. As we can see from equation \eqref{eq:background_as_exp}, the asymptotic expansion at the $z=0$ boundary reveals that our functions will not be infinitely differentiable there. This point immediately excludes the use of spectral methods uniformly in the radial direction. We have checked that a Chebyshev decomposition would still work with a convergence that would only be power law. The same type of convergence is also achieved using finite difference methods. We will use the latter approach since it
is more memory efficient since the linear systems that we have to solve at the iterative steps of Newton's method are much sparser.
More specifically, the results in the paper are obtained using sixth-order finite differences, but we note that
we also made some cross-checks using fourth-order finite differences.

At temperatures which are not too low, we have found that a simple finite difference patch is enough to accurately describe the solutions we are interested in. As we lower the temperature 
we find that we need to increase the resolution in the radial direction. In fact we find that as $T\to 0$
the near horizon limit of our black holes approach $AdS_2\times\mathbb{R}^2$ and this is changing
the analytic behaviour near the horizon.
Therefore, instead of increasing the number of points uniformly in our computational domain we can divide it into different regions and consider higher resolution or higher order finite differences\footnote{One can also take one or both of the patches to be spectral.} for the ones closer to the horizon. Some care is required at the interface between two such regions, as one needs to ensure 
that the solution will have a continuous first derivative\footnote{Continuity of the second normal derivative is a result of satisfying the second order equations of motion at the interface from both sides.}. In more detail, consider two such sets of uniformly distributed points $z_{i_{1}}$ and $z_{i_{2}}$ with $i_{1}=1,\ldots ,N_{1}$, $i_{2}=N_{1}+1,\ldots,N_{2}+N_{1}$ and with $z_{N_{1}}=z_{N_{1}+1}$. The simplest way to patch these two grids together is to require that
\begin{align}\label{effcalc}
\mathcal{F}\left(z_{N_{1}},x\right)=\mathcal{F}\left(z_{N_{1}+1},x\right),\quad \mathcal{F}^{\prime}\left(z_{N_{1}},x\right)=\mathcal{F}^{\prime}\left(z_{N_{1}+1},x\right)\,,
\end{align}
and then check that the equations of motion, which are second order in $z$, are satisfied at $z=z_{N_{1}}$ in the continuum limit.

We take the total number of lattice points in the $z$ direction, $N$, to be sufficiently high to ensure that
we achieve a resolution of at least $\xi^{2}<10^{-19}$ for all of the background geometries. For most lattices and
temperatures that we have considered this is achieved for $N\sim 350$. In appendix \ref{convgtest} we discuss
in more detail our convergence tests, where we also achieve resolutions of  $\xi^{2}\sim 10^{-24}$ for larger values of $N$.
We also note that for our high precision numerics at very low temperatures we used $N\sim 5000$ distributed non-uniformly in three patches, in order to achieve  
$\xi^{2}<10^{-19}$ resolution, as described in appendix \ref{floppy}.

\subsection{AC conductivity}\label{sec:ACConductivity}

In this sub-section we describe the numerical strategy we use to extract the AC electric conductivity in the $x$ direction, $\sigma(\omega)$, 
for the class of black holes described in \ref{sec:background}. As usual we need to perturb 
the background geometry by an oscillating electric field in the $x$ direction
of the form $e^{-i\omega t}E$. 
A consistent ansatz for the perturbation that describes the response of the bulk geometry to such an oscillating electric field is given by
\begin{align}\label{eq:opt_cond_ans}
\delta ds^{2}&=\frac{1}{z^{2}}\,\Big[-f\,Q_{tt}\,\hat{h}_{tt}\,dt^{2}+Q_{xx}\, \hat{h}_{xx}\,\left(dx+Q_{zx}\,dz\right)^{2}+Q_{yy}\,\hat{h}_{yy}\,dy^{2}+\nn
&\qquad\qquad\qquad\qquad\qquad\qquad\qquad\qquad\qquad2 f\,Q_{tt}\,\hat{h}_{tx}\,dt\,\left(dx+Q_{zx}\,dz\right) \Big]\,,\nn
\delta A&=\left(1-z \right)\,\hat{a}_{t}\,dt+\hat{a}_{x}\,\left( dx+Q_{zx}\,dz\right)\,,
\end{align}
where 
$\hat{\mathcal{W}}\equiv \left\{ \hat{h}_{tt},\, \hat{h}_{tx},\, \hat{h}_{xx},\,\hat{h}_{yy},\,\hat{a}_{t},\,\hat{a}_{x}\right\}$ are
six functions of $\left\{t,\,z,\,x \right\}$. 
We note that here we have chosen a gauge with
\begin{align}\label{gchce}
\delta g^{\mu z}=\delta A^{z}=0\,.
 \end{align}
 It is convenient to also define
\begin{align}
\hat{h}_{xx}&=\left(1-z \right)\,\hat{h}_{+}+\hat{h}_{-},\qquad \hat{h}_{yy}=\left(1-z \right)\,\hat{h}_{+}-\hat{h}_{-}\,.
\end{align}
Note that we have pulled out some factors of $\left( 1-z\right)$ for convenience arising from regularity considerations and using the equations of motion. As we will elaborate upon below,
we note that regularity implies that $\hat{h}_{tt}\sim \mathcal{O}\left(1-z \right)$ close to the horizon or, more precisely, that the $tt$ component of the metric perturbation should vanish as $\mathcal{O}\left( \left(1-z \right)^{2}\right)$ in this gauge. Thus, the perturbation is not changing the behaviour of
the black hole horizon. We also need to impose in-falling boundary conditions on the Killing horizon of the black hole at $z=1$. By introducing 
 \begin{align}\label{tildefns}
\hat{\mathcal{W}}=e^{-i\omega t}\,\left(1-z^{3} \right)^{\frac{i\omega}{4\pi T}}\,\tilde{\mathcal{W}}\,,
 \end{align}
the in-falling boundary conditions translate into analyticity conditions for the time independent functions $\tilde{\mathcal{W}}$.
 
The equations of motion consist of six second order equations in $z$ as well
as four constraint equations which are first order in $z$. 
The six second-order equations of motion arise from the $\left\{tt,\,t x,\,xx,\,yy \right\}$ components of Einstein's equations and the $\left\{t,\,x\right\}$ components of Maxwell equations. These six equations are all second order with respect to the coordinate $z$ 
 in the domain of the coordinates $z$ and $x$.
 These constraint equations can be characterised by considering the foliation of the spacetime by surfaces with constant $z$. The unit normal one form to these surfaces, $n\propto dz$, has a dual
vector field $n^\mu$ with non-vanishing components $n^z$ and $n^{x_1}$. The
constraint equations are then obtained by contracting this vector field with 
Einstein equations, written in conventional form, and with the Maxwell equations:
$C_{\mu}\equiv n^\nu\left( E_{\mu \nu}-\frac{1}{2}g_{\mu \nu} E^{\rho}{}_{\rho}\right)=0$ and $D\equiv  n^\nu\nabla_{\mu}F^{\mu}{}_{\nu}=0$. This
provides a total of four constraints since $C_{y}=0$ trivially for our background and perturbation ansatz \eqref{eq:opt_cond_ans}.
 
Following the standard ADM type analysis, one can show that the six second-order equations of motion imply that if $C_{\mu}=D=0$ on any constant $z$ slice then we also have $\partial_{z}C_{\mu}=\partial_{z}D=0$ on that slice.
In other words, we only need to impose the constraints on any constant $z$ surface and we will choose  
to impose them on the expansion near the horizon at $z=1$. Note that 
if we had chosen this surface to be the $AdS_{4}$ boundary at $z=0$ it would involve imposing boundary conditions on third order derivatives of fields and this is less accurate.
 
It is worth emphasising that in contrast to the background black holes, for the perturbations we are solving Einstein's equations rather than the equations modified by the DeTurck term. This is because the perturbations involve time-dependence and the DeTurck term does not turn the problem into an elliptic one. However, checking that the constraints are satisfied in the continuum limit is one of the convergence checks that we perform, as discussed in appendix \ref{convgtest}.

We will now turn to the question of boundary conditions that we impose on the 
functions $\tilde{\mathcal{W}}$ defined in \eqref{tildefns}. 
Expanding the six second-order equations in $z$ along with the four constraint equations we find that
a total of ten boundary conditions must be imposed on the horizon at $z=1$. Amongst these we find that we must impose $\tilde{h}_{tt}=0$, as we mentioned earlier.

We are now left with two more conditions that need to be imposed in order to obtain a unique solution to the six second-order equations. 
As we will now show, these come from boundary conditions imposed at the $AdS_{4}$ boundary at $z=0$. We first note that the second-order system of equations implies that we can develop an expansion of the six fields in $\tilde{\mathcal{W}}$ in terms of non-normalisable and normalisable data of the form
\begin{align}\label{nnp}
\tilde{h}_{\mu\nu}\left(z,x \right)&=\tilde{h}_{\mu\nu}^{(0)}\left(x\right)+\cdots+\tilde{h}_{\mu\nu}^{(3)}\left(x\right)\,z^{3}+\cdots\,,\nn
%\tilde{h}_{\pm}\left(z,x \right)&=\tilde{h}_{\pm}^{(0)}\left(x\right)+\cdots+\tilde{h}_{\pm}^{(3)}\left(x\right)\,z^{3}+\cdots\nn
\tilde{a}_{\mu}\left(z,x\right)&=\tilde{a}_{\mu}^{(0)}\left(x\right)+\tilde{a}_{\mu}^{(1)}\left(x\right)\,z+\cdots\,.
\end{align}
Now 
the four first-order constraints can be used to express four of these functions in 
terms of the remaining ones as well as the background fields, but, as mentioned above, this will automatically be taken
into account by the ten boundary conditions that we imposed at the horizon. 
These conditions correspond to the two non-trivial components of
stress-energy conservation, current conservation, and the tracelessness of the stress-energy tensor.

Proceeding, we now find ourselves in a situation very similar to the one discussed in \cite{Donos:2013eha} regarding the UV boundary conditions of the perturbation. We have a total of six non-normalisable fall-offs in \eqref{nnp} 
but only two boundary conditions left to impose and furthermore we only want to source a single field on the boundary - a time oscillating electric field in the $x$ direction. 
At first sight this seems to lead to an over-defined boundary value problem.
 
The simple resolution to this puzzle is that the requirement of sourcing only one of the perturbation fields is actually weaker than setting the remaining non-normalisable pieces all equal to zero. This can be seen in detail as follows.
Suppose that we allow all of the non-normalisable pieces in \eqref{nnp} to be switched on in
such a way that there exists a combination of boundary reparametrisations, $x^{\mu}\to x^{\mu}+\xi^{\mu}$, and gauge transformations, $A_{\mu}\to A_{\mu}+\partial_{\mu}\Lambda$, where
\begin{align}
\xi&=e^{-i\omega t}\,\left(\xi^{t}\left(x\right)\partial_{t}+\xi^{z}\left(x\right)\,z\,\partial_{z} +\xi^{x}\left(x\right)\,\partial_{x}\right)+\cdots\,,\nn
\Lambda&=e^{-i\omega t}\,\lambda\left(x\right)+\cdots\,,
\end{align}
such that close to the $AdS_{4}$ boundary we have
\begin{align}\label{ggeconds}
z^{2}\,\left[\delta g_{\mu\nu}+\mathcal{L}_{\xi}g_{\mu\nu} \right]&\to 0\,,\nn
\delta A +\mathcal{L}_{\xi}A+d\Lambda & \to e^{-i\omega t}\,\mu_{J}\,dx\,.
\end{align}
This would imply that we are actually only sourcing our boundary theory by an oscillating electric field
and all of the other non-normalisable fall-offs of the functions are just gauge artefacts.

Conversely, if we demand that the asymptotic behaviour in \eqref{nnp} is such that there is a combination of coordinate and gauge transformations satisfying \eqref{ggeconds} we deduce that we must have
\begin{align}\label{inters}
&\xi^{x}=-\frac{i}{\omega}\,\left( \tilde{h}^{(0)}_{tx}-\xi^{t}{}^{\prime}\right),\qquad \xi^{t}=\frac{i}{2\omega}\,\left( -\tilde{h}^{(0)}_{-}+\tilde{h}^{(0)}_{+}+\tilde{h}^{(0)}_{tt}\right)\,,\nn
&\xi^{x}{}^{\prime}=-\tilde{h}^{(0)}_{-},\qquad \xi^{z}=\frac{1}{2}\,\left( -\tilde{h}^{(0)}_{-}+\tilde{h}^{(0)}_{+}\right)\,,\nn
%&\omega\,\lambda=-\omega\,a_{t}^{(0)}\,\xi^{t}-i\,\left( \tilde{a}^{(0)}_{t}+\xi^{x}\,a_{t}^{(0)}{}^{\prime}\right),\qquad \tilde{a}^{(0)}_{x}=\mu_{J}-a_{t}^{(0)}\,\xi^{t}{}^{\prime}-\lambda'\,,\nn
&\omega\,\lambda=-\omega\,\mu(x)\,\xi^{t}-i\,\left( \tilde{a}^{(0)}_{t}+\xi^{x}\,\mu(x)^{\prime}\right),\qquad \tilde{a}^{(0)}_{x}=\mu_{J}-\mu(x)\,\xi^{t}{}^{\prime}-\lambda'\,,
\end{align}
where we notice the appearance of the background function $\mu(x)$ of the holographic lattice.
%\begin{align}
%a_{t}^{(0)}\left(x\right)=\mu\left(x\right)\,.
%\end{align}
This gives a total of six equations that should be satisfied. However, the general reparametrisation and gauge transformation is parametrised by only four functions $\left\{\xi^{t},\xi^{r},\,\xi^{x},\,\lambda \right\}$. Examining the integrability conditions of the six equations \eqref{inters}
we are lead to two constraints that our non-normalisationle fall-offs should satisfy:
\begin{align}
2\,\omega^{2}\,\tilde{h}_{-}^{(0)}+\tilde{h}_{-}^{(0)}{}^{\prime\prime}-2i\,\omega\,\tilde{h}_{tx}^{(0)}{}^{\prime}-\tilde{h}_{+}^{(0)}{}^{\prime\prime}-\tilde{h}_{tt}^{(0)}{}^{\prime\prime}&=0\,,\nn
\omega^{3}\,\left( \tilde{a}_{x}^{(0)}-\mu_{J}\right)+\frac{i\,\omega^{2}}{2}\,\left(\left( 3\,\tilde{h}^{(0)}_{-}-\tilde{h}_{+}^{(0)}-\tilde{h}_{tt}^{(0)}\right) \mu(x){}^{\prime}-2\,\tilde{a}_{t}^{(0)}{}^{\prime}\right)\nn
\qquad\qquad\qquad\quad+\frac{1}{2}\mu(x)^{\prime\prime}\,\left(-2\omega\,\tilde{h}_{tx}^{(0)}-i\,\left( \tilde{h}_{-}^{(0)}{}^{\prime}-\tilde{h}_{+}^{(0)}{}^{\prime}-\tilde{h}_{tt}^{(0)}{}^{\prime}\right) \right)&=0\,.
\end{align}

These two conditions are precisely the remaining two boundary conditions that we need to impose on the $AdS_{4}$ boundary in order that we are only sourcing an oscillating electric field on the boundary. Moreover, the current can be read off after performing the above transformation and then using \eqref{defcurrent} or, equivalently, from the sub-leading fall-off of the gauge field perturbation, and we find
\begin{align}\label{genjcurrent}
\mathcal{J}=\tilde{a}_{x}^{(1)}+\frac{i}{2\,\omega}\,\left( \mu(x)-q_{t}\right)\,\left(\tilde{h}_{-}^{(0)}-\tilde{h}_{+}^{(0)} -\tilde{h}_{tt}^{(0)} \right)^{\prime}\,.
\end{align}
The uniform mode of the current is now given by a simple integration over a period:
\begin{align}
%J=\frac{k}{2\,\pi}\,\int_{0}^{2\pi/k}\,\mathcal{J}\,dx\,,
J=\frac{1}{L}\,\int_{0}^{L}\,\mathcal{J}\,dx\,,
\end{align}
and the AC electric conductivity in the $x$ direction is given by the Kubo formula:
\begin{align}\label{defacc}
\sigma(\omega)=\frac{J}{i\,\omega\,\mu_{J}}.
\end{align}

Finally, we comment that for numerically obtaining the optical conductivity we used the same 
computational grid that we used for the background lattice black holes.

\subsection{Numerical results}

In this section we will present the results that we extracted from the numerical setup we outlined in the previous sub-sections. Our implementation can handle a very general class of periodic lattices. 
The class that we have analysed in greatest detail are monochromatic lattices of the form
\begin{align}
\mu(x)=\mu+A\mu \cos\left(k \,x\right)\,,
\end{align}
where $\mu\ne 0$ is a constant (note that for clarity, we have dropped the subscript in \eqref{genmu} here and in the remainder of this section), as is $A$ and $k$.
Such lattices were first constructed in \cite{Horowitz:2012gs} and, as we shall discuss, 
while we find some agreement with their results we find some important differences too. We have also looked in some detail at second class of lattices are
dichromatic lattices of the form
\begin{align}\label{dichrom}
\mu(x)=\mu+A\mu \cos\left(k\,x\right)+B\mu \cos\left(2k\,x\right)\,,
\end{align}
which have similar properties but also exhibit some new features. Finally, we have briefly considered a single
example of a dirty lattice that is constructed from a larger number of modes, specifically ten, and random phases in section \ref{sec:dirty}

\subsubsection{Drude peaks and DC conductivity}
\begin{figure}
\centering
\includegraphics[height=5.5cm]{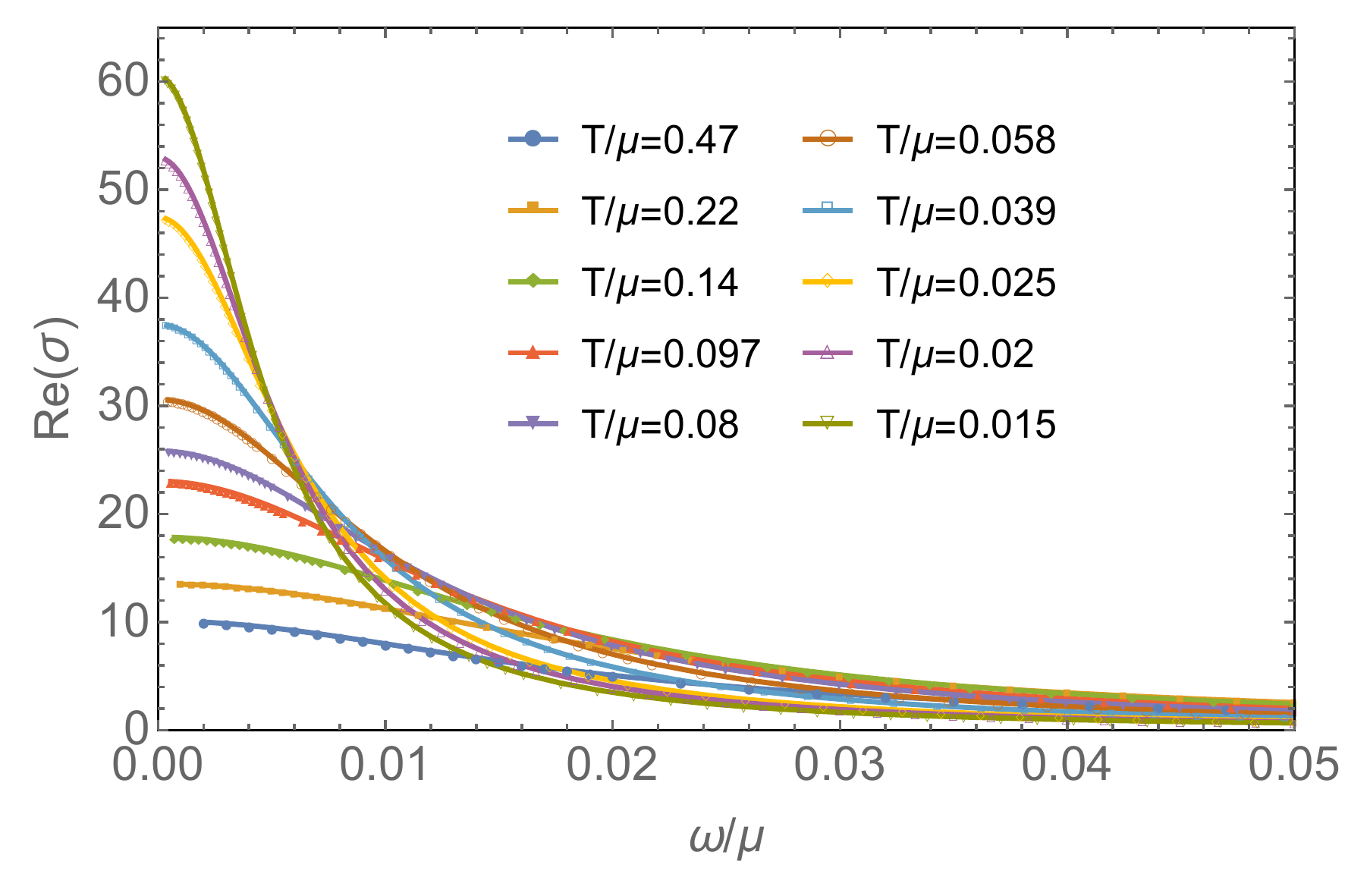}\includegraphics[height=5.5cm]{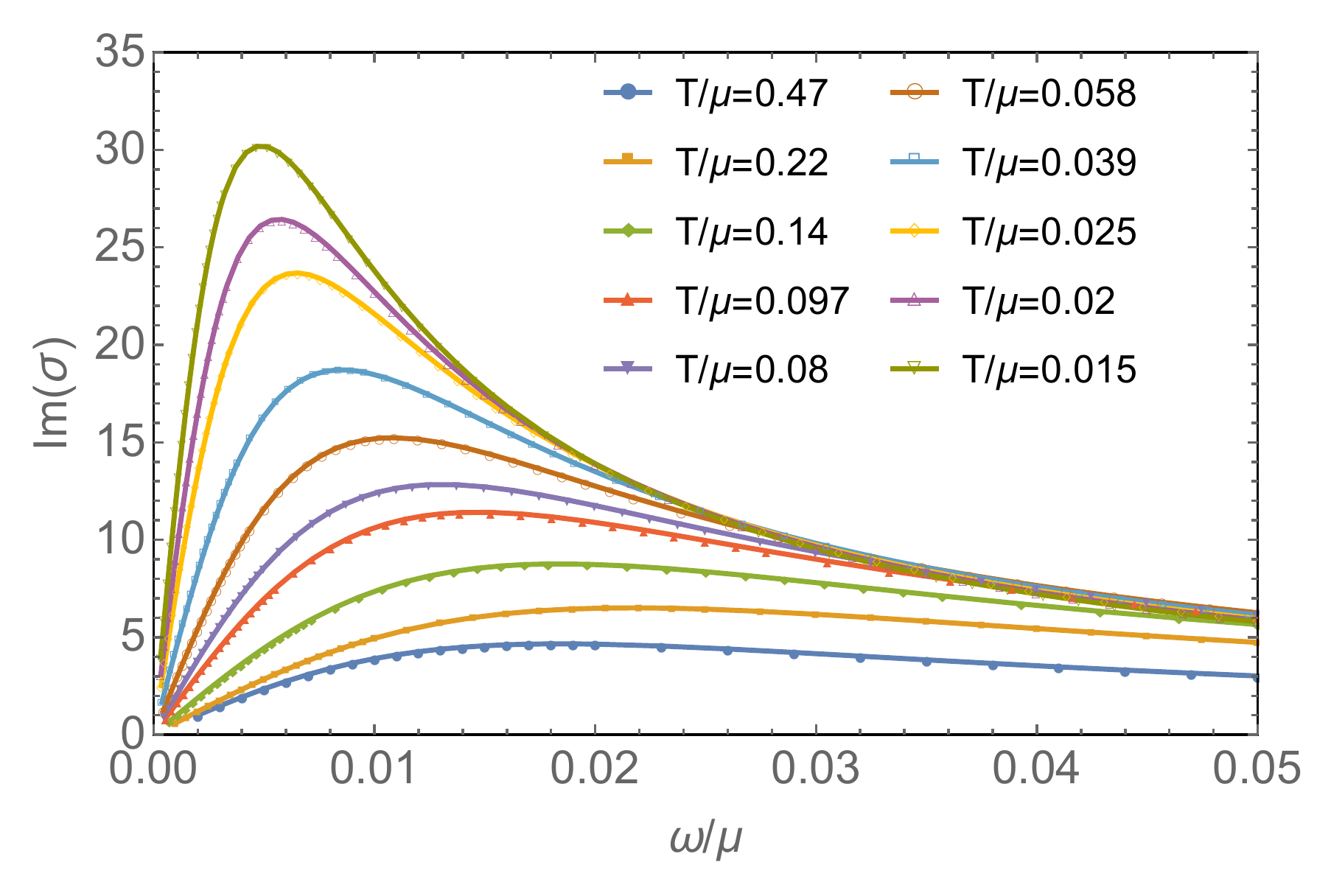}\\
\includegraphics[height=5.5cm]{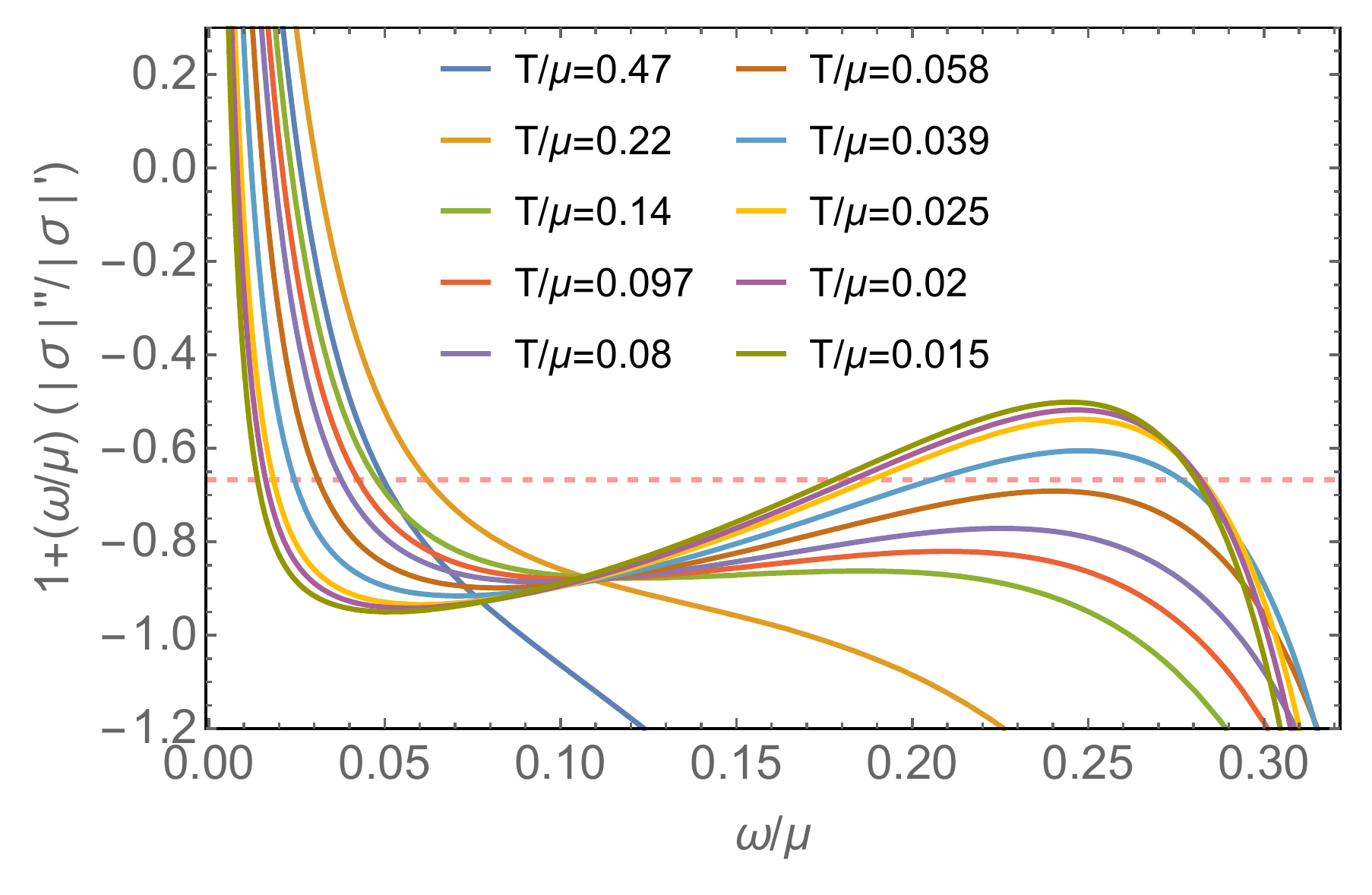}
\caption{The real (top left) and the imaginary (top right) parts of the optical conductivity $\sigma$ as a function of 
$\omega/\mu$ for a monochromatic lattice $\mu(x)/\mu=1+A\cos\left(k\,x\right)$, with $A=1/2$, $k/\mu=1/\sqrt{2}$, and various $T/\mu$ close to the origin. The conductivity clearly shows a Drude-like peak developing at low temperatures. 
The bottom figure shows the corresponding behaviour of $1+(\omega/\mu)\,\left|\sigma\right|^{\prime\prime}/\left|\sigma\right|^{\prime}$ 
and there is no evidence of a mid-frequency intermediate scaling with exponent $-2/3$. 
Note the different horizontal scale in the top and bottom figures.
}\label{fig:S2}
\end{figure}

In figure \ref{fig:S2} we show the real and imaginary parts of the optical conductivity for a monochromatic lattice with
$A=1/2$, $k/\mu=1/\sqrt{2}$ and for various temperatures.
We have only plotted low temperatures and small values of frequency in order to highlight some important features. In particular, we see a 
Drude-type peak emerging at low-temperatures, as also seen in \cite{Horowitz:2012gs}. 
Indeed, for low frequency we find an excellent
two-parameter fit\footnote{As in \cite{Donos:2013eha}, one can also
make a four parameter fit: $1/\sigma= (a_1+a_2\omega^2)-i\omega(a_3+a_4\omega^2)$, for constants $a_i$, where one uses the fact
that $\sigma^*(\omega)=\sigma(-\omega)$, and it leads to very similar results.} of the form
\begin{align}\label{drude}
\sigma\sim \frac{K\tau}{1-i\omega\tau}\,.
\end{align}
This fit is carried out for $\omega<<T$; in practise
in the region $10^{-4}\lesssim\omega/\mu\lesssim 10^{-2}$ and only for values of $\omega/\mu$
significantly smaller than the maximum in Im$(\sigma)$ (see figure \ref{fig:S2}).
This leads to the results, including a numerical result for the DC conductivity given by $K\tau$, which we summarise in table \ref{tabone}.
Comparing this quantity with the result that is obtained
from our formulae \eqref{finform} in the last section,
we find excellent agreement for both monochromatic and dichromatic lattices; see figure \ref{fig:DC}.
We also note for comparison that the 
value of $K/\mu$ for the AdS-RN black hole (i.e. with no lattice deformation) is given by
$(K/\mu)_{RN}=q^2/(\mu(T^{tt}+T^{xx}))$ \cite{Hartnoll:2007ih} and hence
\begin{align}\label{krn}
(K/\mu)_{RN}= \frac{1}{2\sqrt{3+(4\pi T/\mu)^2}-4\pi T/\mu}
\end{align}
which differs a little from the lattice results.
\begin{table}[!th]
\begin{center}
\setlength{\tabcolsep}{0.45em}
\begin{tabular}{|c|c|c|c|c|}
\hline
$T/\mu$ & $\tau\mu$ &$K/\mu$&$K\tau$& $2<\omega\tau< 8$\\
\hline
0.14 & 53.31 & 0.33 & 17.86&$0.036<\omega/\mu< 0.15$\\
0.080 & 75.99 & 0.34 & 25.84 &$0.026<\omega/\mu< 0.11$\\
0.039 & 117.05 & 0.32 & 37.53&$0.017<\omega/\mu< 0.068$\\
0.020 & 175.58 & 0.30 & 52.98&$0.011<\omega/\mu< 0.046$\\
0.015& 204.66 & 0.30 & 60.52&$0.0098<\omega/\mu< 0.039$\\
\hline
\end{tabular}
\end{center}
\caption{Parameters after fitting to the Drude behaviour \eqref{drude} for small $\omega$,
for the monochromatic lattices displayed in figure \ref{fig:S2}.Note that $K\tau$ gives a numerical estimate for the DC conductivity which can be compared with the analytic result; see figure \ref{fig:DC}.}
\label{tabone}
\end{table}

We now make some specific comparisons with the results of \cite{Horowitz:2012gs}. 
To do so, we need to take into account 
a relative scaling of the chemical potential: $\mu=\sqrt{2}\mu^{there}$. Then the monochromatic lattices that we have been considering for the specific temperature $T/\mu=0.039$ correspond to those in figures 6-9 of \cite{Horowitz:2012gs}.
We find very good agreement with figure 6 which shows the charge density of the background black holes. We find
less good agreement (of the order of a couple of percent) with the plot of the AC conductivity in figure 8 of \cite{Horowitz:2012gs}. Furthermore, in figure 9 of
\cite{Horowitz:2012gs} distinct kinks are found in the AC conductivity which we do not see for these or in fact any of our lattices. We discuss the issue of intermediate scaling claimed in \cite{Horowitz:2012gs} in the next subsection.

\subsubsection{Absence of intermediate scaling}
The next feature that we would like to report on is the possibility of an intermediate frequency scaling behaviour of the form 
\begin{align}\label{fitform}
|\sigma(\omega)|\sim B\omega^{-2/3}+C\,,
\end{align}
where $B,C$ are frequency independent constants. 
Such a scaling was reported in \cite{Horowitz:2012ky,Horowitz:2012gs}, based on log-log plots,
for the approximate range $2<\omega\tau< 8$, where $\tau$ is obtained from the fit to the Drude peak. 
If this scaling is present then a sharp diagnostic is that
$1+(\omega/\mu)|\sigma|''/|\sigma|'\sim -2/3$. 
As illustrated in figure \ref{fig:S2}, we find no evidence for such scaling (the relevant range of $\omega/\mu$ is given in table \ref{tabone}). 
Moreover, we find similar behaviour to what we saw for a homogeneous Q-lattice in \cite{Donos:2013eha}. 
Finally, we highlight that at very low temperatures, where the Drude peak becomes sharper, the function  
$1+(\omega/\mu)|\sigma|''/|\sigma|'$ approaches 2 as $\omega\to 0$ and, in addition, there is a scaling region with exponent $-1$, visible in 
figure \ref{fig:S2}; both of these features arise from \eqref{drude}.

We make a final comparison with \cite{Horowitz:2012gs} for the specific case of $T/\mu=0.039$.
In figure 9 of \cite{Horowitz:2012gs} a log-log plot suggested a scaling with exponent $-2/3$ for the approximate range
$0.02\lesssim\omega/\mu\lesssim 0.07$. However, there is no evidence for this scaling in 
the bottom panel in figure \ref{fig:S2}. In fact, for this range of $\omega/\mu$ we can see from the top panels in figure \ref{fig:S2}
that we are not too far from the Drude peak. Indeed, we have checked that our fit to the Drude behaviour is in fact rather reasonable\footnote{As an aside, 
if instead one fits to the form \eqref{drude} over this
entire range of $\omega/\mu$ (leading to different non-physical values of $K,\tau$ than those given in table 1), then on a 
log-log plot one finds excellent agreement with the data. 
This underscores the difficulties in deducing power-law behaviour from a log-log plot.}
over this entire range of $\omega/\mu$.

\subsubsection{Scaling behaviour and $AdS_2\times\mathbb{R}^2$ in the IR as $T\to 0$}
We next discuss how the black holes behave as $T\to 0$. When the constant part of
$\mu(x)$ is non-vanishing, i.e. $\mu\ne 0$, for the monochromatic lattices we find that in the far IR the solutions
all seem to approach $AdS_2\times \mathbb{R}^2$. More precisely, as we discuss in the next paragraph, we find that the 
black hole solutions exhibit a low temperature scaling behaviour that are consistent with the $T=0$ solutions
interpolating between
$AdS_4$ in the UV and $AdS_2\times \mathbb{R}^2$ 
perturbed by an irrelevant operator in the locally quantum critical theory in the IR.
In particular, for the values of $k$ that we have looked at and for the temperatures we have looked at, we find no evidence for the ``floppy" ground states discussed in \cite{Hartnoll:2014gaa}. Some additional comparisons with this work are made in appendix \ref{floppy}.

In figure \ref{fig:Snew}
we show the behaviour of the
DC conductivities $\sigma$ and $\bar\kappa$, obtained from \eqref{finform},
as well as their log-derivatives,
as a function of temperature for four different
monochromatic lattices. At low temperatures we see that the conductivities approach 
the scaling behaviour 
given in \eqref{dcscal} and \eqref{defDel} as depicted by the dashed red lines.
Note that the low-temperature scaling
is obtained by taking the lowest temperature black hole to deduce the approximate value of the renormalised wave-number $\bar k$ given in \eqref{defDel}. The renormalisation factor $\bar \lambda$, defined in \eqref{renscale} is actually very small:
for example it is $\bar\lambda=1.027$, for the case $A=1/2$, 
$k/\mu=1/\sqrt{2}$ (red) and for other cases it is given in table \ref{green}.
Note the former case has $\bar\kappa\to0$ while for the latter case
$\bar\kappa\to\infty$ as $T\to 0$. At high temperatures we see that $\sigma\to 1+2/A^2=9$ in agreement with \eqref{hightsc}.
\begin{figure}
\centering
\includegraphics[height=5.cm]{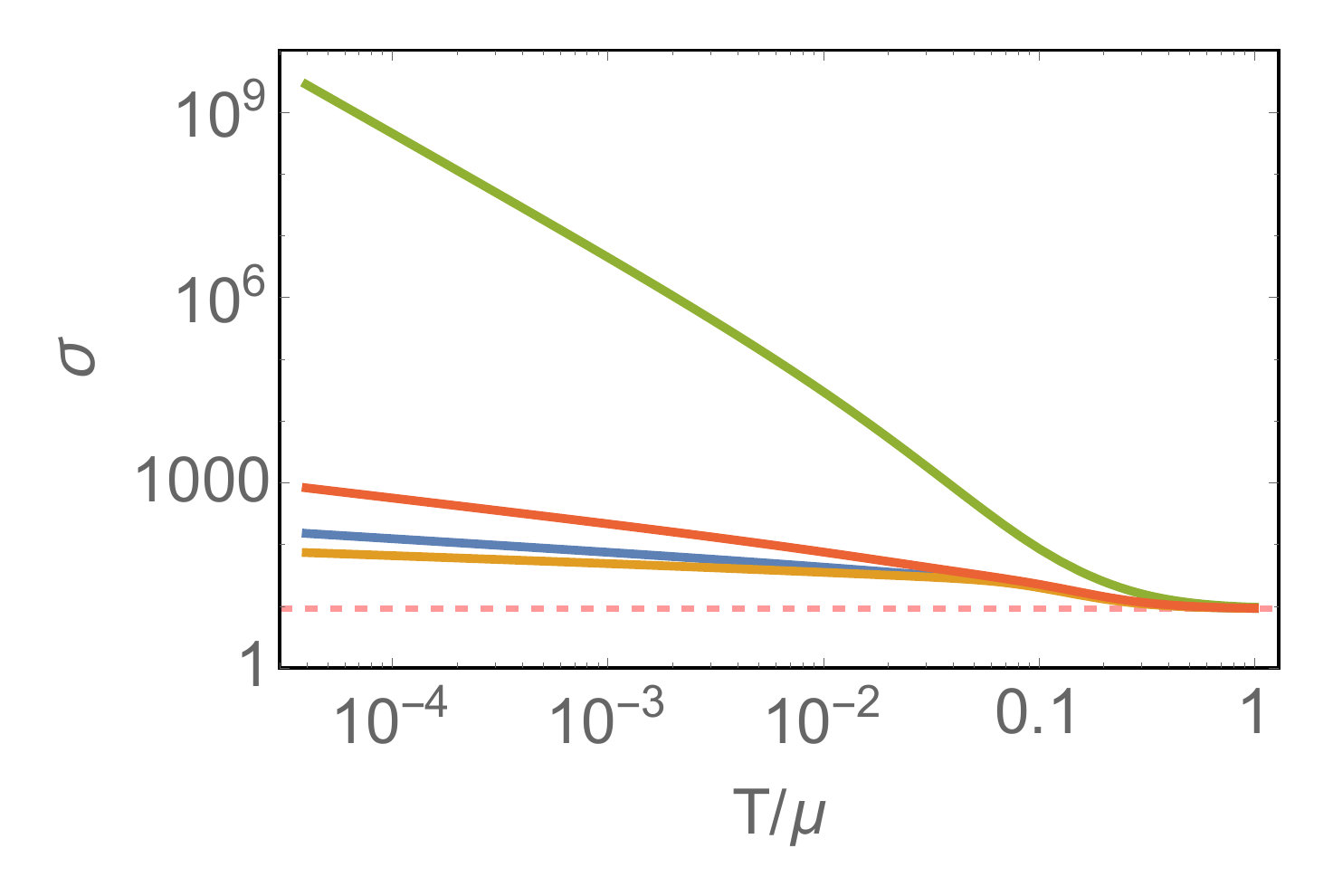}
\includegraphics[height=5.cm]{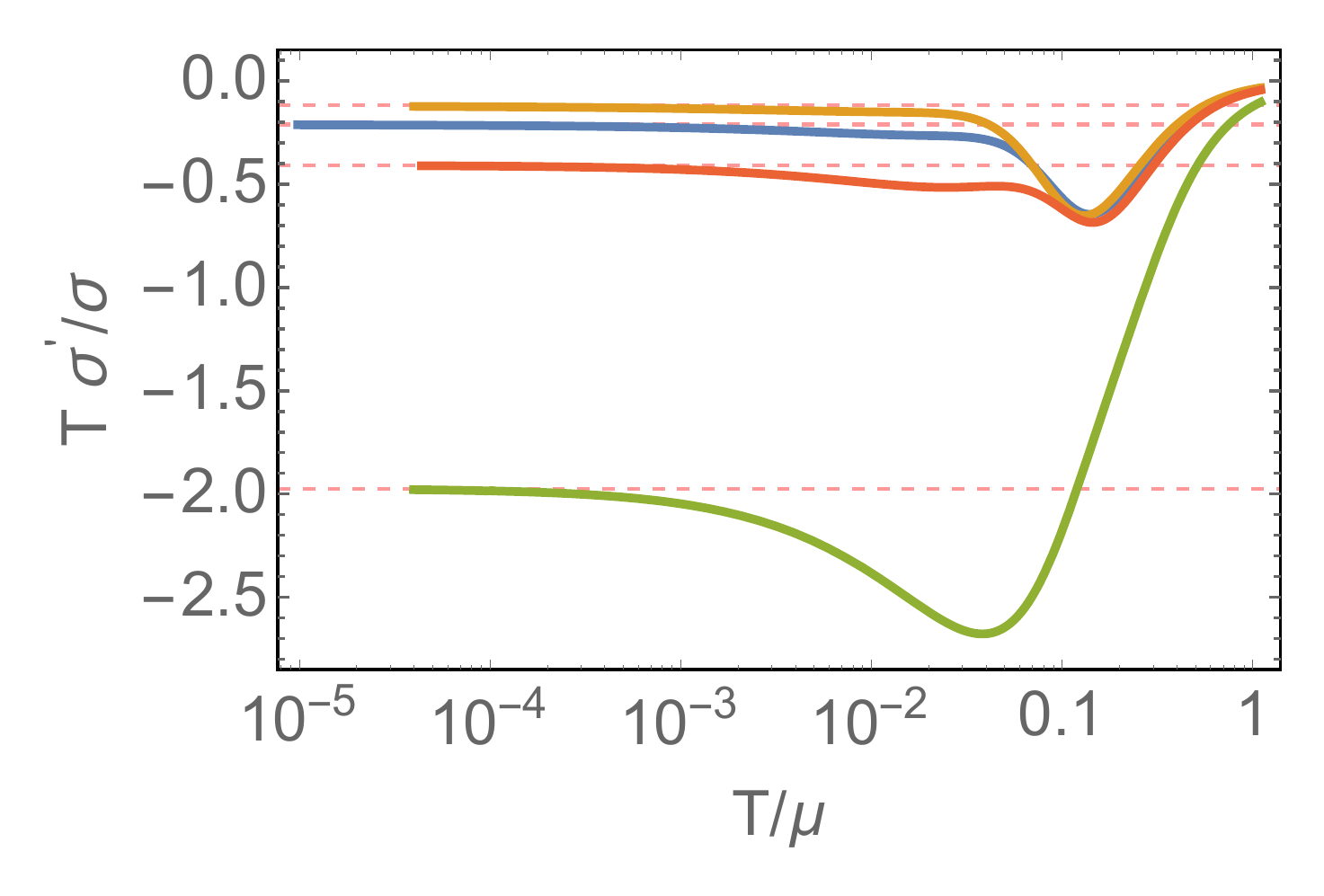}\\
\begin{picture}(0.1,0.25)(0,0)
\put(12,80){\makebox(0,0){\line(0,1){5}}}
\end{picture}
\includegraphics[height=5.cm]{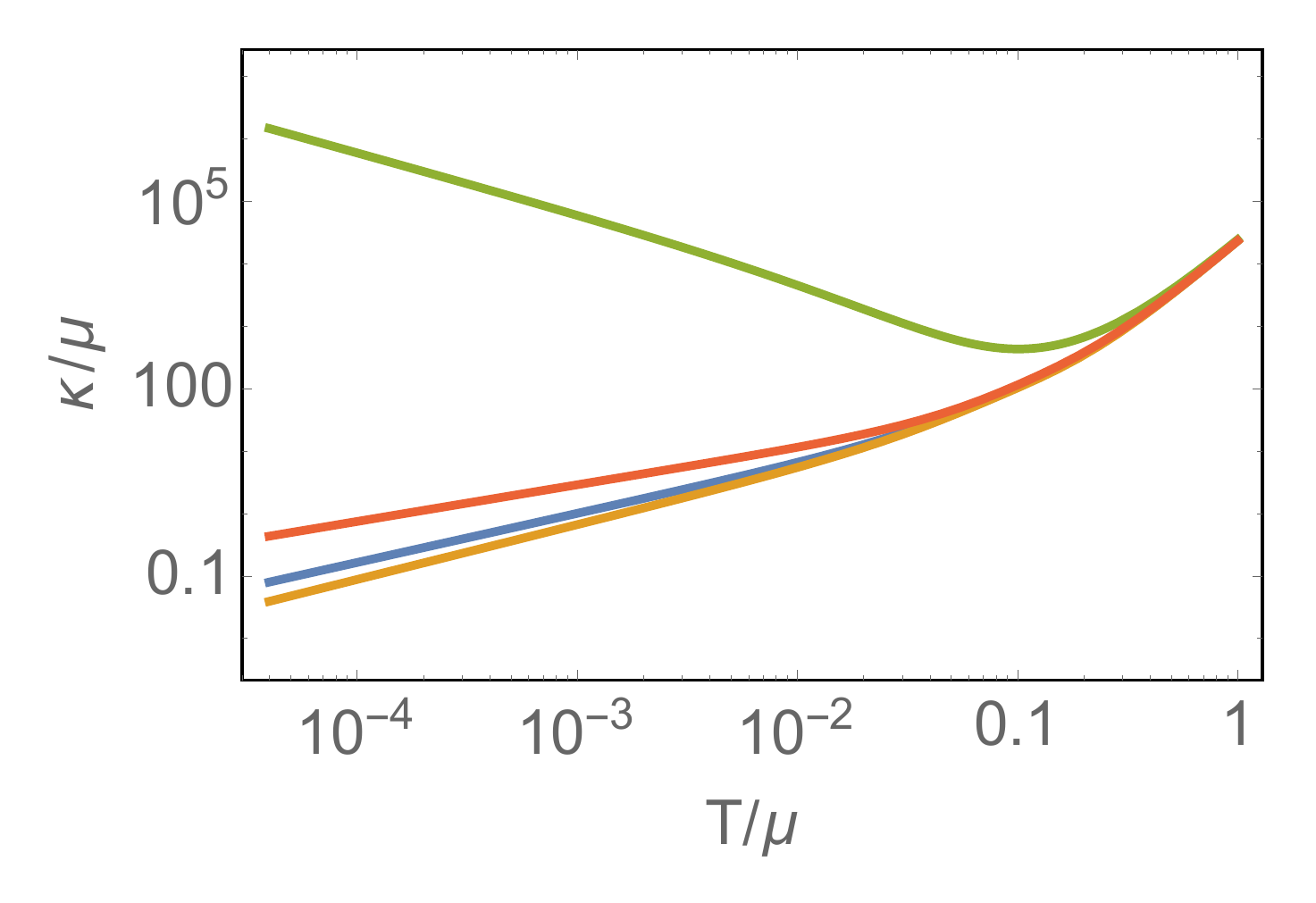}
\begin{picture}(0.1,0.25)(0,0)
\put(13,83){\makebox(0,0){\line(0,1){5}}}
\put(13,95){\makebox(0,0){\line(0,1){5}}}
\end{picture}
\includegraphics[height=5.cm]{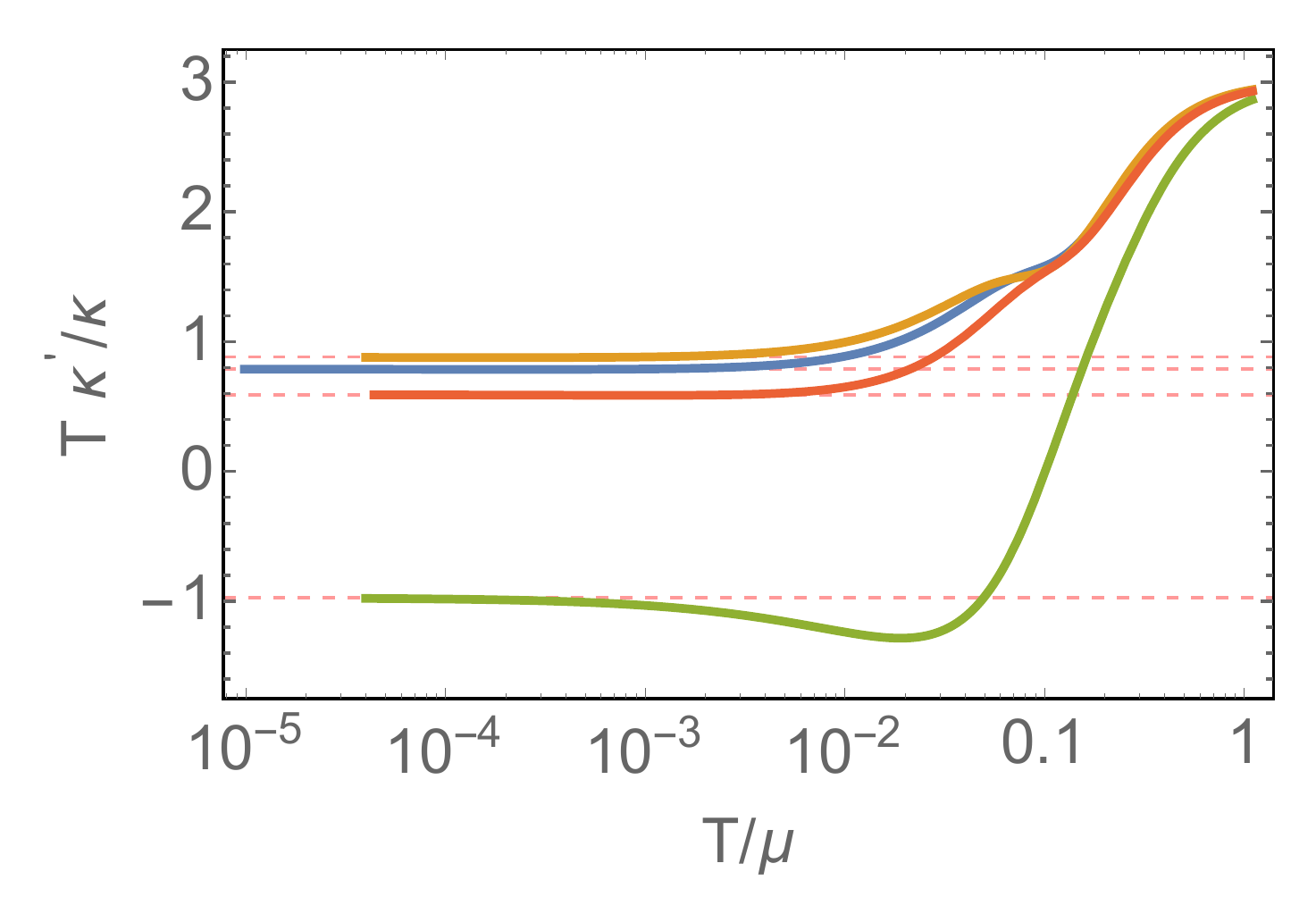}\\
\caption{
Plots of the DC conductivity for $\sigma$ (top left) and $\bar\kappa$ (bottom left), obtained from \eqref{finform},
against temperature for four monochromatic lattices of the form
$\mu(x)/\mu=1+A\,\cos\left(k\,x\right)$, all with $A=1/2$ and $k/\mu=\sqrt{2}/3$ (orange), $2\sqrt{2}/5$ (blue), $1/\sqrt{2}$ (red)
and $\sqrt{2}$ (green). The red dashed lines on the right hand plots indicate the low-temperature scaling behaviour,
given in \eqref{dcscal} expected for black holes approaching $AdS_2\times\mathbb{R}^2$ in the far IR. 
The $k/\mu=\sqrt{2}$ case provides an example where  $\bar\kappa$ diverges as $T\to 0$,
while the other cases are examples where $\bar\kappa$ vanishes 
as $T\to 0$. In all cases $\kappa$ vanishes linearly with $T$. As $T\to \infty$ we see that $\sigma\to 1+2/A^2=9$, marked with a red dashed line 
in the top left figure, in agreement with \eqref{hightsc}.
}\label{fig:Snew}
\end{figure}
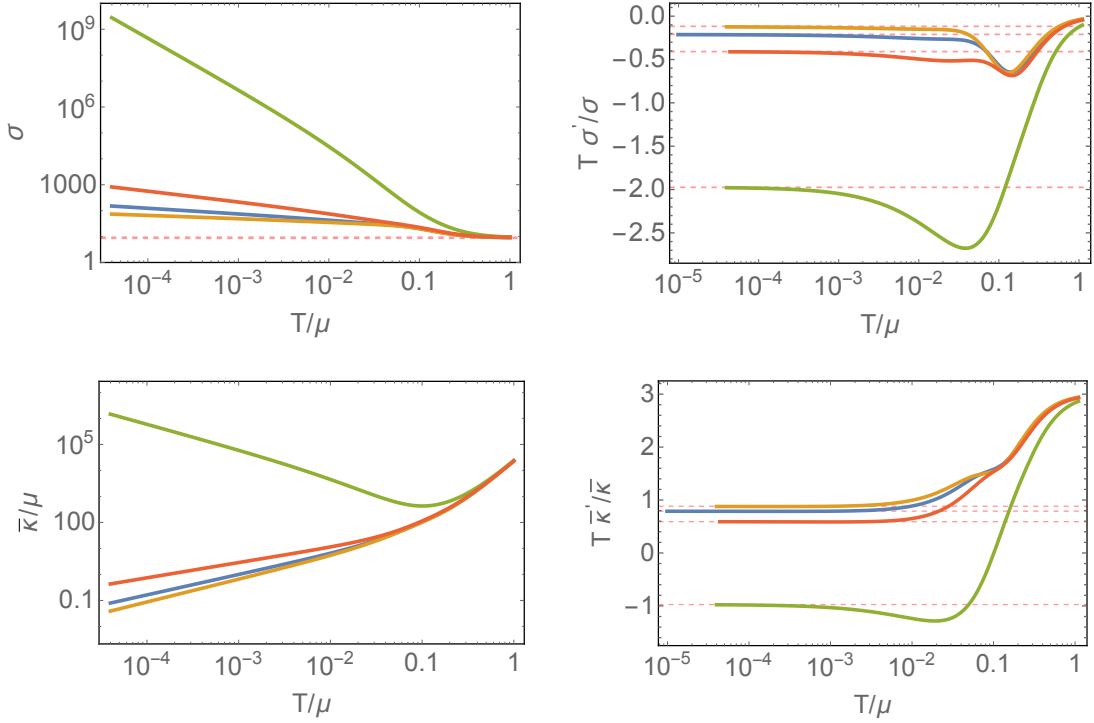

\subsubsection{Sum rules on conductivity}
\begin{figure}
\centering
\includegraphics[height=4.5cm]{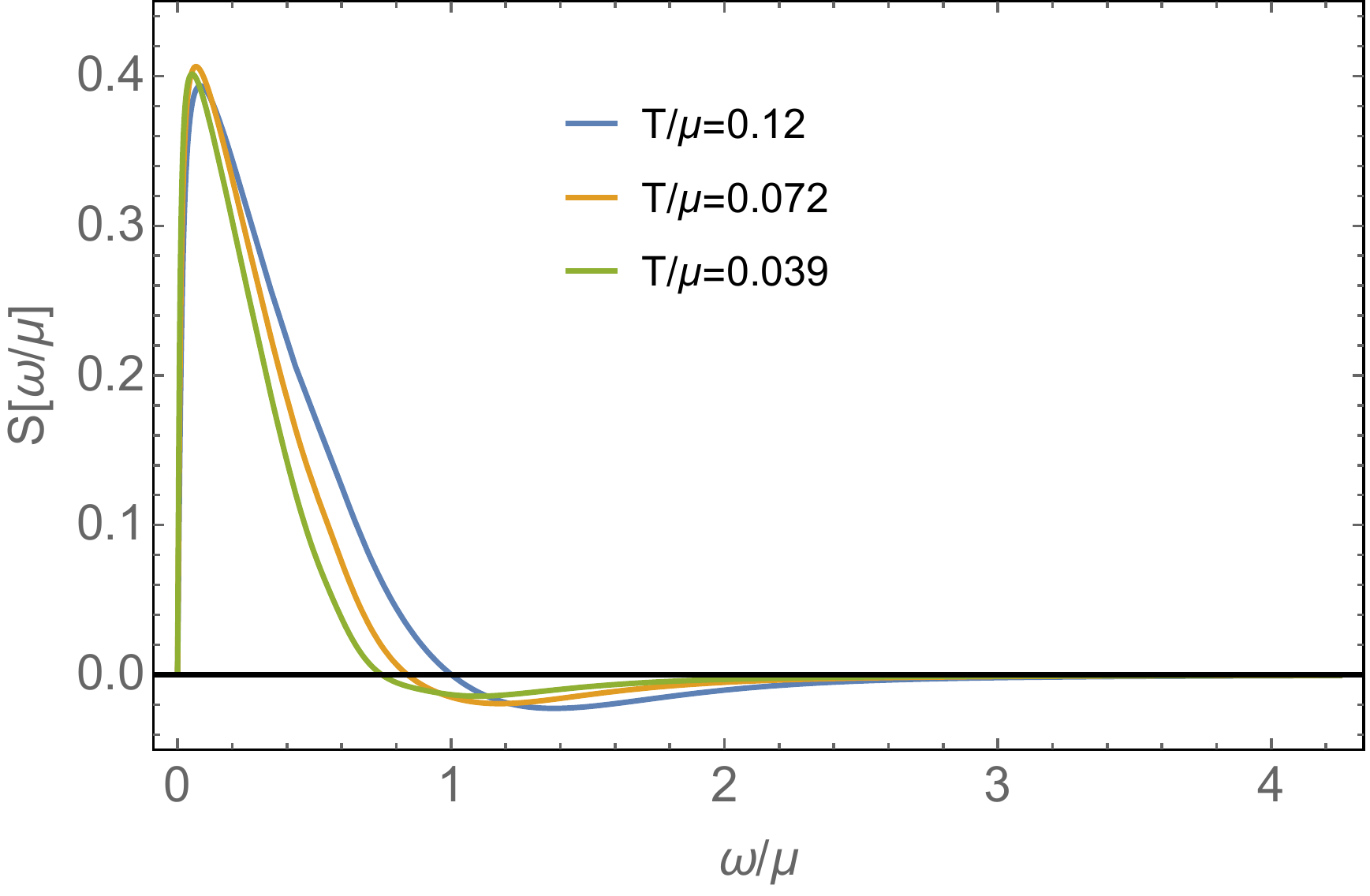}\quad\includegraphics[height=4.5cm]{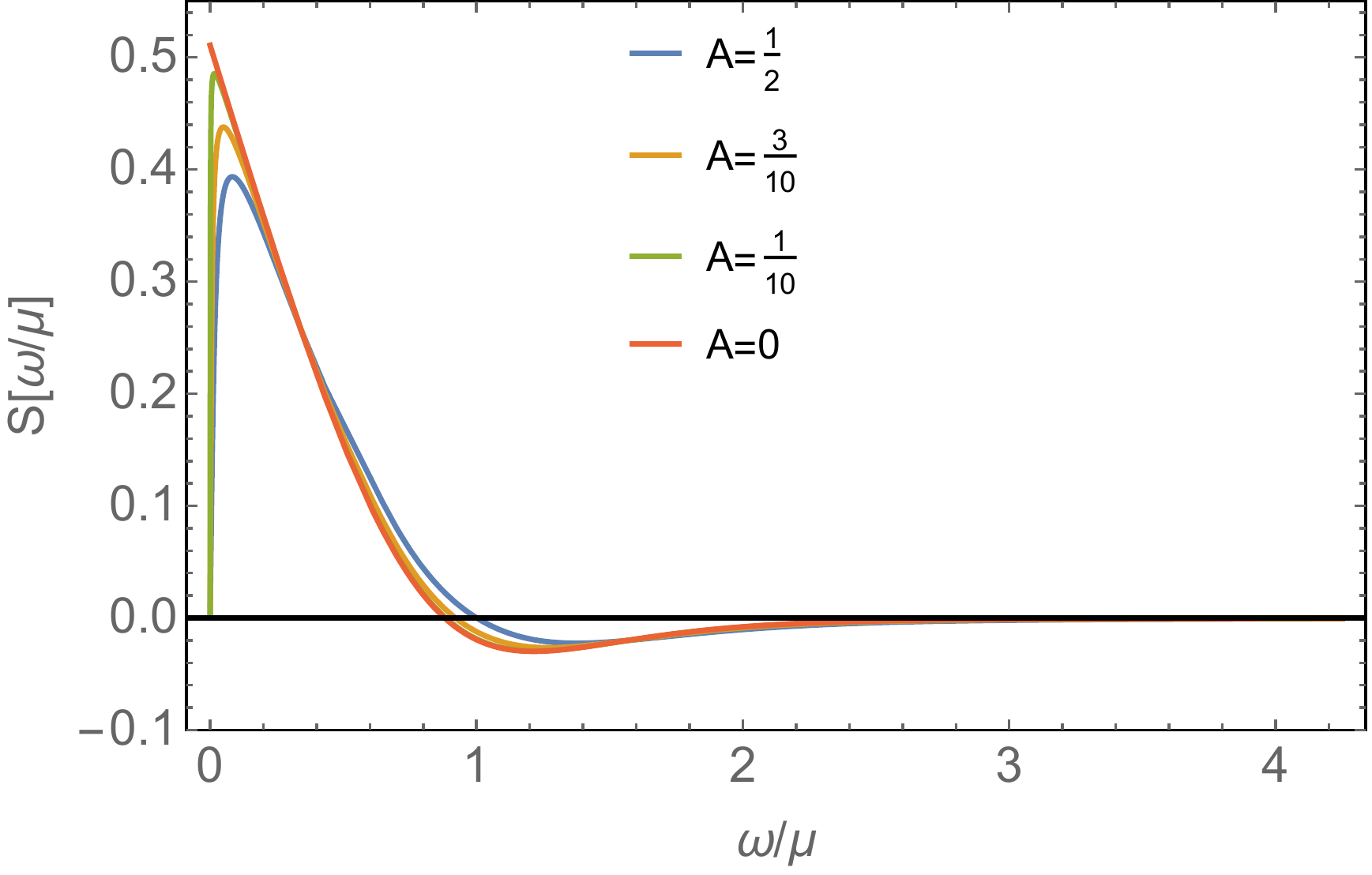}\\
\includegraphics[height=4.5cm]{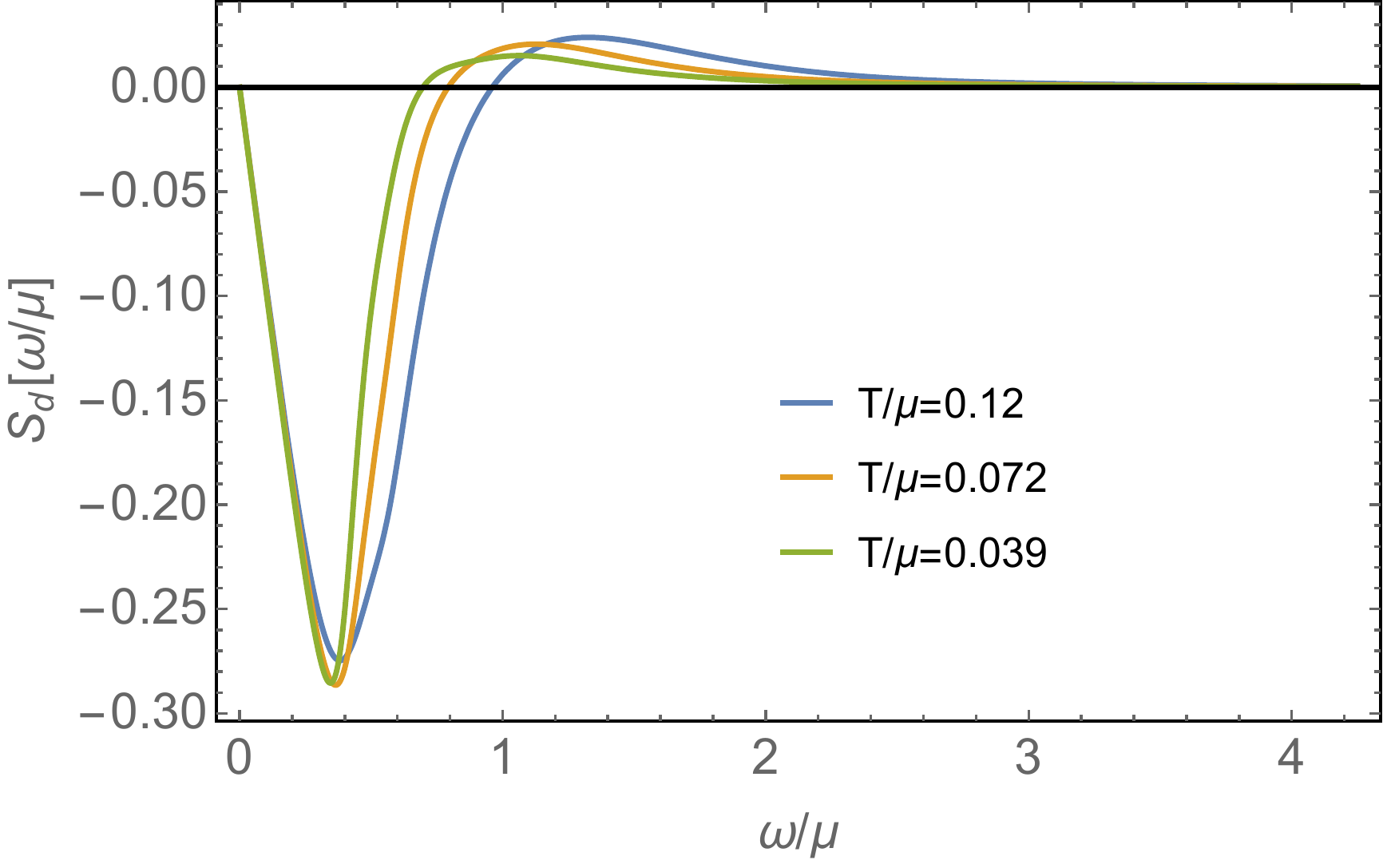}
\caption{Sum rules for monochromatic lattices. The top left panel 
plots the integrated spectral weight $S(\omega/\mu)$, defined in \eqref{sumrulefn}, for 
a monochromatic lattice $\mu(x)/\mu=1+A\cos\left(k\,x\right)$, with $A=1/2$, $k/\mu=1/\sqrt{2}$ (as in figure \ref{fig:S2}) for
three different temperatures, and we see it vanishes when $\omega/\mu\to\infty$ as expected from the first sum rule.
As $T/\mu\to 0$ we see that $S(\omega/\mu)$ is developing a step-like behaviour corresponding to the appearance of
a delta function with weight smaller than the $T=0$ AdS-RN black hole (which has the value $\sim0.45$).
The top right panel considers monochromatic lattices with $k/\mu=1/\sqrt{2}$ and fixed $T/\mu=0.12$ and various $A$. As $A\to 0$
we see that the $S(\omega/\mu)$ is developing a step-like behaviour corresponding to the appearance of
a delta function with the same weight as the AdS-RN black hole at the same temperature (which for this case has the value $\sim 0.51$). The bottom panel plots
$S_d(\omega/\mu)$, defined in \eqref{sumrulefnd}, for the same monochromatic lattices
as in the top left panel and we see that the second sum rule is also satisfied.
}\label{fig:sumrule}
\end{figure}

It is illuminating to check sum-rules associated with the AC electrical conductivity. 
As $\omega\to\infty$ we have $\sigma(\omega)\to 1$ and hence after defining
the integrated spectral weight as:
\begin{align}\label{sumrulefn}
S(\omega/\mu)\equiv \int_{0}^{\omega/\mu} (Re[\sigma(\omega')]-1)d\omega'\,,
\end{align}
following  \cite{Gulotta:2010cu} we might expect that $\lim_{\omega\to\infty}S(\omega)\to 0$.
We briefly discuss the proof highlighting the underlying assumptions. 
In the absence of instabilities the retarded Greens function $G_{J^x J^x}(\omega)$ is analytic in the upper half plane
and we assume this includes the real axis. We also need to assume that for $Im(\omega)>0$ that 
the function $G_{J^x J^x}(\omega)-i\omega$ vanishes faster than $1/|\omega|$ 
as $|\omega|\to\infty$. The Kramers-Kr\"onig relations then imply that $\lim_{\omega\to \infty} S(\omega)= \lim_{\omega\to 0^+} \pi/2\text{Re}[G_{J^x J^x}(\omega)-i\omega]= 0$, provided that 
$\text{Re}[G_{J^x J^x}(0)]=0$ which is satisfied in our case (see figure \ref{fig:S2}).
This sum rule is manifest in figure \ref{fig:sumrule} for monochromatic lattices. 

We also note from figure \ref{fig:sumrule}
that as $T\to 0$, the function $S(\omega)$ is developing a step like behaviour near $\omega\to 0$, corresponding to the
Drude-peak becoming a delta function exactly at $T=0$. It would appear that the weight of the delta function is slightly smaller than that of
the AdS-RN black hole.
To see this, and to make an additional comparison, we note that the electrical conductivity of the AdS-RN black hole has a delta function
for all temperatures with $\sigma^{RN}(\omega)=1+\sigma_0(\omega)+ K_{RN}\left(\frac{i}{\omega}+\pi \delta(\omega)\right)$ where
$\sigma_0(\omega)$ is an analytic function that falls off faster than $1/|\omega|$ at infinity and $K_{RN}$ is given in \eqref{krn}.  
Thus for AdS-RN black holes, as $\omega/\mu \to 0$ we should have $S(\omega/\mu)\to (\pi/2)K_{RN}/\mu$. 
At $T=0$ we have $K_{RN}/\mu=q/\mu^2=1/(2\sqrt{3})$ and hence $S(\omega/\mu)\sim 0.45$ as $\omega/\mu\to 0$, which is slightly bigger than the weight of the delta function
appearing at $T=0$ for the lattice black holes. We can also consider lattice black holes at finite temperature with fixed $k/\mu$ and then take the lattice
strength $A\to 0$. In this limit we should find that as $\omega\to 0$, $S(\omega/\mu)$ should approach the 
AdS-RN result at the same temperature; this is also confirmed in figure \ref{fig:sumrule} for the case of $T/\mu=0.12$ for which
$(\pi/2)K_{RN}/\mu\sim 0.51$.

We can also consider a different sum rule first discussed in \cite{WitczakKrempa:2012gn}. Defining
\begin{align}\label{sumrulefnd}
 S_d(\omega/\mu)\equiv \int_{0}^{\omega/\mu} (Re[\frac{1}{\sigma(\omega')}]-1)d\omega'\,,
\end{align}
the sum rule is $\lim_{\omega\to\infty}\tilde S_d(\omega)\to 0$. This arises from the
electromagnetic duality of the $D=4$ Einstein-Maxwell theory, with the dual gauge-field
being associated with a second CFT arising from an alternative quantisation scheme \cite{Witten:2003ya}
(for related discussion see also \cite{Herzog:2007ij,Hartnoll:2007ip,Myers:2010pk,Jokela:2013hta}.) In our setup the lattice deformation
with chemical potential $\mu(x)$ gets mapped to a magnetic field that is spatially modulated in the $x$ direction. 
We have verified this sum rule as shown in figure \ref{fig:sumrule}.

\subsubsection{Intermediate resonances}
\begin{figure}
\centering
\includegraphics[height=5.5cm]{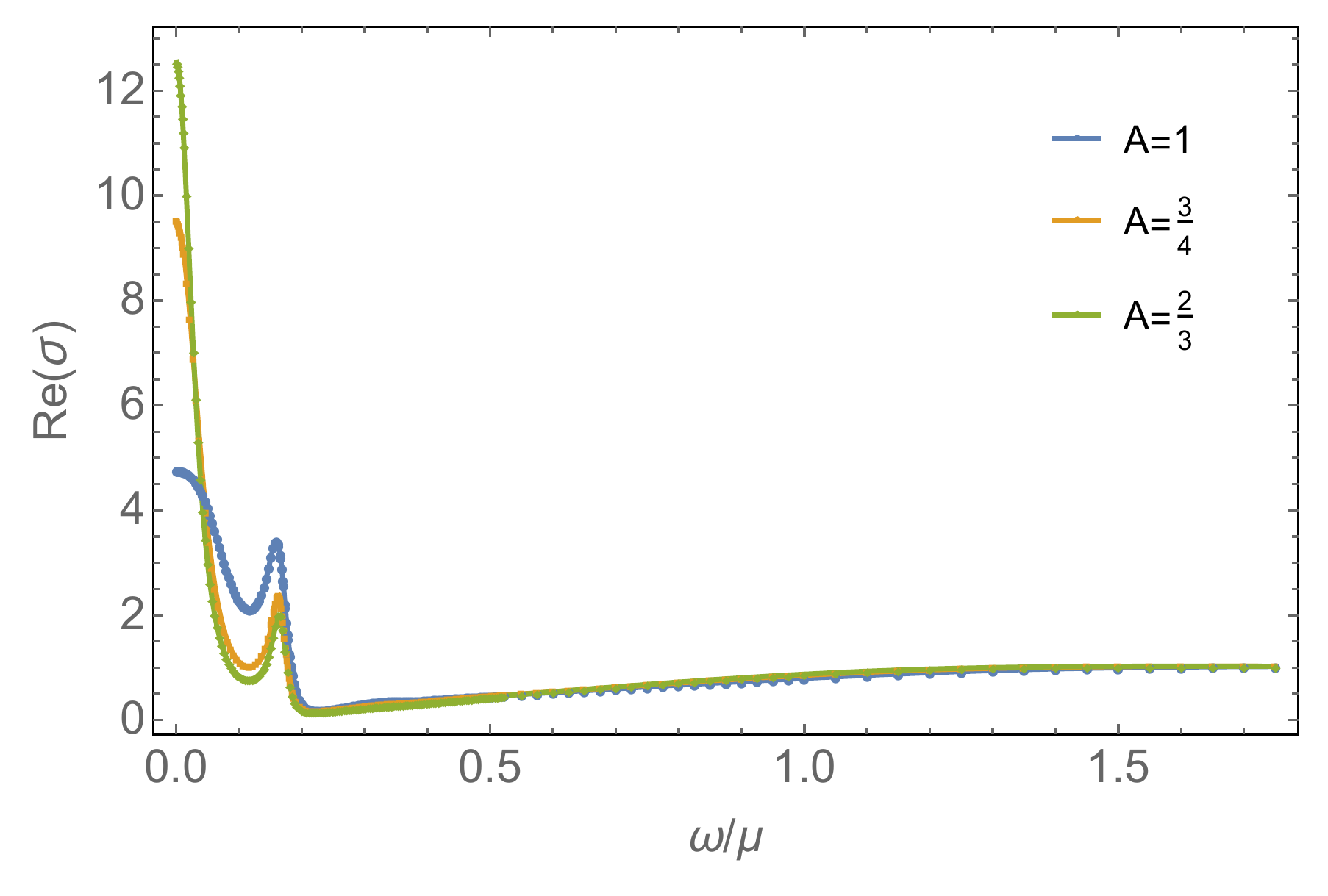}\includegraphics[height=5.5cm]{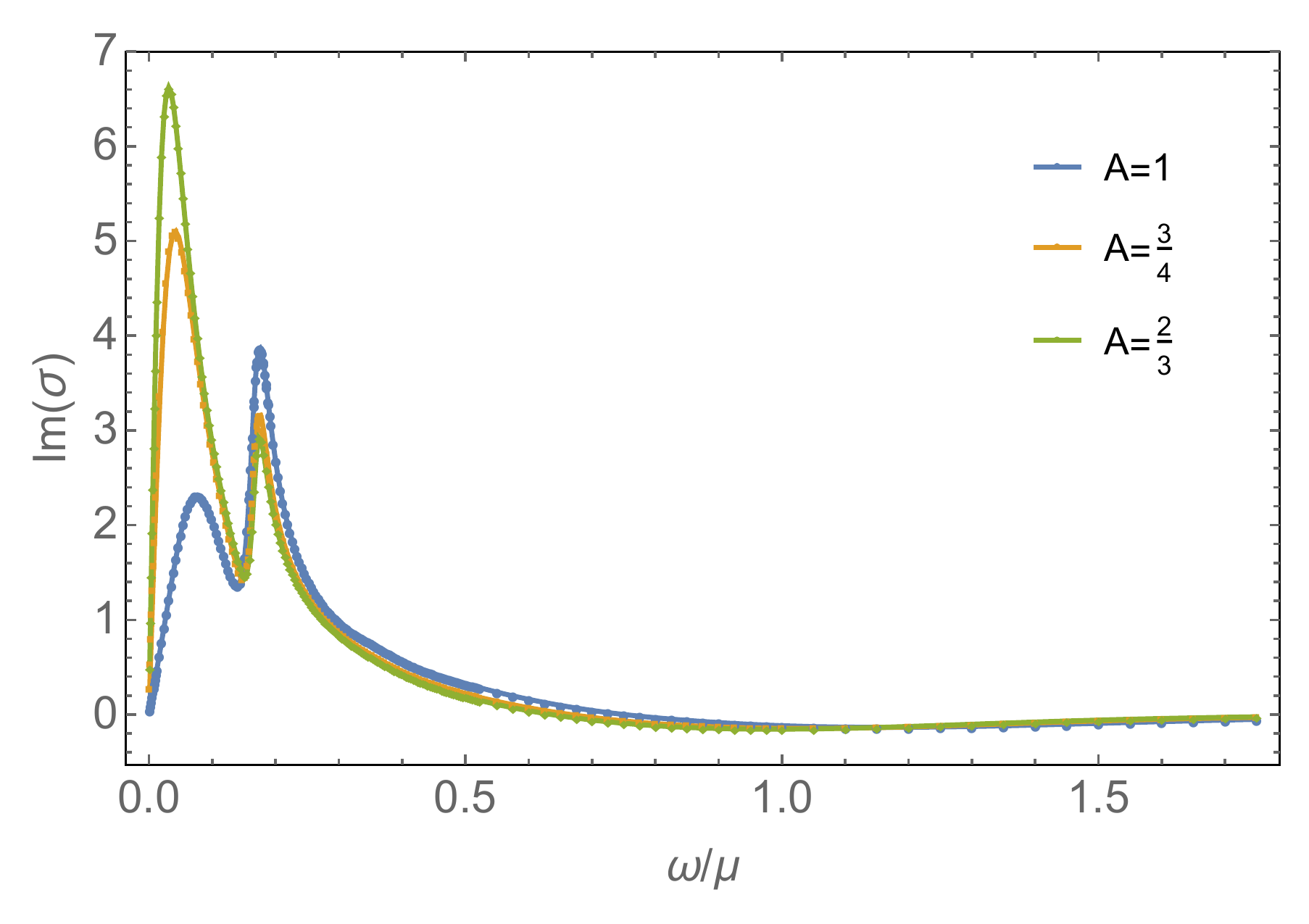}
\caption{
The real (left) and the imaginary (right) parts of the optical conductivity as a function of $\omega$ for various monochromatic lattices $\mu(x)/\mu=1+A\,\cos\left(k\,x\right)$. The three different cases have fixed temperature $T/\mu\approx 0.0795$ and period $k/\mu=\left(3\,\sqrt{2}\right)^{-1}$ but varying lattice strength $A$. We clearly see the appearance of a resonance associated with the sound mode frequency $\omega/\mu\sim v_sk/\mu\sim 1/6$.
}\label{fig:S1}
\end{figure}

Next, we highlight some interesting features of the optical conductivity that appear at intermediate frequencies, as illustrated in
figure \ref{fig:S1}. In particular for the monochromatic lattices with $k/\mu=1/(3\sqrt{2})$ and various lattice strengths $A$, we find that there is a bump in the 
optical conductivity just outside the Drude-peak. Now the Drude peak arises because there is a pole near $\omega =0$ in the
$T^{tx}T^{tx}$ correlator. One might expect that there could be additional features due to contributions from the 
sound modes. From the analysis of \cite{Edalati:2010pn} 
at $T=0$ we have $v_s=1/\sqrt{2}$ and furthermore it was shown that $v_s$ has only a weak dependence on temperature in \cite{Davison:2011uk}.
Thus we might expect to see a resonance appear near $\omega/\mu\sim v_s (k/\mu)\sim1/6$, 
and
this is what is seen in figure \ref{fig:S1}.
Note that such resonances are also seen for the lattices with $A=1/2$, $k/\mu=1/\sqrt{2}$ at $\omega/\mu\sim v_sk/\mu\sim 1/2$, but these lie outside
the range plotted in figure \ref{fig:S2}.

\begin{figure}
\centering
\includegraphics[height=5.5cm]{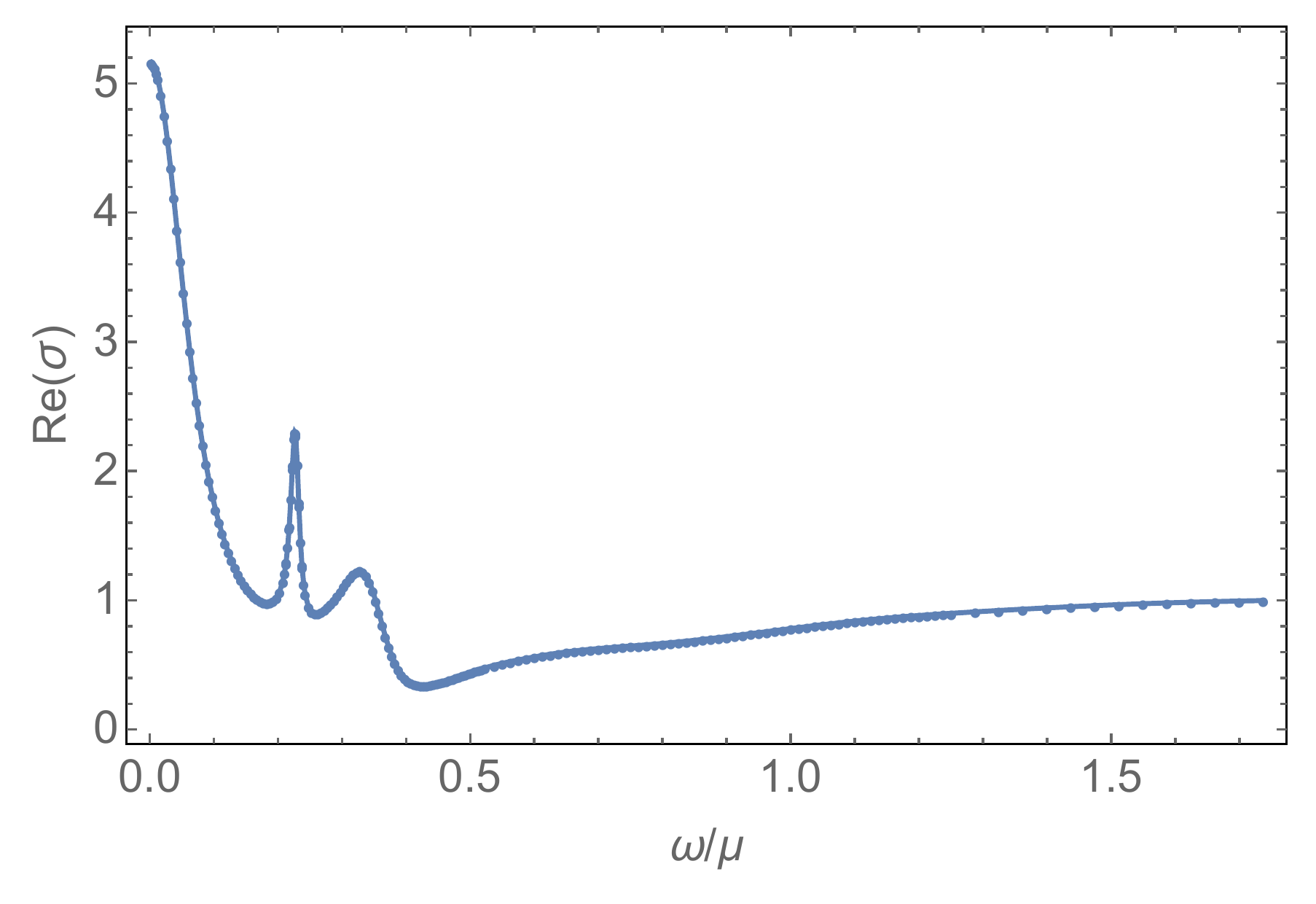}\includegraphics[height=5.5cm]{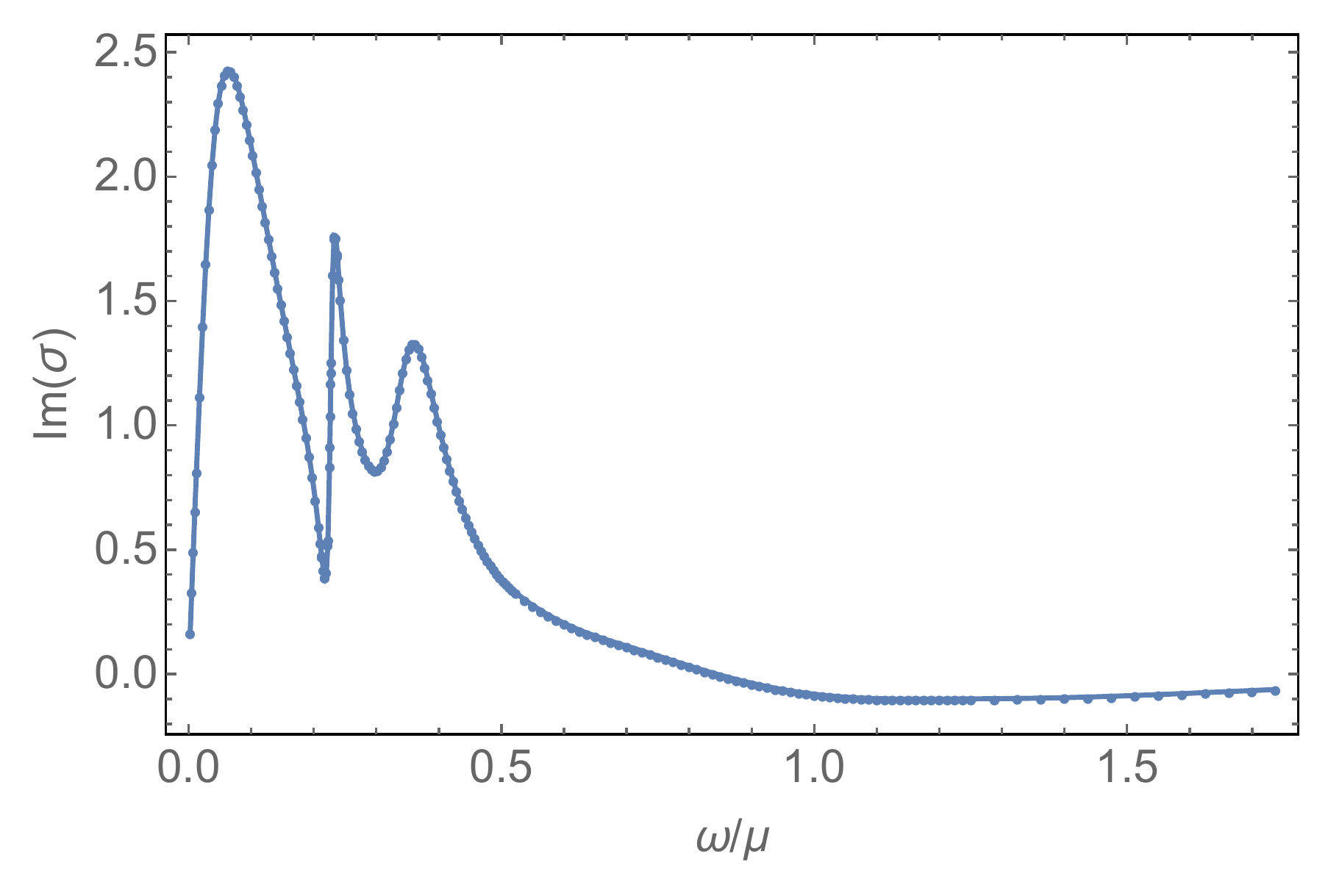}
\caption{The real (left) and the imaginary (right) parts of the optical conductivity as a function of $\omega/\mu$ for the dichromatic lattice
$\mu(x)/\mu=1+A\cos\left(k\,x\right)+B\cos\left(2k\,x\right)$, with $A=1/2$, $B=1$, $k/\mu=1/\left(3\sqrt{2}\right)$ and $T/\mu\approx 0.0796$. In this case we see two resonances 
associated with sound modes at 
$\omega/\mu\sim v_sk/\mu\sim1/6$ and also $\omega/\mu\sim v_s(2k)/\mu\sim 1/3$.}\label{fig:S3}
\end{figure}

\begin{figure}
\centering
\includegraphics[height=5.5cm]{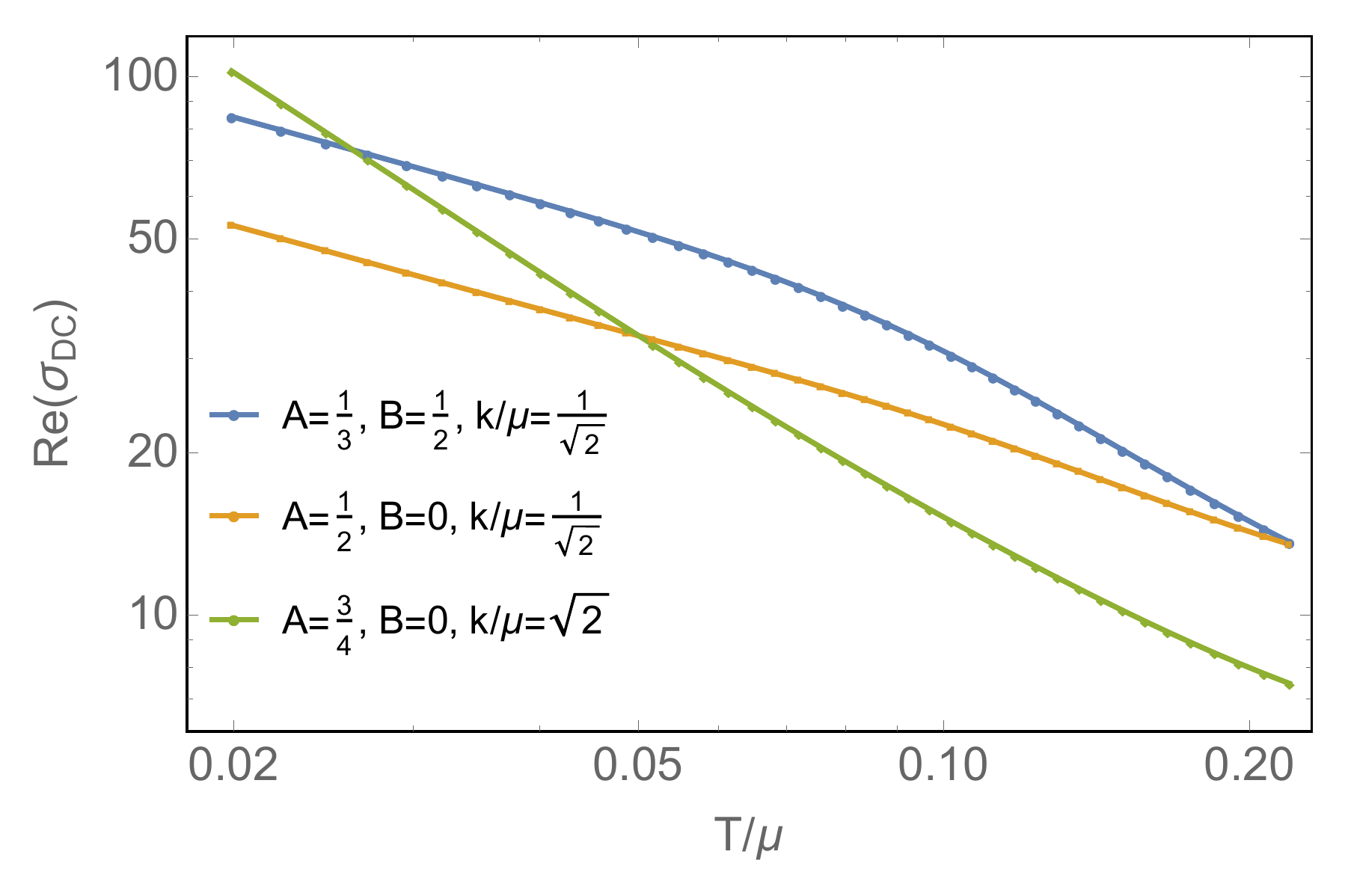}\includegraphics[height=5.5cm]{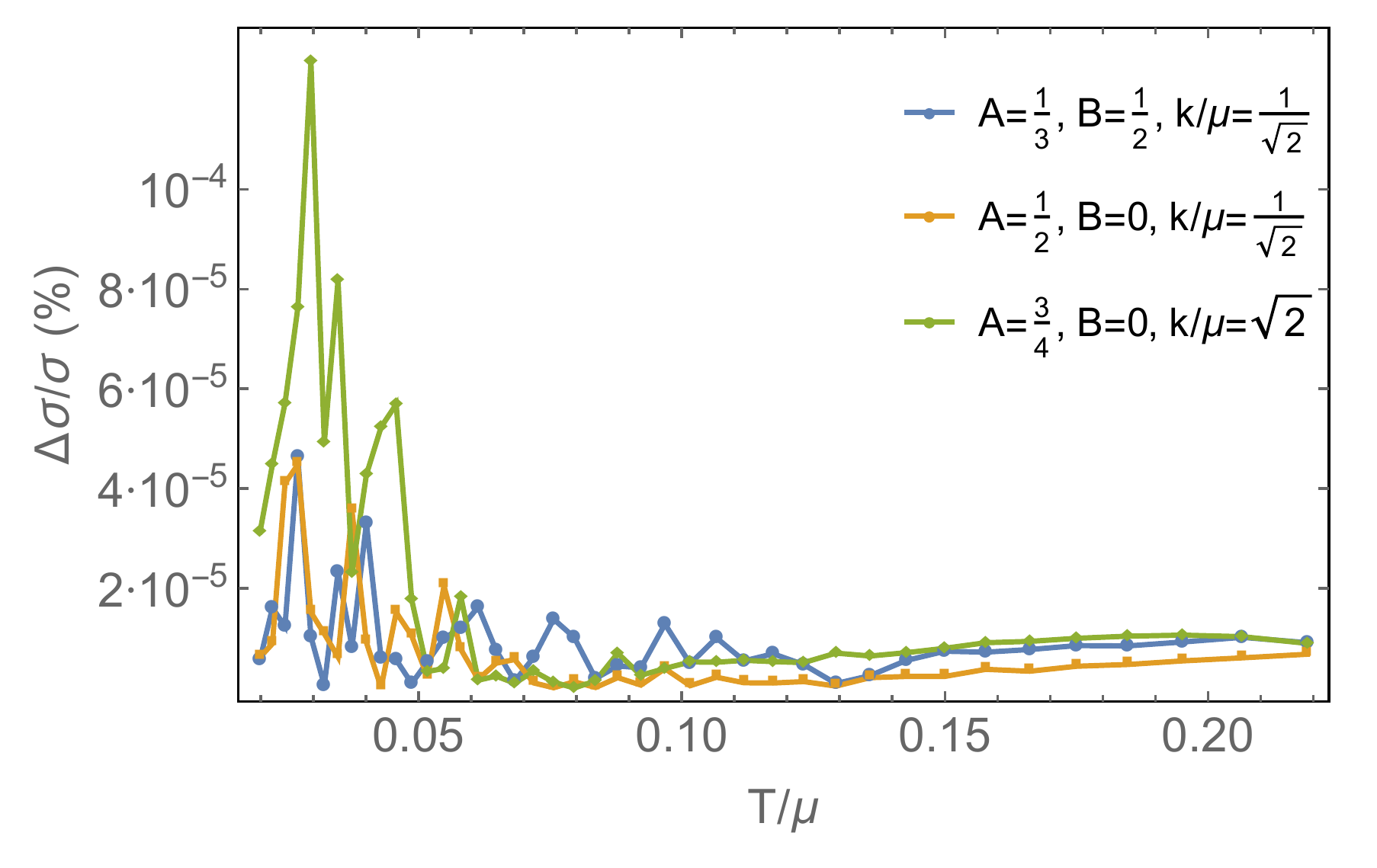}
\caption{Comparison of the two numerical results for the DC conductivity for three different lattices of the form
$\mu(x)/\mu=1+A\,\cos\left(k\,x\right)+B\,\cos\left(2k\,x\right)$. 
The first is obtained from using the analytic formulae involving black hole horizon data in equation \eqref{eq:DC_analytic}.
The second is obtained from the $\omega\rightarrow 0$ limit of the AC conductivity after fitting to a Drude-peak form.  The data is superimposed in the figure on the left and the difference is undetectable to the naked eye.
The relative difference is shown in the figure on the right hand and we see agreement at a level better than $10^{-4}\%$.}\label{fig:DC}
\end{figure}

Similarly, for the dichromatic lattices \eqref{dichrom}, containing wave-numbers $k$ and $2k$, 
we might expect to see structure in the optical conductivity at frequencies $\omega/\mu\sim v_s(k/\mu)$
and also twice this frequency. Such behaviour is illustrated in figure \ref{fig:S3}.

\subsubsection{Conductivities for higher Fourier modes}
\begin{figure}
\centering
\includegraphics[height=5.5cm]{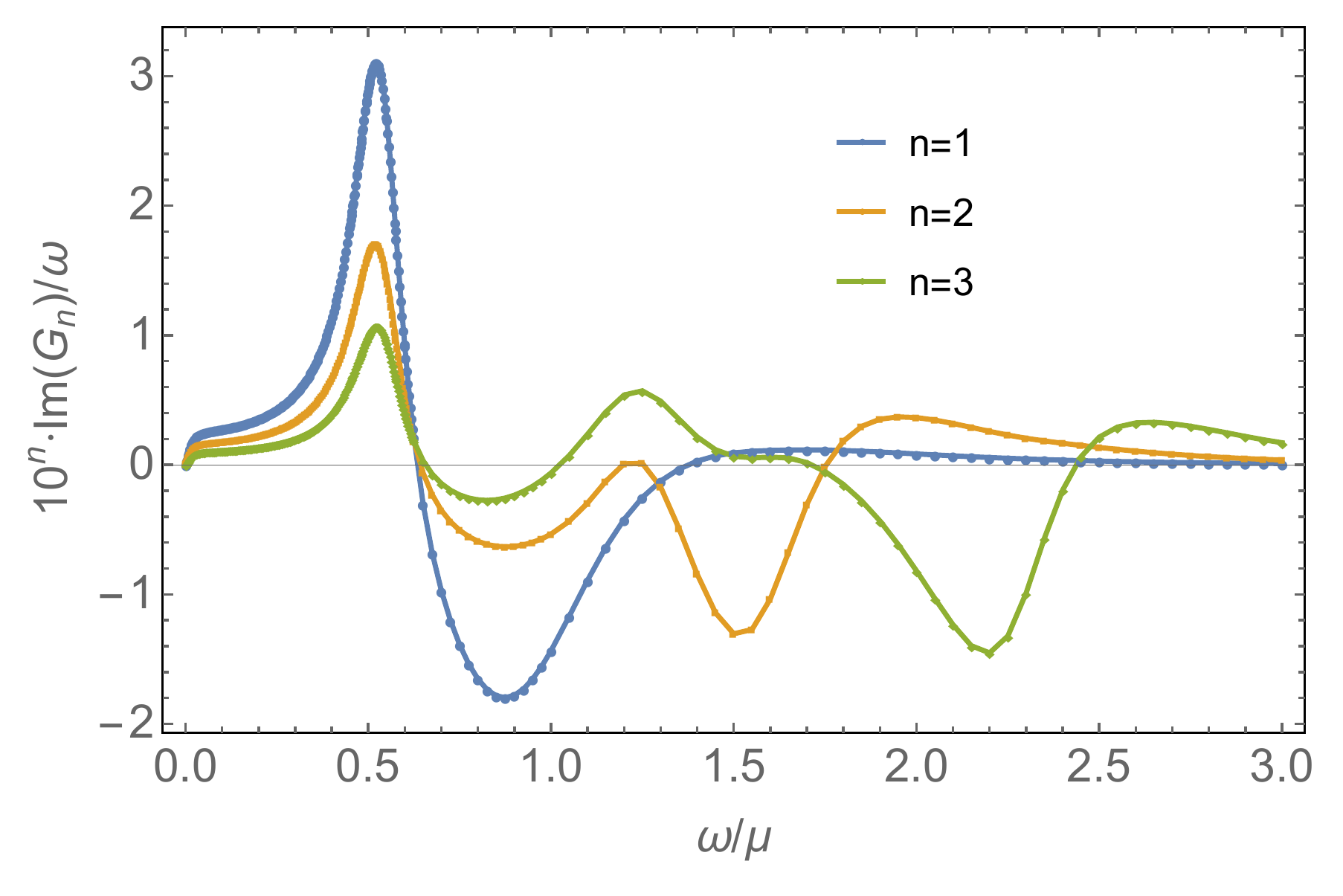}\includegraphics[height=5.5cm]{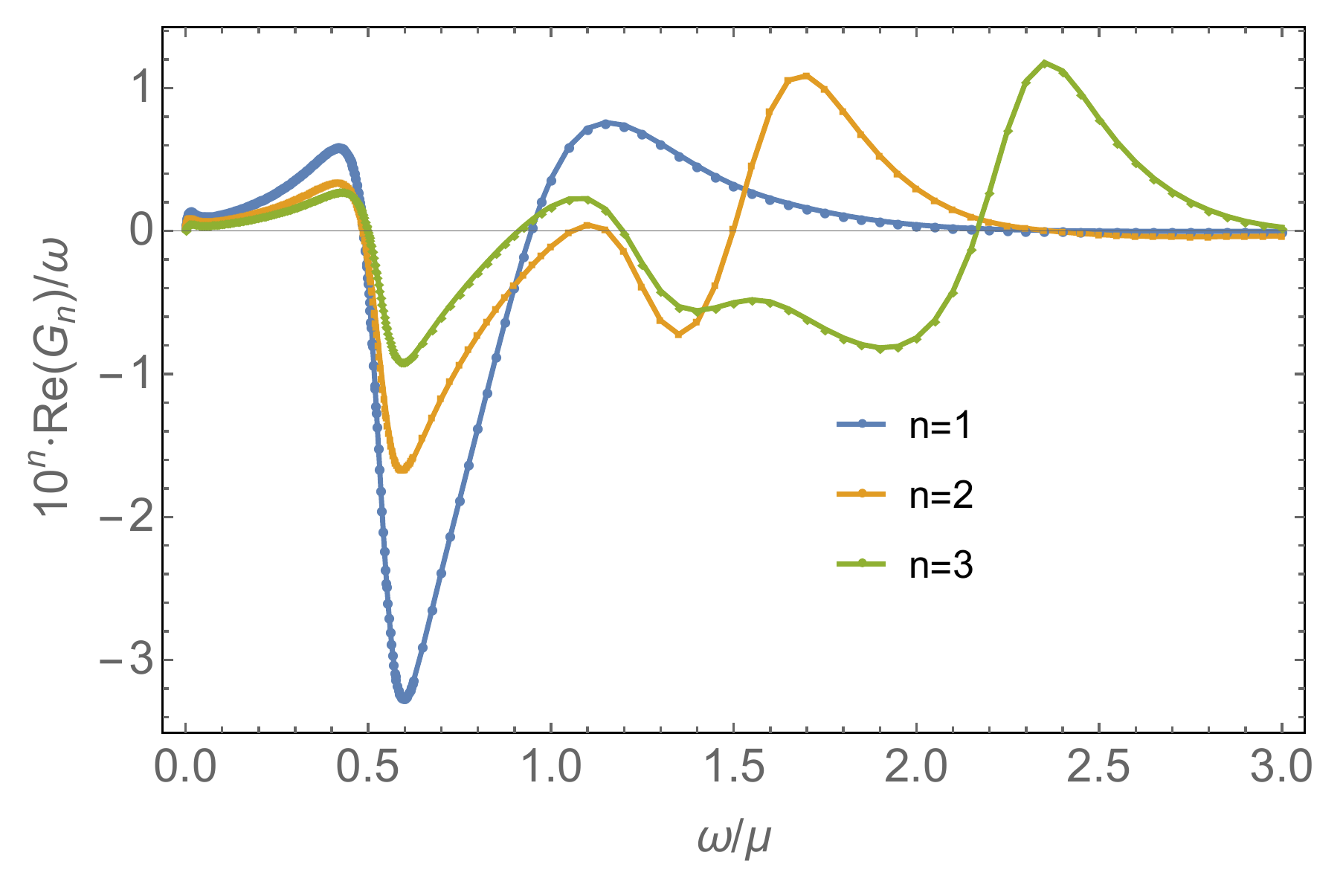}
\caption{A plot of the imaginary (left) and real (right) parts of $G_n/\omega$, where $G_n$ is the Green's function for the $n$th Fourier modes of the current as in \eqref{highfmodes}. The plots are
for the monochromatic lattice in figure 1 with $T/\mu=0.08$ and for
Fourier modes $n=1,2$ and $3$. Notice that different vertical scales are used for each $n$.
Observe they all vanish at $\omega=0$, as expected from current conservation, and the feature at 
$\omega/\mu\sim v_sk/\mu\sim 1/2$, associated with the sound mode.}\label{highmodes}
\end{figure}

Until this point we have focussed on the zero-mode of the current $\mathcal{J}$ appearing in
\eqref{genjcurrent} in order to extract the optical conductivity as in \eqref{defacc}.
We can also extract the higher Fourier modes of $\mathcal{J}$ and construct the corresponding Greens function.
If we write the $n$th Fourier mode as $\mathcal{J}_n$, then we can define
\begin{align}\label{highfmodes}
G_n=\frac{\mathcal{J}_n}{\mu_{J}}\,,
\end{align}
which defines the current two point correlator $G_{J_x J_x}\left(k_{1}=n\,k_{L},k_{2}=0, \omega\right)$. It is worth emphasising that
these correlators with $k_{2}\ne k_{1}$ are non-vanishing as a consequence of the broken translation invariance of
the backgrounds. For the monochromatic lattice of figure 1, with $A=1/2$, $k=1/\sqrt{2}$ and $T/\mu=0.08$, in figure 
\ref{highmodes} we have plotted the real
and imaginary parts of  $G_n/\omega$ for $n=1,2$ and $3$. Notice that the conservation of the current $\partial_a\mathcal{J}^a=0$ implies that at $\omega=0$ we have $\mathcal{J}_n=0$ if $n\ne 0$, as we see in the plot. Observe that there is a peak in the imaginary part at $\omega/\mu\sim 0.5$, which is associated with the sound mode at $\omega/\mu\sim v_sk/\mu\sim 1/2$. 

\subsubsection{A dirty lattice}\label{sec:dirty}
\begin{figure}
\centering
\includegraphics[height=5.5cm]{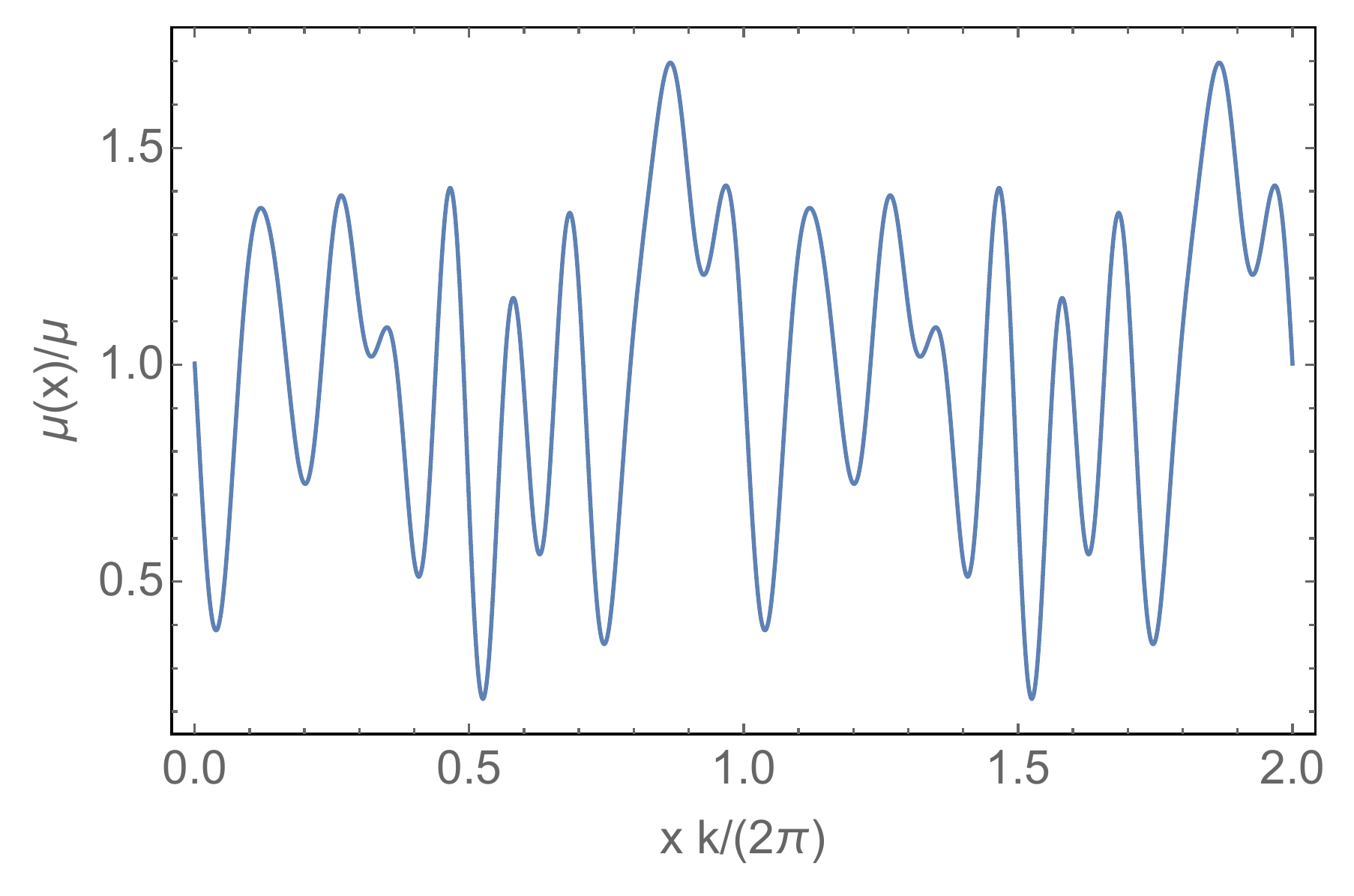}\\
\includegraphics[height=5.5cm]{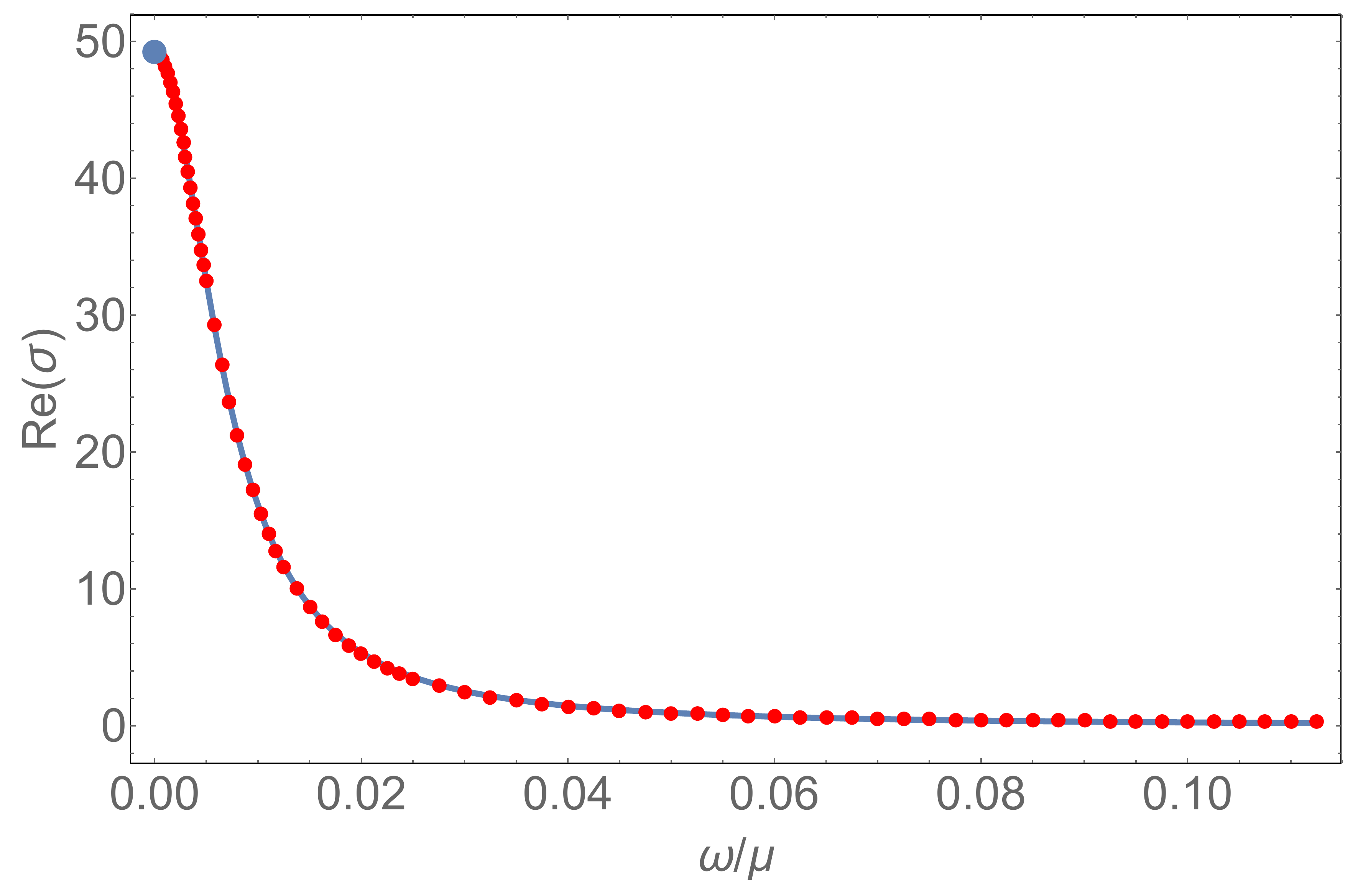}\includegraphics[height=5.5cm]{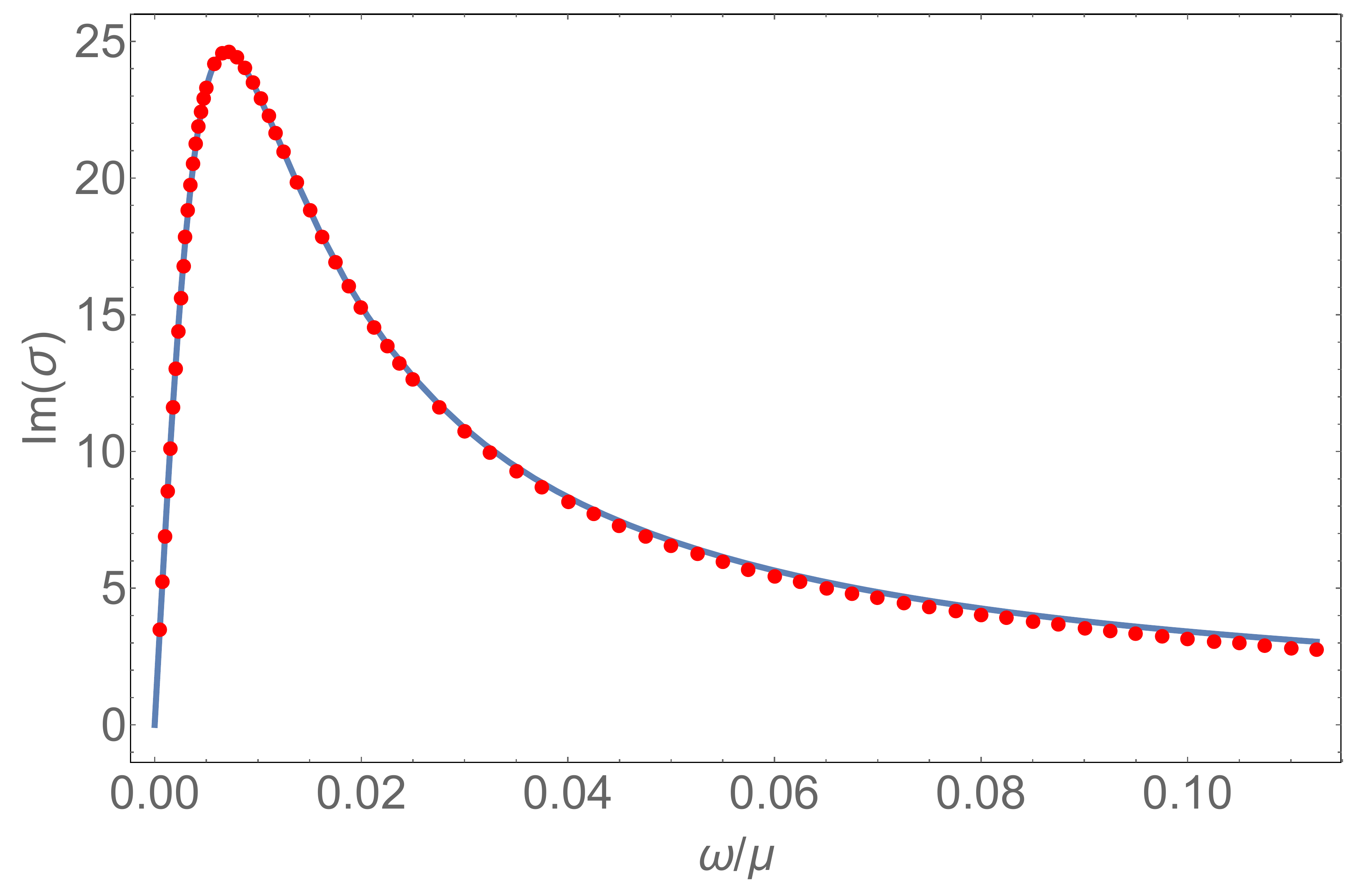}
\caption{The top panel shows the chemical potential  for a ``dirty lattice" constructed from 
ten different wave numbers as in \eqref{dldef} with $A=1/2$ and $k/\mu=1/(4/\sqrt{2})$.
In the bottom panels we show the real (bottom left) and imaginary (bottom right) parts of the optical conductivity for this lattice at 
$T/\mu=0.08$. The real part exhibits a Drude peak with a DC conductivity that agrees precisely with
the analytic result obtained from the black hole horizon, indicated by a blue dot.
}\label{dirty}
\end{figure}

We have also constructed black holes for ``dirty lattices", comprising of many wave-numbers and random phases, 
with a view to modelling disorder (see e.g. \cite{foll:2008hs,Adams:2011rj,Adams:2012yi,Arean:2013mta,Lucas:2014zea,Hartnoll:2014cua,Arean:2014oaa}.)
Specifically, we consider a truncated version of Gaussian white noise given by
\begin{align}\label{dldef}
\mu(x)=1+\frac{A}{\sqrt{n_{m}}}\,\sum_{n=1}^{n_{m}}\,\cos(n\,k\,x+\theta_{n})\,,
\end{align}
for a random collection of phases $\theta_{n}$ sampled over a uniform distribution. 
The maximum wavenumber, $n_{m}\,k$, represents a UV cutoff 
while the overall period, $2\pi/k$, is the IR cutoff.  A specific example
%\footnote{For this case the number of points we used for both our backgrounds and the optical conductivity perturbation were $N_{x}=150$ and $N=350$, in the $x$ and $z$-directions,
%respectively. The memory requirement of our numerical computation was significantly higher than that for
%the monochromatic or dichromatic cases.} 
that we analysed has $A=1/2$, $k/\mu=1/(4/\sqrt{2})$, $T/\mu=0.08$ and $n_{max}=10$, with the corresponding local chemical potential plotted in figure \ref{dirty}. In figure \ref{dirty} we also show the optical conductivity: it is manifest
that these lattices continue to exhibit a Drude-peak with a DC electrical conductivity
that is in precise agreement with our analytic result. We also expect resonances at mid-frequencies arising
from sound modes, and we have verified the existence of the first peak at $\omega/\mu\sim v_s k/\mu\sim0.125$, 
as well as the next two at roughly twice and three times this value.

 \section{Final Comments}
We have found a remarkably compact analytic expression for the thermoelectric DC conductivity for a class of inhomogeneous black hole lattices, for
all temperatures,
generalising the results for homogeneous lattices in \cite{Donos:2014uba,Donos:2014cya}. Our results provide strong evidence that this approach can be generalised to arbitrary lattices. 
It would
be interesting to next examine an inhomogeneous lattice with a UV deformation
that depends on more than
one spatial dimension, as in the recent construction of holographic checkerboards in \cite{Withers:2014sja}.

Our results provide a powerful way to obtain the low-temperature scaling behaviour 
of the DC conductivity. For translationally invariant ground states, such  
as black hole solutions which approach irrelevant deformations of $AdS_2\times\mathbb{R}^2$ in the far IR, when the lattice strength is small
one can also use\footnote{Although the renormalisation of length scales from the UV to the IR needs to be put in as an extra ingredient.}
 the memory matrix formalism \cite{Hartnoll:2012rj}, and in this case we find precise agreement. On the other hand, if one approaches a ground state which breaks translations, as in \cite{Donos:2014uba,Gouteraux:2014hca}, then the memory matrix formalism cannot be used and so our analytic results provide a particularly powerful tool to study the properties of these novel holographic ground states.

Here and in \cite{Donos:2014cya}, we have seen that as $T\to\infty$  the UV lattice deformation leads to a modification of the DC conductivity away from the value of the optical
conductivity $\sigma(\omega)$ in the limit $\omega\to\infty$. In this paper, we have seen
that a periodic chemical potential $\mu(x)$ leads to a saturation of the electric DC conductivity to a constant value\footnote{By contrast, recall that for the Q-lattices of [14] the DC conductivity saturates at high temperature to a 
divergent result. It is interesting that this implies that there is an associated minimum value of the DC conductivity for Q-lattice metals which, by definition, also have a divergent DC conductivity as $T\to 0$.}
as $T\to\infty$, with the value depending on $\mu(x)$ as in \eqref{hightsc}.
This is a kind of generalised Mott-Ioffe-Regel bound \cite{Gunnarsson:2003zz,takmir} without quasi-particles. 
It is also worth noting that as $T\to\infty$ we find that the optical conductivity approaches unity for all
values of $\omega/\mu$, except at $\omega/\mu\to 0$ where it jumps to the higher DC value.

We have also made detailed constructions of the inhomogeneous black holes arising
in Einstein-Maxwell theory for various periodic chemical potentials. We focussed in most detail
on monochromatic lattices, associated with a single wave-number, $k$, but we also considered
dichromatic lattices, with wave-numbers $k$ and $2k$ and the same phase. In addition we constructed
black holes that model a dirty lattice which were built from ten sequential wave-numbers with random phases.
The black holes, as well as the optical conductivity 
were obtained by numerically solving PDEs.
We have found Drude peaks in the optical conductivity at finite temperature, as in \cite{Horowitz:2012ky}, 
but, in contrast to \cite{Horowitz:2012ky}, we do not find any intermediate scaling for the monochromatic lattices.
At low temperatures our ground states for the monochromatic lattices all
seem to approach $AdS_2\times\mathbb{R}^2$ in the
far IR and, specifically, we find the DC scaling behaviour is precisely consistent with this. While it is possible that lowering the temperature
of the black holes further will reveal some exotic new ground states, as in \cite{Hartnoll:2014gaa}, we feel this is unlikely.
It would be interesting to know if exotic ground states appear for 
stronger lattice deformations and/or for other deformations of the chemical potential.

We have shown that the monochromatic, dichromatic and dirty lattices naturally give rise to mid frequency resonances that
can be associated with sound modes. It will be interesting to consider this issue in more detail for the dirty lattices since
in the limit where the number of modes and the characteristic wave-number is going to zero (i.e. $n_m\to\infty$, $k\to 0$ in 
\eqref{dldef}) the resonances may coalesce and change the analytic structure of the Greens functions in the mid-frequency region.

  \section*{Acknowledgements}
We thank Pau Figueras, Sean Hartnoll, Gary Horowitz, Elias Kiritsis, Jorge Santos,
David Tong, Toby Wiseman and Jan Zaanen for helpful discussions. 
The work is supported by STFC grant ST/J0003533/1, EPSRC programme grant EP/K034456/1
and also by the European Research Council under the European Union's Seventh Framework Programme (FP7/2007-2013), ERC Grant agreement STG 279943 and ADG 339140.

\appendix
 
\section{The stress tensor and heat current}\label{stressheat}
For the perturbed black holes of interest we can obtain the heat current from the stress tensor following the approach of \cite{Donos:2014cya}. 
From \cite{Balasubramanian:1999re} we can write the stress tensor and the current as
\begin{align}\label{genpertap}
\frac{1}{2}\tilde T^{\mu\nu}&=-K^{\mu\nu}+K\gamma^{\mu\nu}-2\gamma^{\mu\nu}+ G^{\mu\nu}\,,\nn
\tilde J^\nu&=-n_\mu F^{\mu\nu} \,,
\end{align}
where $n^\mu$ is the unit normal to the boundary, $K_{\mu\nu}=(\delta_\mu{}^\rho+n_\mu n^\rho)\nabla_{\rho}n_\nu$, $G^{\mu\nu}$ is the Einstein tensor
of the boundary metric $\gamma$ and
expressions are to be evaluated at the boundary $r\to\infty$.

For all of the black holes that we considered in calculating the DC conductivities, including the perturbation, we have
\begin{align}
n_\mu=\left(\frac{H_{rr}}{U}\right)^{1/2}(1+\frac{U\delta g_{rr}}{2 H_{rr}})\,(0,1,0,0)\,.
\end{align}
For the black hole backgrounds (with vanishing perturbation) we obtain the following expressions
\begin{align}
\tilde T^{tt}&=\frac{1}{UH_{tt}}\Bigg[4-\frac{U^{1/2}}{H_{rr}^{1/2}}\partial_r \ln \Sigma\Bigg]+2G^{tt}\,,\nn
\tilde T^{xx}&=\frac{1}{e^B\Sigma H_{tt}(UH_{rr})^{1/2}}\Bigg[\partial_r(U H_{tt})-UH_{tt}\left(\partial_r\ln\frac{e^B}{\Sigma}+4\frac{H_{rr}^{1/2}}{U^{1/2}}\right)\Bigg]\nn
&\qquad\qquad\qquad -\frac{1}{2\Sigma^2H_{tt}e^{2B}}\left(\partial_x\ln\frac{e^B}{\Sigma}\right)\left(\partial_x H_{tt}\right)\,,\nn
\tilde T^{yy}&=\frac{e^B}{\Sigma H_{tt}(UH_{rr})^{1/2}}\Bigg[\partial_r(U H_{tt})-UH_{tt}\left(\partial_r\ln\frac{e^{-B}}{\Sigma}+4\frac{H_{rr}^{1/2}}{U^{1/2}}\right)\Bigg]+2G^{yy}\,,\nn
\tilde J^t&=\frac{1}{(UH_{rr})^{1/2}H_{tt}}\partial_r a_t\,,
\end{align}
where we have omitted the explicit expressions for $G^{tt}$ and $G^{yy}$ for brevity.
As $r\to\infty$ we have $\tilde T^{ab}\sim r^{-5}$ and $\tilde J^a \sim r^{-3}$
so it is convenient to define
\begin{align}\label{endtwo}
T^{ab}=r^5\tilde T^{ab}\,,\qquad
J^a=r^3 \tilde J^{a}\,.
\end{align}

We next consider the perturbation about the black holes backgrounds discussed in section 3, but with
a general time dependence in $\delta g_{tx}$ for the moment, finding
\begin{align}\label{bingop}
\tilde T^{tx}
=&\frac{1}{e^B\Sigma H_{tt}(UH_{rr})^{1/2}}\Bigg[-\delta g_{tx}(t,r,x)\left(\partial_r\frac{\ln e^B}{\Sigma}+\frac{H_{rr}^{1/2}}{U^{1/2}}\right)\nn
&+\partial_r\delta g_{tx}(t,r,x)-H_{tt}\partial_x\frac{\delta g_{tr}}{H_tt}\Bigg]-\frac{\delta g_{tx}(t,r,x)}{2 e^{2B}\Sigma^2H_{tt}^2 U}
\left(\partial_x\frac{\ln e^B}{\Sigma}\right)\left(\partial_x H_{tt}\right)\,,
\end{align}
where we have included the argument of $\delta g_{tx}$, here and below, for clarity.
It will be convenient, shortly, to note that
\begin{align}\label{convexp}
U^{1/2}H_{tt}^{1/2}\Sigma\left(UH_{tt} \tilde T^{tx}-\delta g_{tx}(t,r,x)\tilde T^{xx}\right)=
\frac{U^2H_{tt}^{3/2}}{e^B H_{rr}^{1/2}}\Bigg[\partial_r\left(\frac{\delta g_{tx}(t,r,x)}{UH_{tt}}\right) -\partial_x\frac{\delta g_{tr}}{U H_{tt}}  \Bigg]\,.
\end{align}

We now consider the particular linearised time-dependence for the perturbation given in sections 3.2 and 3.3:
\begin{align}\label{pertap}
\delta A&=-tEd x+t\zeta a_t  +{\delta a_{\mu}}(r,x)dx^\mu\,,\nn
\delta ds^2&= -2t\zeta UH_{tt} dt dx+ \delta g_{\mu\nu}(r,x)dx^\mu dx^\nu\,,
\end{align}
with falloffs of ${\delta a_{\mu}}(r,x)$ and $\delta g_{\mu\nu}(r,x)$ as $r\to \infty$ chosen so
that the only sources are parametrised by $E$ and $\zeta$.
Now these time-dependent sources give rise to a time-independent expression
 for $\tilde J^{x}$:
 \begin{align}\label{jex2}
 \tilde J^x=\frac{e^{-B}}{\sqrt{H_{rr}H_{tt}}}\,\left[  \delta g_{tr}\partial_{x}a_{t}- \delta g_{tx}(r,x)\partial_{r}a_{t}+ H_{tt} U \left( \partial_{x}\delta a_{r}-\partial_{r}\delta a_{x}\right)\right]\,,
\end{align}
which, when evaluated at $r\to\infty$ and using \eqref{endtwo},
agrees with the expression for $J$ in \eqref{jex}.
By contrast we obtain a time-dependent component in $T^{tx}$. Explicitly, from \eqref{bingop}
we immediately obtain 
\begin{align}
\tilde T^{tx}
&=\frac{1}{e^B\Sigma H_{tt}(UH_{rr})^{1/2}}\Bigg[-\delta g_{tx}(r,x)\left(\partial_r\frac{\ln e^B}{\Sigma}+\frac{H_{rr}^{1/2}}{U^{1/2}}\right)\nn
&+\partial_r\delta g_{tx}(r,x)-H_{tt}\partial_x\frac{\delta g_{tr}}{H_{tt}}\Bigg]-\frac{\delta g_{tx}(r,x)}{2 e^{2B}\Sigma^2H_{tt}^2 U}
\left(\partial_x\frac{\ln e^B}{\Sigma}\right)\left(\partial_x H_{tt}\right)-\zeta t \tilde T^{xx}\,,\nn
&\equiv \tilde T^{tx}_0-\zeta t \tilde T^{xx}\,.
\end{align}
Returning now to \eqref{convexp} and 
substituting in \eqref{pertap} we find that all of the time dependence drops out and hence we can conclude that
\begin{align}
U^{1/2}H_{tt}^{1/2}\Sigma\left(UH_{tt} \tilde T^{tx}_0-\delta g_{tx}(r,x)\tilde T^{xx}\right)=
\frac{U^2H_{tt}^{3/2}}{e^B H_{rr}^{1/2}}\Bigg[\partial_r\left(\frac{\delta g_{tx}(r,x)}{UH_{tt}}\right) -\partial_x\frac{\delta g_{tr}}{U H_{tt}}  \Bigg]\,.
\end{align}
Evaluating both sides at $r\to\infty$ we deduce that
\begin{align}\label{btwelve}
r^5\tilde T^{tx}_0=\lim_{r\to\infty}   \frac{U^2H_{tt}^{3/2}}{e^B H_{rr}^{1/2}}\Bigg[\partial_r\left(\frac{\delta g_{tx}(r,x)}{UH_{tt}}\right) -\partial_x\frac{\delta g_{tr}}{U H_{tt}}  \Bigg]          \,.
\end{align}
Recalling the expression for $Q$ given in \eqref{pexp}, we deduce that
\begin{align}
T^{tx}-\mu J^{x}=Q-\zeta t T^{xx}\,.
\end{align}
Now as explained in appendix C of \cite{Donos:2014cya}, the time dependent piece is associated with a static susceptibility for the $Q$ $Q$ correlator, which we see is explicitly given by $T^{xx}$ of the background black holes. 
On the other hand
the time independent piece is associated with the DC conductivity.

 \section{Convergence tests}\label{convgtest}

 In this section we will provide a few details on three different convergence tests that we carried out 
 for the numerical methods which we discussed in sections \ref{sec:background} and \ref{sec:ACConductivity}.
 
For the black hole backgrounds, in the continuum limit, which is approached as the number of grid points is taken to infinity, we expect that the norm of the
DeTurck vector, $\xi^{2}$, should approach zero uniformly everywhere on our computational grid. Checking that this happens is the first test that we performed.
Along the same lines, our backgrounds should satisfy the equations of motion \eqref{eq:eom} without the additional DeTurck term. Correspondingly,
our second convergence test is to check the absolute value of the trace of Einstein's equations in \eqref{eq:eom}.

 Our third check concerns the convergence properties of the perturbation about the background
 black holes in order to extract the optical conductivity as described in section \ref{sec:ACConductivity}. As we pointed out in the main text, the six functions that we used in the perturbation ansatz \eqref{eq:opt_cond_ans} should solve ten equations of motion, of which four are constraints that 
we impose on the black hole horizon. As a non-trivial check of our numerics, we check that the constraints are satisfied everywhere in the bulk in the continuum limit. As an illustration of this we
can take the trace of Einstein's equations, expand it to first order in the perturbation and then
examine the absolute value of the leading term.
 
Let us present some results of these tests for the specific monochromatic lattice
$\mu(x)=\mu\,\left(1+\frac{1}{2}\,\cos\left(\frac{\mu}{\sqrt{2}}\,x\right)\right)$ for three different temperatures $T/\mu\approx 0.035$, $T/\mu\approx 0.02$ and $T/\mu\approx 0.015$. These 
black holes have been discussed in the main text and some of their properties are presented in figure \ref{fig:S2}. In order to give a more detailed treatment, we divide our computational grid into two halves: the ``boundary'' half, defined by  $0< z <1/2$,  and the ``horizon'' half, defined by $1/2 < z<1$. In all of our tests we have fixed the number of points in the periodic, field theory direction to be $N_{x}=45$ and then we vary the number of points, $N$, in the radial direction. For the perturbation convergence tests we have fixed the frequency $\omega/\mu\approx 0.0008$ which for the three temperatures is very close to the top of the Drude peak; we do this because it is a region in $\omega/\mu$ which is challenging numerically.

\begin{figure}
\center
\includegraphics[height=5 cm]{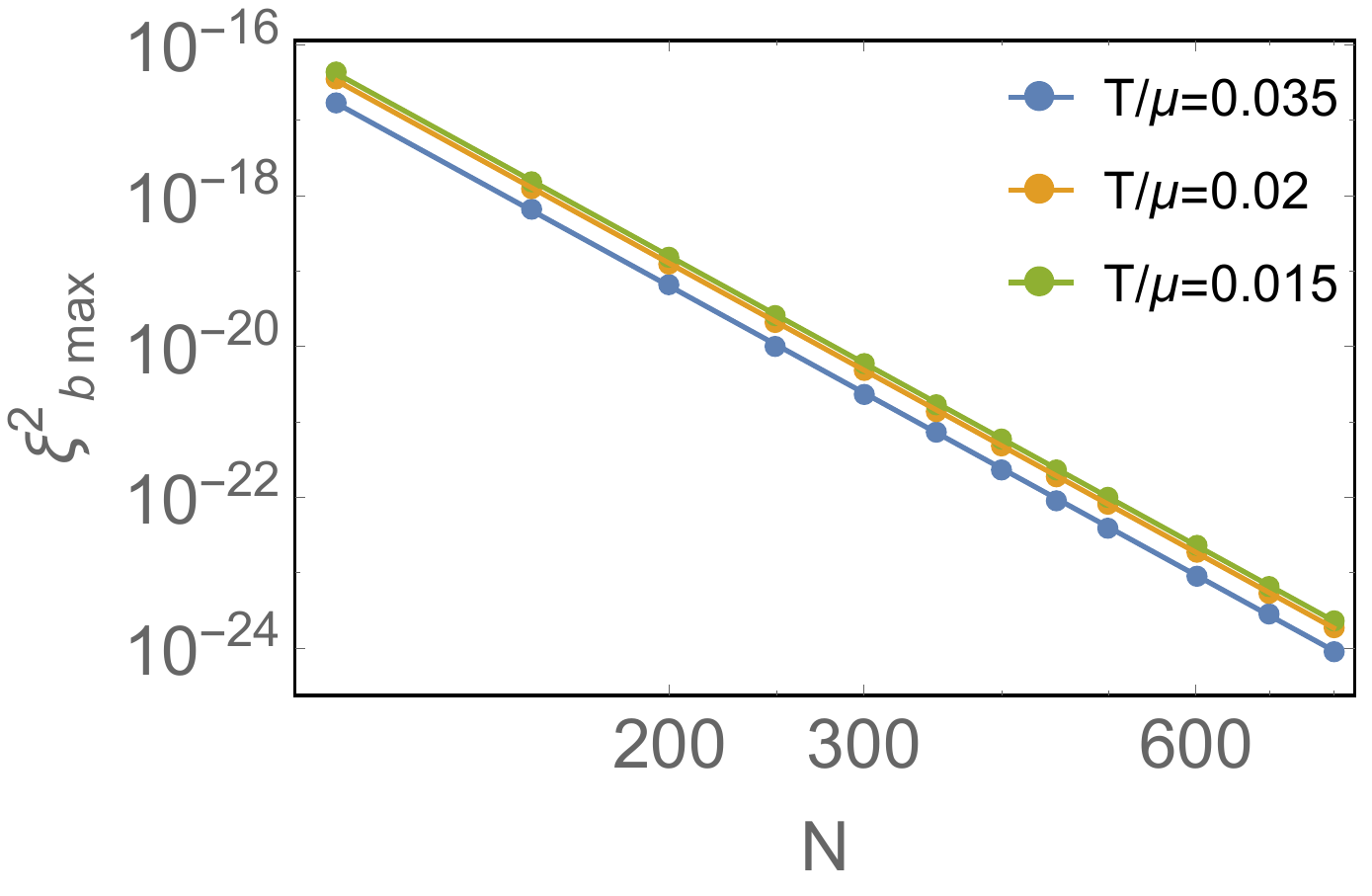}\includegraphics[height=5 cm]{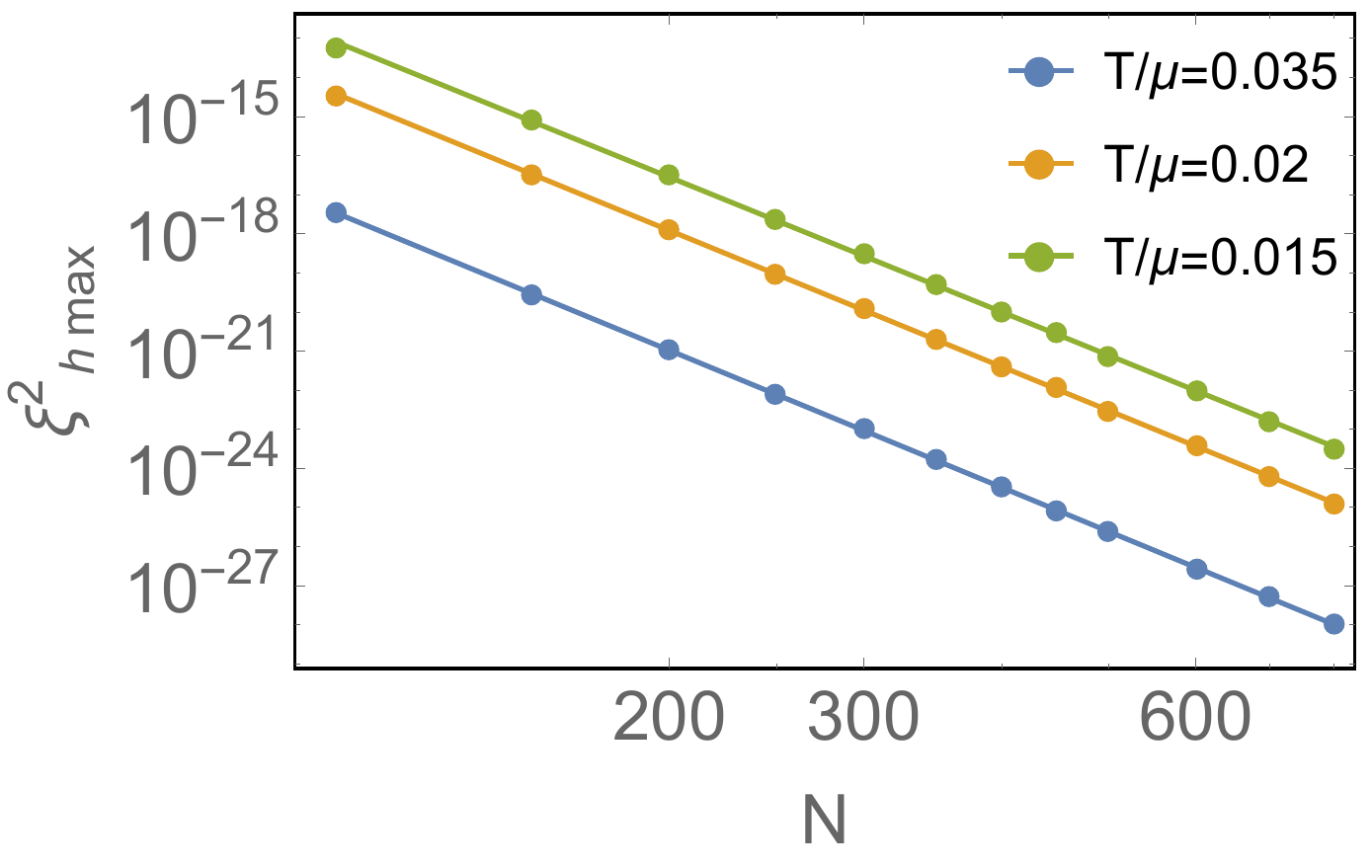}\\
\includegraphics[height=5 cm]{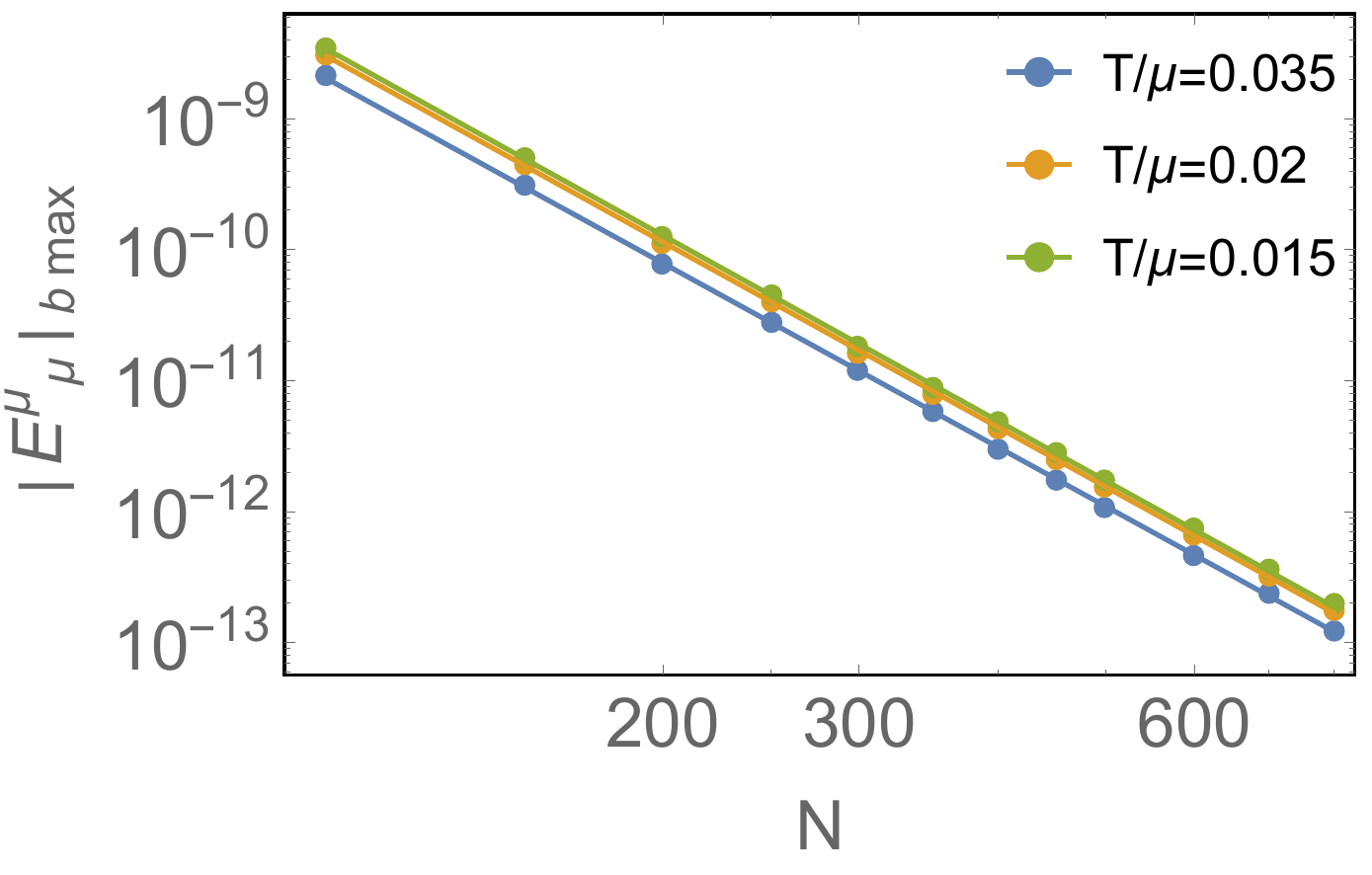}\includegraphics[height=5 cm]{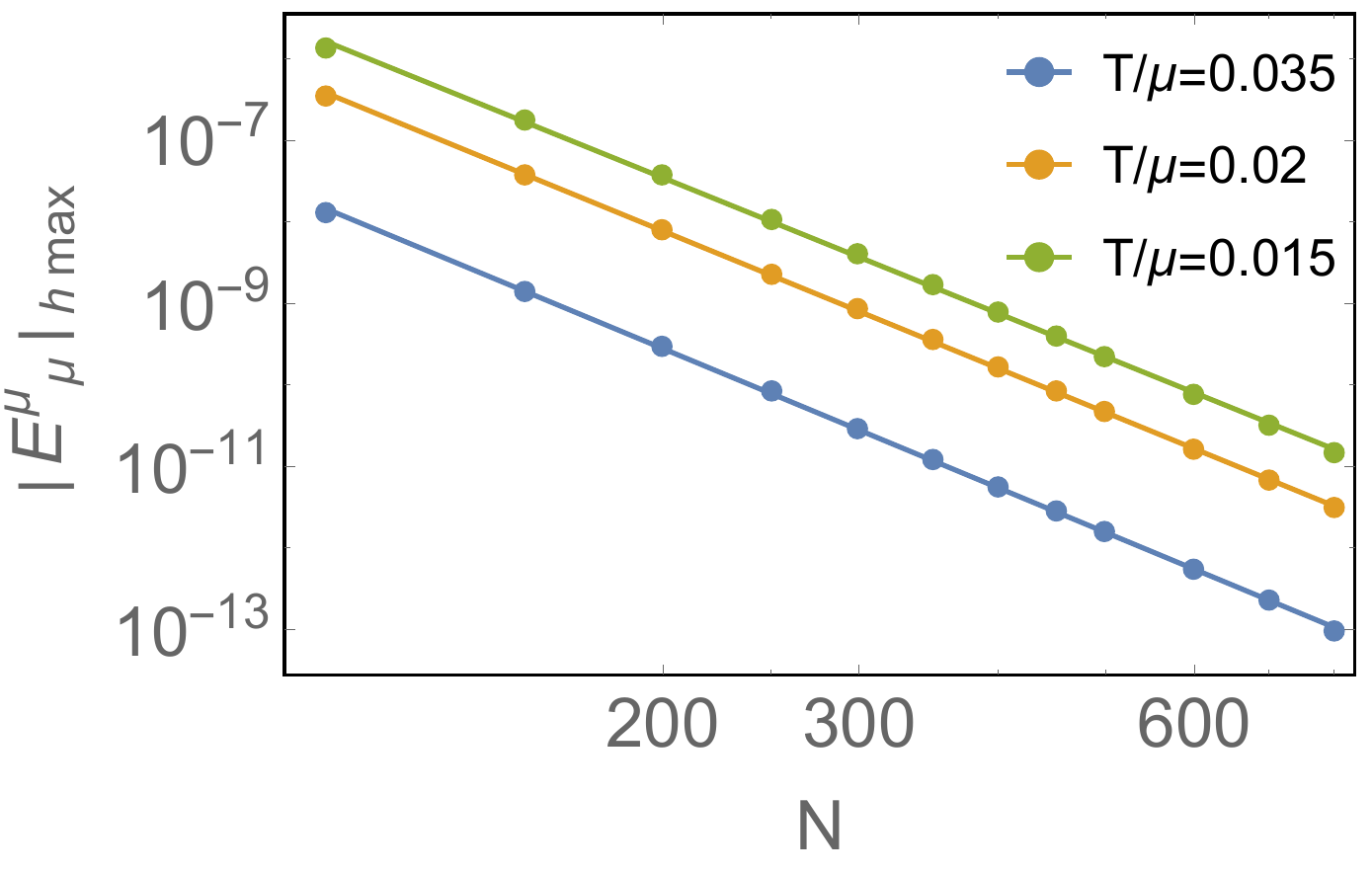}
\caption{Convergence tests for the numerical construction of monochromatic lattice black holes of figure \ref{fig:S2}
for three different temperatures. The figures on the left denote convergence tests in the boundary region, while those on the right correspond to the horizon region. We have plotted the norm of the 
de Truck vector, $\xi^2$ and the absolute value of the trace of Einstein's equations $E^\mu{}_\mu$ against the number of radial points in the grid, $N$, with a fixed number $N_x=45$
points in the periodic spatial $x$-direction.}\label{fig:back_converg}
\end{figure}

In figure \ref{fig:back_converg} we show the results of the two convergence tests for the black hole
solutions, discussed above, for the boundary and horizon regions. The boundary expansion \eqref{eq:background_as_exp} suggests that we should have convergence not better than fifth order for the boundary region and indeed we find that while $\xi_{b}^{2}$ converges as $N^{-8}$, the trace of Einstein's equations converges as $N^{-4.6}$. On the other hand, close to the horizon, we have an analytic expansion and we find convergence of the same quantities of the form $N^{-11.7}$ and $N^{-5.7}$, respectively.

\begin{figure}
\center
\includegraphics[height=5cm]{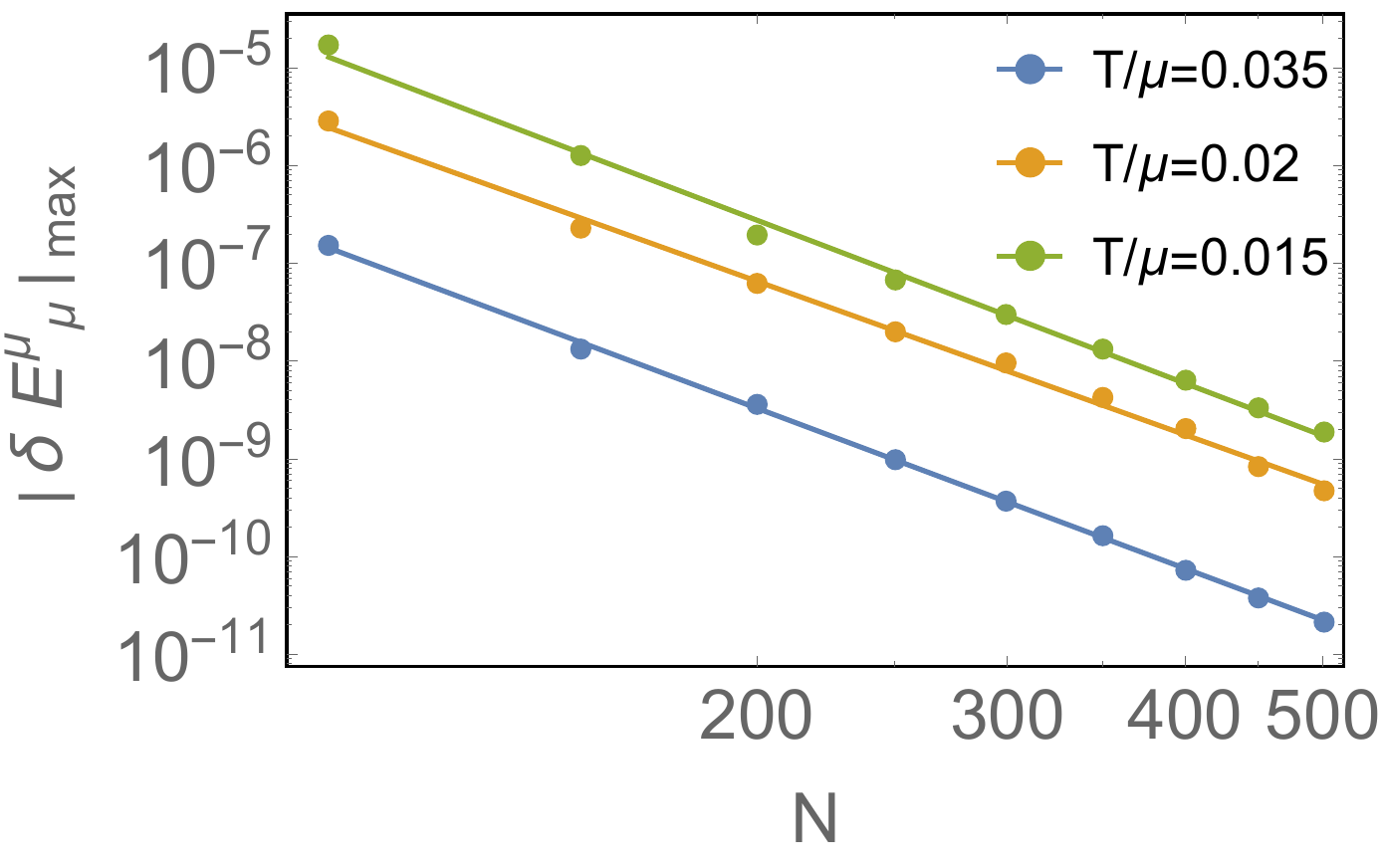}
\caption{Convergence tests for the numerical construction of the perturbation about the black holes considered in figure \ref{fig:back_converg} that is needed to obtain the AC conductivity. We have plotted $\delta E^\mu{}_\mu$, which is obtained by considering the trace of Einstein's equations, expanding it to first order in the perturbation and then
taking the absolute value of the leading term, against the number of radial points in the grid, $N$.
Again, $N_x=45$.}\label{fig:pert_converg}
\end{figure}

In figure \ref{fig:pert_converg} we show a plot of the convergence test for the perturbation that we discussed above. For the range of resolutions shown in the figure we find a convergence rate scaling like $N^{-5.4}$ which is suggestive that all the error comes from the horizon and, moreover, from the fact that our background satisfies the DeTurck modified equations instead of Einstein's.

The numerical schemes outlined in section \ref{sec:numerics} were implemented in \verb!C++!. The facility of class templates has been particularly helpful to accommodate the various numerical precisions we have used at low temperatures and in the convergence tests. At certain key points of our code
we have specialised our templates to \verb!double!, \verb!long double!, Intel's \verb!_QUAD! and \verb!MPFR! \cite{mpfrcpp} data types\footnote{These allowed us to work with 53, 80, 113 and arbitrary bits of significand precision, respectively.}. In particular this was necessary
for the relevant sparse linear solver that we used both in Newton's method and for the linear perturbations for the optical conductivity. For our \verb!double! precision numerics we have chosen \verb!UMFPACK! from the \verb!SuiteSparse! library \cite{SuiteSparse} compiled with Intel's MKL BLAS which takes advantage of multicore systems when combined with \verb!OpenMP!. For the three remaining data types we have chosen to use the \verb!SparseLU! solver from the Eigen3 template library \cite{eigenweb}. In writing our code we have greatly benefited from the \verb!float128! wrapper class of the Boost \verb!C++! library \cite{Boost}.

Concerning the plots appearing in figures \ref{fig:back_converg} and \ref{fig:pert_converg}, we have found that \verb!double! precision is saturated when we reach 450 points in the radial direction after which we need to use \verb!long double! precision numerics for the backgrounds. As far as the conductivity perturbation is concerned, we found that \verb!_QUAD! precision had to be used when we reach 300 points in the radial direction. For these cases the corresponding black hole background was computed using the same \verb!_QUAD! precision.

\section{Further comments on scaling behaviour}\label{floppy}
The black holes that we have constructed numerically, described in section 4, are consistent with the $T=0$ limits approaching
domain wall solutions interpolating
from $AdS_2\times\mathbb{R}^2$ in the IR to $AdS_4$ in the UV. This can be contrasted with the conclusion of \cite{Hartnoll:2014gaa} where it was argued that the $T=0$ ground states have an inhomogeneous IR behaviour. Here we would like to provide a possible explanation for the disparity.

\begin{figure}
\center
\includegraphics[height=5 cm]{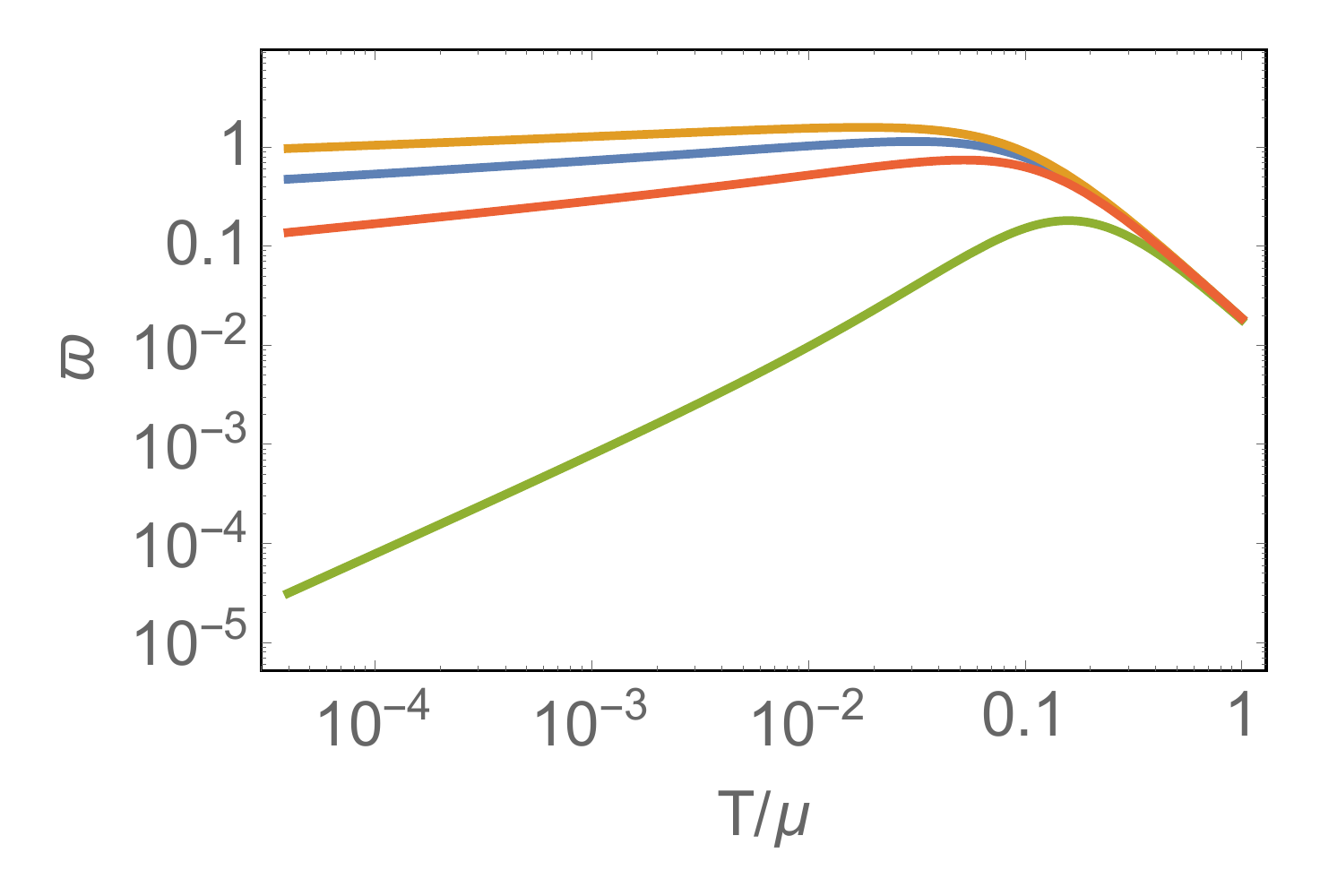}\includegraphics[height=5 cm]{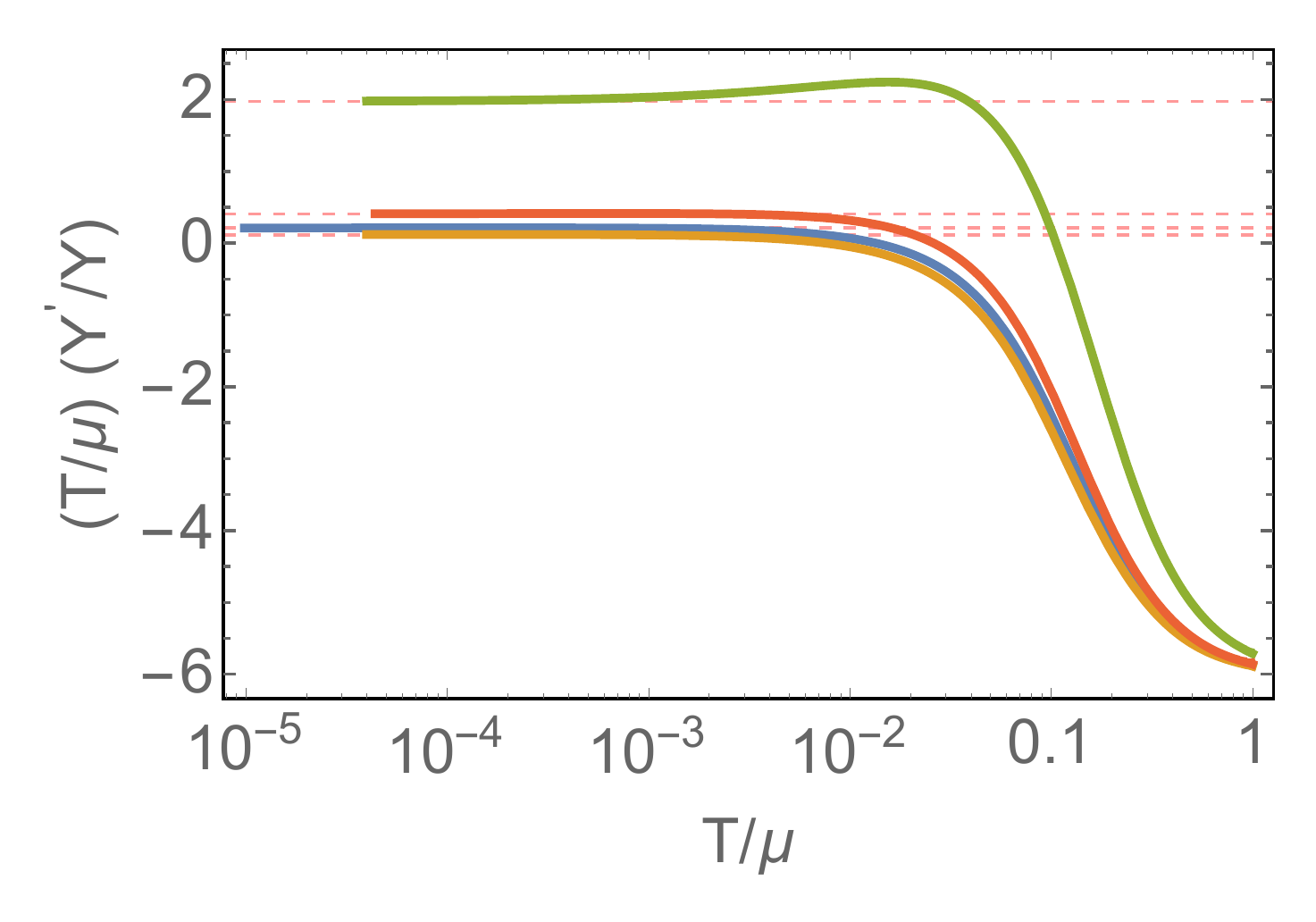}
\caption{The behaviour of $\varpi$, defined in \eqref{varpidef}, and $\Upsilon$, defined in \eqref{updef},
with temperature for monochromatic lattices with $A=1/2$ and $k=\sqrt{2}/3$ (orange)), $k=2\sqrt{2}/5$ (blue),
$1/\sqrt{2}$ (red) and
$\sqrt{2}$ (green).
The red dashed lines on the right hand plots indicate the low-temperature scaling behaviour,
given in \eqref{bscale2} expected for black holes approaching $AdS_2\times\mathbb{R}^2$ in the far IR. 
The left plot shows that in the scaling regime, the value of $\varpi$, can be parametrically larger than the scale set by the temperature
if the scaling exponent is suitably small. This situation arises for small lattice wave-numbers $k$.}\label{bfloppy}
\end{figure}

As illustrated in figure \ref{fig:Snew} we have seen that for temperatures as low as $T/\mu\sim 4\times10^{-5}$, and $T/\mu\sim 9.8\times 10^{-6}$ for one specific case, the electrical and thermal DC
conductivities
exhibit a clear scaling behaviour, exactly consistent with
 \eqref{dcscal} and \eqref{defDel}, predicted from the dimension of the least irrelevant operator about $AdS_2\times\mathbb{R}^2$.
It is also illuminating to consider a quantity $\varpi$ introduced in \cite{Hartnoll:2014gaa}.
Let $\mathcal{W}=||\partial_y||^2_{r=r_+}$ and then, by considering the variation along the $x$ direction, define
\begin{align}\label{varpidef}
\varpi=\frac{\mathcal{W}_{max}}{\mathcal{W}_{min}}-1\,.
\end{align}
If the $T=0$ ground states have $AdS_2\times\mathbb{R}^2$ in the IR, then
this should approach $0$ at $T=0$. More precisely, it should approach $0$ with a specific scaling behaviour which can be extracted from
the analysis of \cite{Hartnoll:2012rj,Edalati:2010pn}:
\begin{align}\label{bscale}
\varpi\sim T^{\Delta(\bar k)-1}\,.
\end{align} 
It is worth restating here that $\bar k$ is related to the UV lattice factor as in \eqref{renscale}, which involves
a renormalisation scale $\bar \lambda$ that depends on the UV data. 
In figure \ref{bfloppy}, for four monochromatic lattices with\footnote{In the notation of \cite{Hartnoll:2014gaa} these correspond to $A_0=1/2$ and $k_0=2/3,4/5, 1$ and $2$, respectively.} $A=1/2$ and $k/\mu=\sqrt{2}/3,2\sqrt{2}/5, 1/\sqrt{2}$ and $\sqrt{2}$, we see that $\varpi$ scales exactly\footnote{As before, we deduce the value of $\bar\lambda$ from our lowest temperature solution, which is an approximation to the $T=0$ value. The value of $\bar \lambda$ is small for the black holes we have considered, and moreover, we have checked that it is changing very slowly with $T$ once we are in the scaling regime. Thus the approximation is a very good one.}
 as expected for an $AdS_2\times\mathbb{R}^2$ ground state.
Now, a simple but key observation is that if the scaling exponent is small, then the value of $\varpi$ can still be parametrically large, compared
to the temperature scale, even when one is in the scaling regime as we see in figure \ref{bfloppy} and also in table \ref{green}. This situation occurs when $\bar k$ is small which arises, in practise when $k$ is small.

Recalling \eqref{defM}, another quantity we can consider is
\begin{align}\label{updef}
\Upsilon= \int \frac{1}{\Sigma^{(0)}} \left[\partial_{x}\ln{\frac{e^{B^{(0)}}}{\Sigma^{(0)}}} \right]^{2}\,.
\end{align}
If the black holes approach $AdS_2\times\mathbb{R}^2$ in the far IR as $T\to 0$, then we should also have
\begin{align}\label{bscale2}
\Upsilon\sim T^{2\Delta(\bar k)-2}\,.
\end{align}
In fact we find that this quantity approaches the scaling behaviour slightly quicker
than $\varpi$ and we have illustrated this in figure \ref{bfloppy}.

For these constructions, in order to keep the error small at low temperatures we used three patches in combination with \verb!long double! precision. As an indicative example, in lowest temperature black hole for the case $k/\mu=\frac{2\sqrt{2}}{{5}}$ of table \ref{green} we partitioned the coordinate $z$ interval $\left(0,\,1\right)$ into three patches as $\left(0, \, \frac{94}{100} \right] \cup\left[\frac{94}{100},\,\frac{997}{1000} \right]\cup \left[\frac{997}{1000},\, 1 \right)$. Following the discussion preceding equation \eqref{effcalc}, we took $N_{z}^{1}=1200$, $N_{z}^{2}=2500$ and $N_{z}^{3}=1500$ points in the corresponding intervals while for the $x$ direction we took $N_{x}=45$ points. We used sixth order finite differences in the radial direction while Fourier basis differentiation in the $x$ direction. The resulting geometry turned out to have a maximum $\xi^{2}\sim 10^{-20}$, where $\xi^2$ is the norm of DeTurck vector.
\begin{table}[!th]
\begin{center}
\setlength{\tabcolsep}{0.45em}
\begin{tabular}{|c|c|c|c|c|c|c|}
\hline
$k/\mu$&$T/\mu$ & $\varpi$ &$s/(8\pi \mu^2)$&$\bar\lambda$&$\Upsilon$\\
\hline
$\frac{\sqrt{2}}{{3}}$&$4.0\times 10^{-5}$& 0.964& 0.0458&1.04& 0.14 \\
$\frac{2\sqrt{2}}{{5}}$&$9.8\times 10^{-6}$& 0.396& 0.0452&1.03& 0.049 \\
$\frac{1}{\sqrt{2}}$&$4.4\times 10^{-5}$ & 0.140 & 0.0445& 1.03& 0.012\\
$\sqrt{2}$&$4.0\times 10^{-5}$ & $3.12\times 10^{-5}$ & 0.0425&1.01& 2.9$\times 10^{-9}$ \\
\hline
\end{tabular}
\end{center}
\caption{The values of $\varpi$ (see \eqref{varpidef}, entropy density $s$, renormalisation of length scale $\bar\lambda$ (see \eqref{renscale}) and
$\Upsilon$ (see \eqref{updef}) for the
three monochromatic lattices plotted in figure \ref{bfloppy}, for the given temperature.}
\label{green}
\end{table}

\begin{figure}
\center
\includegraphics[height=6 cm]{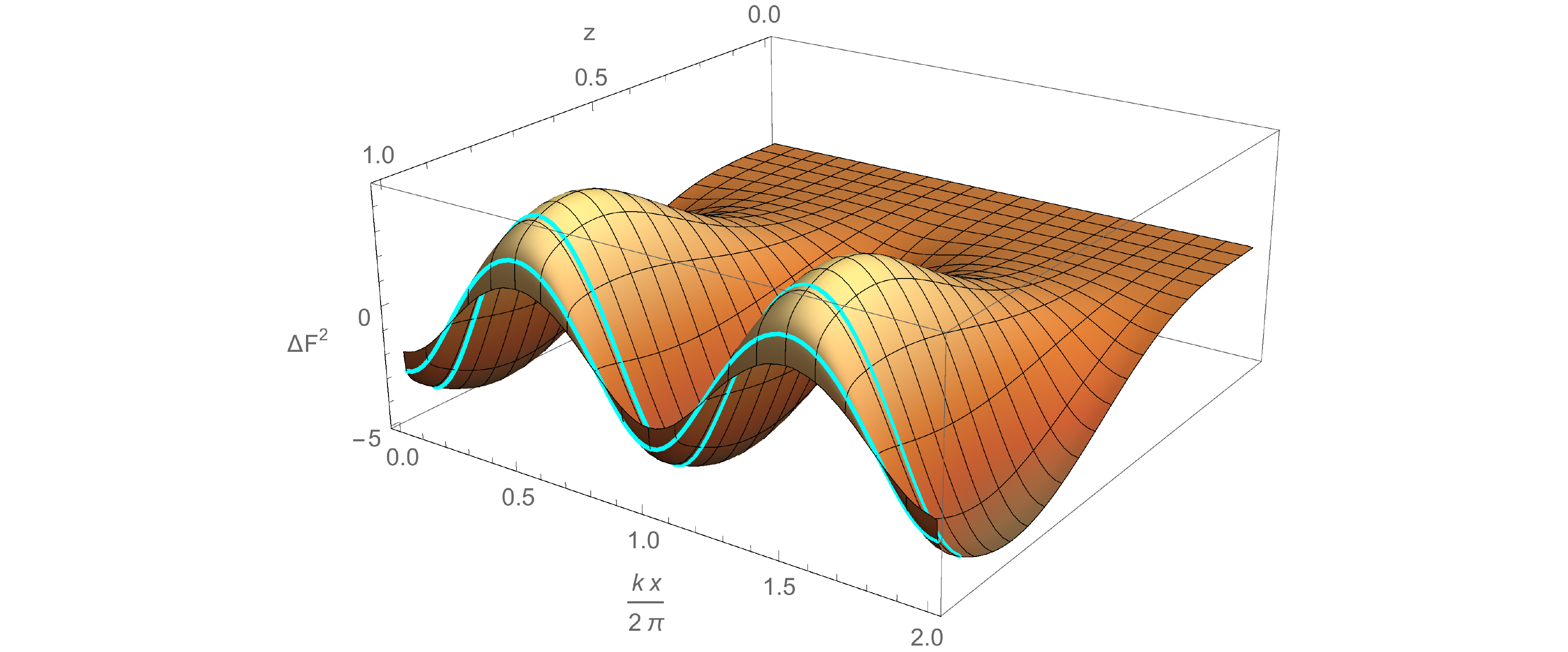}
\caption{The behaviour of $\Delta F^2$ (see \eqref{delfsq}) for the monochromatic lattice with $k/\mu=2\sqrt{2}/5$ at $T/\mu=9.8\times 10^{-6}$.
The behaviour is consistent with it vanishing at $T=0$ at the black hole horizon at $z=1$, consistent with the appearance of $AdS_2\times\mathbb{R}^2$ in the IR.}\label{bfloppy2}
\end{figure}

To further illustrate our results we can consider the quantity
\begin{align}\label{delfsq}
\Delta F^2\equiv F^2-F^2_{RN}
\end{align}
where $F^2=F_{\mu\nu}F^{\mu\nu}$ is the norm of the field strength for the lattice black holes and 
$F^2_{RN}$ is the corresponding quantity for the AdS-RN black hole at the same temperature. 
If the black holes are approaching $AdS_2\times \mathbb{R}^2$ at $T=0$
then this quantity should approach zero at the black hole horizon. It will also vanish at the $AdS_4$ boundary, since each term does separately.
For the monochromatic lattice with $A=1/2$, $k/\mu=2\sqrt{2}/5$ (as in figure \ref{bfloppy}) at the lowest temperature $T/\mu=9.8\times 10^{-6}$
we have plotted $\Delta F^2$ against the spatial coordinate $x$ and the radial coordinate $z$
in figure \ref{bfloppy2}. The behaviour is consistent with the $T=0$ limit
approaching zero at the black hole horizon at $z=1$ followed by a sharp rise to non-trivial behaviour in the bulk, fading to zero at the $AdS_4$ boundary at $z=0$. The cyan lines in figure \ref{bfloppy2} are the location of the boundaries of the patches we 
discussed in the previous paragraph.

We believe our numerical results at finite temperature provide strong evidence that the scaling should continue all the way down to $T=0$ and
that the $T=0$ solutions will approach $AdS_2\times\mathbb{R}^2$ in the IR.
We therefore think it is unlikely that the $T=0$ numerical solutions found in \cite{Hartnoll:2014gaa} are in fact $T=0$ solutions, since, if they were,
it would
imply that there is a sudden discontinuous jump in  the behaviour of the solutions.
One possibility is that they are, instead, solutions at very small temperatures and the observed non-vanishing $\varpi$
for small lattice wave-number that was observed in \cite{Hartnoll:2014gaa} would just correspond to scaling with a small exponent as we have seen for
our finite temperature solutions.

\bibliographystyle{utphys}
\bibliography{helical}{}

\providecommand{\href}[2]{#2}\begingroup\raggedright\begin{thebibliography}{10}

\bibitem{Hartnoll:2007ih}
S.~A. Hartnoll, P.~K. Kovtun, M.~Muller, and S.~Sachdev, ``{Theory of the
  Nernst effect near quantum phase transitions in condensed matter, and in
  dyonic black holes},''
  \href{http://dx.doi.org/10.1103/PhysRevB.76.144502}{{\em Phys. Rev.}
  {\bfseries B76} (2007) 144502},
\href{http://arxiv.org/abs/0706.3215}{{\ttfamily arXiv:0706.3215
  [cond-mat.str-el]}}.
%%CITATION = 0706.3215;%%.

\bibitem{Hartnoll:2009sz}
S.~A. Hartnoll, ``{Lectures on holographic methods for condensed matter
  physics},'' \href{http://dx.doi.org/10.1088/0264-9381/26/22/224002}{{\em
  Class.Quant.Grav.} {\bfseries 26} (2009) 224002},
\href{http://arxiv.org/abs/0903.3246}{{\ttfamily arXiv:0903.3246 [hep-th]}}.
%%CITATION = ARXIV:0903.3246;%%.

\bibitem{Herzog:2009xv}
C.~P. Herzog, ``{Lectures on Holographic Superfluidity and
  Superconductivity},''
  \href{http://dx.doi.org/10.1088/1751-8113/42/34/343001}{{\em J.Phys.}
  {\bfseries A42} (2009) 343001},
\href{http://arxiv.org/abs/0904.1975}{{\ttfamily arXiv:0904.1975 [hep-th]}}.
%%CITATION = ARXIV:0904.1975;%%.

\bibitem{Horowitz:2012ky}
G.~T. Horowitz, J.~E. Santos, and D.~Tong, ``{Optical Conductivity with
  Holographic Lattices},''
  \href{http://dx.doi.org/10.1007/JHEP07(2012)168}{{\em JHEP} {\bfseries 1207}
  (2012) 168},
\href{http://arxiv.org/abs/1204.0519}{{\ttfamily arXiv:1204.0519 [hep-th]}}.
%%CITATION = ARXIV:1204.0519;%%.

\bibitem{Horowitz:2012gs}
G.~T. Horowitz, J.~E. Santos, and D.~Tong, ``{Further Evidence for
  Lattice-Induced Scaling},''
  \href{http://dx.doi.org/10.1007/JHEP11(2012)102}{{\em JHEP} {\bfseries 1211}
  (2012) 102},
\href{http://arxiv.org/abs/1209.1098}{{\ttfamily arXiv:1209.1098 [hep-th]}}.
%%CITATION = ARXIV:1209.1098;%%.

\bibitem{Horowitz:2013jaa}
G.~T. Horowitz and J.~E. Santos, ``{General Relativity and the Cuprates},''
\href{http://arxiv.org/abs/1302.6586}{{\ttfamily arXiv:1302.6586 [hep-th]}}.
%%CITATION = ARXIV:1302.6586;%%.

\bibitem{Donos:2012js}
A.~Donos and S.~A. Hartnoll, ``{Interaction-driven localization in
  holography},'' \href{http://dx.doi.org/10.1038/nphys2701}{{\em Nature Phys.}
  {\bfseries 9} (2013) 649--655},
\href{http://arxiv.org/abs/1212.2998}{{\ttfamily arXiv:1212.2998}}.
%%CITATION = ARXIV:1212.2998;%%.

\bibitem{Ling:2013nxa}
Y.~Ling, C.~Niu, J.-P. Wu, and Z.-Y. Xian, ``{Holographic Lattice in
  Einstein-Maxwell-Dilaton Gravity},''
  \href{http://dx.doi.org/10.1007/JHEP11(2013)006}{{\em JHEP} {\bfseries 1311}
  (2013) 006},
\href{http://arxiv.org/abs/1309.4580}{{\ttfamily arXiv:1309.4580 [hep-th]}}.
%%CITATION = ARXIV:1309.4580;%%.

\bibitem{Chesler:2013qla}
P.~Chesler, A.~Lucas, and S.~Sachdev, ``{Conformal field theories in a periodic
  potential: results from holography and field theory},''
  \href{http://dx.doi.org/10.1103/PhysRevD.89.026005}{{\em Phys.Rev.}
  {\bfseries D89} (2014) 026005},
\href{http://arxiv.org/abs/1308.0329}{{\ttfamily arXiv:1308.0329 [hep-th]}}.
%%CITATION = ARXIV:1308.0329;%%.

\bibitem{Donos:2013eha}
A.~Donos and J.~P. Gauntlett, ``{Holographic Q-lattices},''
  \href{http://dx.doi.org/10.1007/JHEP04(2014)040}{{\em JHEP} {\bfseries 1404}
  (2014) 040},
\href{http://arxiv.org/abs/1311.3292}{{\ttfamily arXiv:1311.3292 [hep-th]}}.
%%CITATION = ARXIV:1311.3292;%%.

\bibitem{Andrade:2013gsa}
T.~Andrade and B.~Withers, ``{A simple holographic model of momentum
  relaxation},'' \href{http://dx.doi.org/10.1007/JHEP05(2014)101}{{\em JHEP}
  {\bfseries 1405} (2014) 101},
\href{http://arxiv.org/abs/1311.5157}{{\ttfamily arXiv:1311.5157 [hep-th]}}.
%%CITATION = ARXIV:1311.5157;%%.

\bibitem{Donos:2014uba}
A.~Donos and J.~P. Gauntlett, ``{Novel metals and insulators from
  holography},'' \href{http://dx.doi.org/10.1007/JHEP06(2014)007}{{\em JHEP}
  {\bfseries 1406} (2014) 007},
\href{http://arxiv.org/abs/1401.5077}{{\ttfamily arXiv:1401.5077 [hep-th]}}.
%%CITATION = ARXIV:1401.5077;%%.

\bibitem{Balasubramanian:2013yqa}
K.~Balasubramanian and C.~P. Herzog, ``{Losing Forward Momentum
  Holographically},''
  \href{http://dx.doi.org/10.1088/0264-9381/31/12/125010}{{\em
  Class.Quant.Grav.} {\bfseries 31} (2014) 125010},
\href{http://arxiv.org/abs/1312.4953}{{\ttfamily arXiv:1312.4953 [hep-th]}}.
%%CITATION = ARXIV:1312.4953;%%.

\bibitem{Donos:2014cya}
A.~Donos and J.~P. Gauntlett, ``{Thermoelectric DC conductivities from black
  hole horizons},''
\href{http://arxiv.org/abs/1406.4742}{{\ttfamily arXiv:1406.4742 [hep-th]}}.
%%CITATION = ARXIV:1406.4742;%%.

\bibitem{Iqbal:2008by}
N.~Iqbal and H.~Liu, ``{Universality of the hydrodynamic limit in AdS/CFT and
  the membrane paradigm},''
  \href{http://dx.doi.org/10.1103/PhysRevD.79.025023}{{\em Phys.Rev.}
  {\bfseries D79} (2009) 025023},
\href{http://arxiv.org/abs/0809.3808}{{\ttfamily arXiv:0809.3808 [hep-th]}}.
%%CITATION = ARXIV:0809.3808;%%.

\bibitem{Davison:2013jba}
R.~A. Davison, ``{Momentum relaxation in holographic massive gravity},''
  \href{http://dx.doi.org/10.1103/PhysRevD.88.086003}{{\em Phys.Rev.}
  {\bfseries D88} (2013) 086003},
\href{http://arxiv.org/abs/1306.5792}{{\ttfamily arXiv:1306.5792 [hep-th]}}.
%%CITATION = ARXIV:1306.5792;%%.

\bibitem{Blake:2013bqa}
M.~Blake and D.~Tong, ``{Universal Resistivity from Holographic Massive
  Gravity},'' \href{http://dx.doi.org/10.1103/PhysRevD.88.106004}{{\em
  Phys.Rev.} {\bfseries D88} (2013) 106004},
\href{http://arxiv.org/abs/1308.4970}{{\ttfamily arXiv:1308.4970 [hep-th]}}.
%%CITATION = ARXIV:1308.4970;%%.

\bibitem{Blake:2013owa}
M.~Blake, D.~Tong, and D.~Vegh, ``{Holographic Lattices Give the Graviton a
  Mass},'' \href{http://dx.doi.org/10.1103/PhysRevLett.112.071602}{{\em
  Phys.Rev.Lett.} {\bfseries 112} (2014) 071602},
\href{http://arxiv.org/abs/1310.3832}{{\ttfamily arXiv:1310.3832 [hep-th]}}.
%%CITATION = ARXIV:1310.3832;%%.

\bibitem{Davison:2013txa}
R.~A. Davison, K.~Schalm, and J.~Zaanen, ``{Holographic duality and the
  resistivity of strange metals},''
  \href{http://dx.doi.org/10.1103/PhysRevB.89.245116}{{\em Phys.Rev.}
  {\bfseries B89} (2014) 245116},
\href{http://arxiv.org/abs/1311.2451}{{\ttfamily arXiv:1311.2451 [hep-th]}}.
%%CITATION = ARXIV:1311.2451;%%.

\bibitem{Gouteraux:2014hca}
B.~Gout\'eraux, ``{Charge transport in holography with momentum dissipation},''
  \href{http://dx.doi.org/10.1007/JHEP04(2014)181}{{\em JHEP} {\bfseries 1404}
  (2014) 181},
\href{http://arxiv.org/abs/1401.5436}{{\ttfamily arXiv:1401.5436 [hep-th]}}.
%%CITATION = ARXIV:1401.5436;%%.

\bibitem{Mefford:2014gia}
E.~Mefford and G.~T. Horowitz, ``{Simple holographic insulator},''
  \href{http://dx.doi.org/10.1103/PhysRevD.90.084042}{{\em Phys.Rev.}
  {\bfseries D90} no.~8, (2014) 084042},
\href{http://arxiv.org/abs/1406.4188}{{\ttfamily arXiv:1406.4188 [hep-th]}}.
%%CITATION = ARXIV:1406.4188;%%.

\bibitem{Taylor:2014tka}
M.~Taylor and W.~Woodhead, ``{Inhomogeneity simplified},''
\href{http://arxiv.org/abs/1406.4870}{{\ttfamily arXiv:1406.4870 [hep-th]}}.
%%CITATION = ARXIV:1406.4870;%%.

\bibitem{Blake:2014yla}
M.~Blake and A.~Donos, ``{Quantum Critical Transport and the Hall Angle},''
\href{http://arxiv.org/abs/1406.1659}{{\ttfamily arXiv:1406.1659 [hep-th]}}.
%%CITATION = ARXIV:1406.1659;%%.

\bibitem{Donos:2014oha}
A.~Donos, B.~Goutéraux, and E.~Kiritsis, ``{Holographic Metals and Insulators
  with Helical Symmetry},''
  \href{http://dx.doi.org/10.1007/JHEP09(2014)038}{{\em JHEP} {\bfseries 1409}
  (2014) 038},
\href{http://arxiv.org/abs/1406.6351}{{\ttfamily arXiv:1406.6351 [hep-th]}}.
%%CITATION = ARXIV:1406.6351;%%.

\bibitem{Amoretti:2014mma}
A.~Amoretti, A.~Braggio, N.~Maggiore, N.~Magnoli, and D.~Musso, ``{Analytic DC
  thermo-electric conductivities in holography with massive gravitons},''
\href{http://arxiv.org/abs/1407.0306}{{\ttfamily arXiv:1407.0306 [hep-th]}}.
%%CITATION = ARXIV:1407.0306;%%.

\bibitem{MRS:7962323}
T.~M. Tritt and M.~A. Subramanian, ``Thermoelectric materials, phenomena, and
  applications: A bird's eye view,''
  \href{http://dx.doi.org/10.1557/mrs2006.44}{{\em MRS Bulletin} {\bfseries 31}
  (3, 2006) 188--198}.
  \url{http://journals.cambridge.org/article_S088376940000991X}.

\bibitem{Gunnarsson:2003zz}
O.~Gunnarsson, M.~Calandra, and J.~Han, ``{Colloquium: Saturation of electrical
  resistivity},'' \href{http://dx.doi.org/10.1103/RevModPhys.75.1085}{{\em
  Rev.Mod.Phys.} {\bfseries 75} (2003) 1085--1099},
\href{http://arxiv.org/abs/cond-mat/0305412}{{\ttfamily arXiv:cond-mat/0305412
  [cond-mat.str-el]}}.
%%CITATION = RMPHA,75,1085;%%.

\bibitem{takmir}
N.~E. Hussey, K.~Takenaka, and H.~Takagi, ``{Universality of the
  Mott-Ioffe-Regel limit in metals},'' {\em Phil. Mag.} {\bfseries 84} (2004)
  2847--2864,
\href{http://arxiv.org/abs/cond-mat/0404263}{{\ttfamily arXiv:cond-mat/0404263
  [cond-mat.str-el]}}.
%%CITATION = RMPHA,75,1085;%%.

\bibitem{2003Natur.425..271M}
D.~v.~d. {Marel}, H.~J.~A. {Molegraaf}, J.~{Zaanen}, Z.~{Nussinov},
  F.~{Carbone}, A.~{Damascelli}, H.~{Eisaki}, M.~{Greven}, P.~H. {Kes}, and
  M.~{Li}, ``{Quantum critical behaviour in a high-T$_{c}$ superconductor},''
  \href{http://dx.doi.org/10.1038/nature01978}{{\em Nature} {\bfseries 425}
  (Sept., 2003) 271--274},
  \href{http://arxiv.org/abs/arXiv:cond-mat/0309172}{{\ttfamily
  arXiv:cond-mat/0309172}}.

\bibitem{2006AnPhy.321.1716V}
D.~{van der Marel}, F.~{Carbone}, A.~B. {Kuzmenko}, and E.~{Giannini},
  ``{Scaling properties of the optical conductivity of Bi-based cuprates},''
  \href{http://dx.doi.org/10.1016/j.aop.2006.04.012}{{\em Annals of Physics}
  {\bfseries 321} (July, 2006) 1716--1729},
  \href{http://arxiv.org/abs/arXiv:cond-mat/0604037}{{\ttfamily
  arXiv:cond-mat/0604037}}.

\bibitem{Hartnoll:2012rj}
S.~A. Hartnoll and D.~M. Hofman, ``{Locally Critical Resistivities from Umklapp
  Scattering},'' \href{http://dx.doi.org/10.1103/PhysRevLett.108.241601}{{\em
  Phys.Rev.Lett.} {\bfseries 108} (2012) 241601},
\href{http://arxiv.org/abs/1201.3917}{{\ttfamily arXiv:1201.3917 [hep-th]}}.
%%CITATION = ARXIV:1201.3917;%%.

\bibitem{Hartnoll:2014gaa}
S.~A. Hartnoll and J.~E. Santos, ``{Cold planar horizons are floppy},''
\href{http://arxiv.org/abs/1403.4612}{{\ttfamily arXiv:1403.4612 [hep-th]}}.
%%CITATION = ARXIV:1403.4612;%%.

\bibitem{Headrick:2009pv}
M.~Headrick, S.~Kitchen, and T.~Wiseman, ``{A New approach to static numerical
  relativity, and its application to Kaluza-Klein black holes},''
  \href{http://dx.doi.org/10.1088/0264-9381/27/3/035002}{{\em
  Class.Quant.Grav.} {\bfseries 27} (2010) 035002},
\href{http://arxiv.org/abs/0905.1822}{{\ttfamily arXiv:0905.1822 [gr-qc]}}.
%%CITATION = ARXIV:0905.1822;%%.

\bibitem{Gulotta:2010cu}
D.~R. Gulotta, C.~P. Herzog, and M.~Kaminski, ``{Sum Rules from an Extra
  Dimension},'' \href{http://dx.doi.org/10.1007/JHEP01(2011)148}{{\em JHEP}
  {\bfseries 1101} (2011) 148},
\href{http://arxiv.org/abs/1010.4806}{{\ttfamily arXiv:1010.4806 [hep-th]}}.
%%CITATION = ARXIV:1010.4806;%%.

\bibitem{WitczakKrempa:2012gn}
W.~Witczak-Krempa and S.~Sachdev, ``{The quasi-normal modes of quantum
  criticality},'' \href{http://dx.doi.org/10.1103/PhysRevB.86.235115}{{\em
  Phys.Rev.} {\bfseries B86} (2012) 235115},
\href{http://arxiv.org/abs/1210.4166}{{\ttfamily arXiv:1210.4166
  [cond-mat.str-el]}}.
%%CITATION = ARXIV:1210.4166;%%.

\bibitem{Gauntlett:2007ma}
J.~P. Gauntlett and O.~Varela, ``{Consistent Kaluza-Klein Reductions for
  General Supersymmetric AdS Solutions},''
  \href{http://dx.doi.org/10.1103/PhysRevD.76.126007}{{\em Phys. Rev.}
  {\bfseries D76} (2007) 126007},
\href{http://arxiv.org/abs/0707.2315}{{\ttfamily arXiv:0707.2315 [hep-th]}}.
%%CITATION = 0707.2315;%%.

\bibitem{Edalati:2010pn}
M.~Edalati, J.~I. Jottar, and R.~G. Leigh, ``{Holography and the sound of
  criticality},'' \href{http://dx.doi.org/10.1007/JHEP10(2010)058}{{\em JHEP}
  {\bfseries 1010} (2010) 058},
\href{http://arxiv.org/abs/1005.4075}{{\ttfamily arXiv:1005.4075 [hep-th]}}.
%%CITATION = ARXIV:1005.4075;%%.

\bibitem{Adam:2011dn}
A.~Adam, S.~Kitchen, and T.~Wiseman, ``{A numerical approach to finding general
  stationary vacuum black holes},''
  \href{http://dx.doi.org/10.1088/0264-9381/29/16/165002}{{\em
  Class.Quant.Grav.} {\bfseries 29} (2012) 165002},
\href{http://arxiv.org/abs/1105.6347}{{\ttfamily arXiv:1105.6347 [gr-qc]}}.
%%CITATION = ARXIV:1105.6347;%%.

\bibitem{Wiseman:2011by}
T.~Wiseman, ``{Numerical construction of static and stationary black holes},''
\href{http://arxiv.org/abs/1107.5513}{{\ttfamily arXiv:1107.5513 [gr-qc]}}.
%%CITATION = ARXIV:1107.5513;%%.

\bibitem{Figueras:2011va}
P.~Figueras, J.~Lucietti, and T.~Wiseman, ``{Ricci solitons, Ricci flow, and
  strongly coupled CFT in the Schwarzschild Unruh or Boulware vacua},''
  \href{http://dx.doi.org/10.1088/0264-9381/28/21/215018}{{\em
  Class.Quant.Grav.} {\bfseries 28} (2011) 215018},
\href{http://arxiv.org/abs/1104.4489}{{\ttfamily arXiv:1104.4489 [hep-th]}}.
%%CITATION = ARXIV:1104.4489;%%.

\bibitem{Donos:2013cka}
A.~Donos and J.~P. Gauntlett, ``{On the thermodynamics of periodic AdS black
  branes},'' \href{http://dx.doi.org/10.1007/JHEP10(2013)038}{{\em JHEP}
  {\bfseries 1310} (2013) 038},
\href{http://arxiv.org/abs/1306.4937}{{\ttfamily arXiv:1306.4937 [hep-th]}}.
%%CITATION = ARXIV:1306.4937;%%.

\bibitem{Witten:2003ya}
E.~Witten, ``{SL(2,Z) action on three-dimensional conformal field theories with
  Abelian symmetry},''
\href{http://arxiv.org/abs/hep-th/0307041}{{\ttfamily arXiv:hep-th/0307041
  [hep-th]}}.
%%CITATION = HEP-TH/0307041;%%.

\bibitem{Herzog:2007ij}
C.~P. Herzog, P.~Kovtun, S.~Sachdev, and D.~T. Son, ``{Quantum critical
  transport, duality, and M-theory},''
  \href{http://dx.doi.org/10.1103/PhysRevD.75.085020}{{\em Phys.Rev.}
  {\bfseries D75} (2007) 085020},
\href{http://arxiv.org/abs/hep-th/0701036}{{\ttfamily arXiv:hep-th/0701036
  [hep-th]}}.
%%CITATION = HEP-TH/0701036;%%.

\bibitem{Hartnoll:2007ip}
S.~A. Hartnoll and C.~P. Herzog, ``{Ohm's Law at strong coupling: S duality and
  the cyclotron resonance},''
  \href{http://dx.doi.org/10.1103/PhysRevD.76.106012}{{\em Phys. Rev.}
  {\bfseries D76} (2007) 106012},
\href{http://arxiv.org/abs/0706.3228}{{\ttfamily arXiv:0706.3228 [hep-th]}}.
%%CITATION = 0706.3228;%%.

\bibitem{Myers:2010pk}
R.~C. Myers, S.~Sachdev, and A.~Singh, ``{Holographic Quantum Critical
  Transport without Self-Duality},''
  \href{http://dx.doi.org/10.1103/PhysRevD.83.066017}{{\em Phys.Rev.}
  {\bfseries D83} (2011) 066017},
\href{http://arxiv.org/abs/1010.0443}{{\ttfamily arXiv:1010.0443 [hep-th]}}.
%%CITATION = ARXIV:1010.0443;%%.

\bibitem{Jokela:2013hta}
N.~Jokela, G.~Lifschytz, and M.~Lippert, ``{Holographic anyonic
  superfluidity},'' \href{http://dx.doi.org/10.1007/JHEP10(2013)014}{{\em JHEP}
  {\bfseries 1310} (2013) 014},
\href{http://arxiv.org/abs/1307.6336}{{\ttfamily arXiv:1307.6336 [hep-th]}}.
%%CITATION = ARXIV:1307.6336;%%.

\bibitem{Davison:2011uk}
R.~A. Davison and N.~K. Kaplis, ``{Bosonic excitations of the $AdS_4$
  Reissner-Nordstrom black hole},''
  \href{http://dx.doi.org/10.1007/JHEP12(2011)037}{{\em JHEP} {\bfseries 1112}
  (2011) 037},
\href{http://arxiv.org/abs/1111.0660}{{\ttfamily arXiv:1111.0660 [hep-th]}}.
%%CITATION = ARXIV:1111.0660;%%.

\bibitem{foll:2008hs}
S.~A. Hartnoll and C.~P. Herzog, ``{Impure AdS/CFT correspondence},''
  \href{http://dx.doi.org/10.1103/PhysRevD.77.106009}{{\em Phys.Rev.}
  {\bfseries D77} (2008) 106009},
\href{http://arxiv.org/abs/0801.1693}{{\ttfamily arXiv:0801.1693 [hep-th]}}.
%%CITATION = ARXIV:0801.1693;%%.

\bibitem{Adams:2011rj}
A.~Adams and S.~Yaida, ``{Disordered Holographic Systems I: Functional
  Renormalization},''
\href{http://arxiv.org/abs/1102.2892}{{\ttfamily arXiv:1102.2892 [hep-th]}}.
%%CITATION = ARXIV:1102.2892;%%.

\bibitem{Adams:2012yi}
A.~Adams and S.~Yaida, ``{Disordered Holographic Systems II: Marginal Relevance
  of Imperfection},'' \href{http://dx.doi.org/10.1103/PhysRevD.90.046007}{{\em
  Phys.Rev.} {\bfseries D90} (2014) 046007},
\href{http://arxiv.org/abs/1201.6366}{{\ttfamily arXiv:1201.6366 [hep-th]}}.
%%CITATION = ARXIV:1201.6366;%%.

\bibitem{Arean:2013mta}
D.~Arean, A.~Farahi, L.~A. Pando~Zayas, I.~S. Landea, and A.~Scardicchio, ``{A
  Dirty Holographic Superconductor},''
  \href{http://dx.doi.org/10.1103/PhysRevD.89.106003}{{\em Phys.Rev.}
  {\bfseries D89} (2014) 106003},
\href{http://arxiv.org/abs/1308.1920}{{\ttfamily arXiv:1308.1920 [hep-th]}}.
%%CITATION = ARXIV:1308.1920;%%.

\bibitem{Lucas:2014zea}
A.~Lucas, S.~Sachdev, and K.~Schalm, ``{Scale-invariant hyperscaling-violating
  holographic theories and the resistivity of strange metals with random-field
  disorder},'' \href{http://dx.doi.org/10.1103/PhysRevD.89.066018}{{\em
  Phys.Rev.} {\bfseries D89} (2014) 066018},
\href{http://arxiv.org/abs/1401.7993}{{\ttfamily arXiv:1401.7993 [hep-th]}}.
%%CITATION = ARXIV:1401.7993;%%.

\bibitem{Hartnoll:2014cua}
S.~A. Hartnoll and J.~E. Santos, ``{Disordered horizons: Holography of randomly
  disordered fixed points},''
  \href{http://dx.doi.org/10.1103/PhysRevLett.112.231601}{{\em Phys.Rev.Lett.}
  {\bfseries 112} (2014) 231601},
\href{http://arxiv.org/abs/1402.0872}{{\ttfamily arXiv:1402.0872 [hep-th]}}.
%%CITATION = ARXIV:1402.0872;%%.

\bibitem{Arean:2014oaa}
D.~Arean, A.~Farahi, L.~A. Pando~Zayas, I.~S. Landea, and A.~Scardicchio,
  ``{Holographic p-wave Superconductor with Disorder},''
\href{http://arxiv.org/abs/1407.7526}{{\ttfamily arXiv:1407.7526 [hep-th]}}.
%%CITATION = ARXIV:1407.7526;%%.

\bibitem{Withers:2014sja}
B.~Withers, ``{Holographic Checkerboards},''
  \href{http://dx.doi.org/10.1007/JHEP09(2014)102}{{\em JHEP} {\bfseries 1409}
  (2014) 102},
\href{http://arxiv.org/abs/1407.1085}{{\ttfamily arXiv:1407.1085 [hep-th]}}.
%%CITATION = ARXIV:1407.1085;%%.

\bibitem{Balasubramanian:1999re}
V.~Balasubramanian and P.~Kraus, ``{A stress tensor for anti-de Sitter
  gravity},'' \href{http://dx.doi.org/10.1007/s002200050764}{{\em Commun. Math.
  Phys.} {\bfseries 208} (1999) 413--428},
\href{http://arxiv.org/abs/hep-th/9902121}{{\ttfamily arXiv:hep-th/9902121}}.
%%CITATION = HEP-TH/9902121;%%.

\bibitem{mpfrcpp}
P.~Holoborodko, ``Mpfr c++.'' Http://www.holoborodko.com/pavel/mpfr/,
  2008-2014.

\bibitem{SuiteSparse}
T.~Davis {\em et al.}, ``Suitesparse.''
  Http://www.cise.ufl.edu/research/sparse/suitesparse/, 2014.

\bibitem{eigenweb}
G.~Guennebaud, B.~Jacob, {\em et al.}, ``Eigen v3 (developer's branch).''
  Http://eigen.tuxfamily.org, 2014.

\bibitem{Boost}
``Boost c++ libraries.'' Http://www.boost.org/, 2013.

\end{thebibliography}\endgroup
\end{document}